\documentclass[12pt]{iopart}

\usepackage{iopams}

\usepackage[utf8]{inputenc}

\usepackage{natbib}
\usepackage{setstack}
\usepackage[ngerman,english]{babel}
\usepackage{graphicx}
\usepackage[nooneline]{subfigure}
\usepackage{upgreek}
\usepackage{url}
\usepackage{xspace}
\usepackage{paralist}
\usepackage{verbatim}

\newcommand{\bilder}{.}

\newcommand{\hb}{0.485\textwidth}
\newcommand{\db}{0.31\textwidth}

\newcommand{\rs}{r_{\rm S}}
\newcommand{\ud}{\mathrm{d}}
\newcommand{\Ve}{{\boldsymbol{e}}}

\begin{document}

\noindent\today
\bigskip

\title[Sector Models -- Part 1: Curved Spaces and Spacetimes]{%
Sector Models -- A Toolkit for Teaching General Relativity.
Part 1: Curved Spaces and Spacetimes}

\author{C Zahn and U Kraus}

\address{Institut für Physik, Universität Hildesheim, Marienburger
  Platz 22, 31141 Hildesheim, Germany}

\eads{\mailto{corvin.zahn@uni-hildesheim.de},
  \mailto{ute.kraus@uni-hildesheim.de}}

\begin{abstract}

\parindent=0pt

Teaching the general theory of relativity
to high school or undergraduate students
must be based on an approach that is conceptual
rather than mathematical.
In this paper we present
such an approach
that requires no more than elementary mathematics.
The central idea
of this introduction to general relativity
is the use of
so-called sector models.
Sector models describe curved spaces
the Regge calculus way
by 
subdivision into 
blocks with euclidean geometry.
This procedure is similar
to the approximation of a curved surface by flat triangles.
We outline a workshop
for high school and undergraduate students
that introduces the notion of curved space
by means of sector models of black holes.
We further describe 
the extension to sector models
of curved spacetimes.
The spacetime models are
suitable for learners
with a basic knowledge of 
special relativity.
The teaching materials presented in this paper
are available online for teaching purposes,
see
\url{http://www.spacetimetravel.org}.

(Some figures are in colour only in the electronic version.)

\end{abstract}

\maketitle

\section{Introduction}

The general theory of relativity
is one of the two fundamental advancements
in physics in the 20th century,
the other one being quantum theory.
Tested to high accuracy
within the solar system,
the theory is well established.
It belongs to the foundations
of today's physical view of the world;
for the understanding 
of many astrophysical phenomena
it is of major importance.
Also, the theory of relativity
strongly attracts the interest of
the general public,
not least because of its relevance for
understanding exotic phenomena like black holes
and for cosmological questions about 
the beginning and the end of the universe.

However,
teaching this theory
that is so important
and so fascinating to many students
is faced with a basic problem
at the high school and undergraduate levels.
The standard introduction
provides the
necessary mathematical tools,
moves on to the motivation of the
field equations and
to the derivation of analytic
solutions,
and then examines the solutions,
especially with regard to the
paths of particles and light.
The mathematical tools
that this programme is based on
are extensive
and way beyond the elementary mathematics
taught in school.
For high school and beginning undergraduate students
the standard approach is therefore not accessible.
It is likewise out of reach of
a physics minor programme
since there is not enough time
to develop the mathematical background
and then continue all the way to the
astrophysical phenomena.

Consequently, it is a desideratum
to teach general relativity
in a way that is
based on elementary mathematics only.
This objective, already stated by Einstein in 1916
\citep{ein1}
has been pursued in many different ways
both in the development of teaching materials 
and in 
popular science publications.
The following four approaches
are widely-used:

\smallskip

1. Conclusions from the principle of equivalence:
Gravitational light deflection
and time dilation
are introduced 
by means of thought experiments
based on the equivalence principle
\citep[e.g.\ in][]{ein1,gam,sar,tip,stan}.

2. Description of the geometry of curved surfaces:
Using simple curved surfaces,
e.g.\ the surface of a sphere,
geometric concepts that are important for curved spacetimes
are introduced, for instance metric,
geodesics and curvature
\citep[e.g.\ in][]{sar,har,stan,nat}.

3. Computations 
based on Newtonian dynamics:
Newtonian computations
are performed for phenomena
that should in fact
be described in terms of general relativity,
like gravitational light deflection,
black holes and the cosmological
expansion
\citep[e.g.\ in][]{ehl,lot1}.
The 
idea behind this
is to use familiar concepts
like force and energy
in order to 
head straight for
the phenomena
(e.g.\ light deflection, described
as deflection of a classical particle stream)
and on to astrophysical applications
(e.g.\ gravitational lenses).
The conceptual change 
associated with general relativity
is an issue that is not raised
in this approach.

4. Analogies:
Especially in 
popular science publications
analogies are widely used
both to introduce basic concepts
(e.g.\ a ball making a depression in a rubber sheet
to illustrate the concept ``mass curves space'')
and to describe relativistic phenomena
(e.g.: a wine glass base acting 
as a ``gravitational lens'')
\citep[e.g.][]{pri,lot2}.
The use of analogies 
entails a considerable risk
of the formation 
of misconceptions \citep{zah2010}.
\smallskip

Another pictorial
but more sophisticated
description of 
curved spaces and space\-times
uses embedding diagrams.
Embeddings of
two-dimensional subspaces 
in three-di\-men\-sional flat space
have been described both for
the spatial 
\citep{fla1916,eps1994,jon2001,jon2005}
and for the
spatiotemporal
case
\citep{mar1998}.
The significance of embeddings
goes beyond that of analogies
as mentioned in items 2 and 4 above.
Embeddings represent
subspaces of definite spacetimes
so that their geometric properties
have physical significance.
Like the analogies mentioned above,
embeddings are limited to 
two-dimensional subspaces.

This paper presents a novel approach
to teaching general relativity
that like the above-mentioned ones
uses elementary mathematics at most.
Its objective is to convey the basic ideas
of the general theory of relativity
as summarized succinctly by John Wheeler
in his well-known saying \citep{whe}:

\begin{quote}
{\em Spacetime tells matter how to move.\par
Matter tells spacetime how to curve.}%
\end{quote}

This summary shows
that 
the following three fundamental questions should be addressed:

\begin{enumerate}
\item What is a curved spacetime?
\item How does matter move in a curved spacetime?
\item How is the curvature of the spacetime
   linked to the distribution of matter?
\end{enumerate}

The central idea in the approach
presented in this paper is the use of
so-called \emph{sector models}.
A sector model depicts a
two- or three-dimensional subspace
of a curved spacetime.
It is true to scale and
is built according to the
description of spacetimes
in the Regge calculus by way of
subdivision into uncurved blocks
\citep{reg1961}.
In case of a two-dimensional space
the blocks are flat elements of area.
In case of a three-dimensional space
the blocks are bricks, the geometry within
each brick being euclidean.
Spatiotemporal subspaces are represented by
sectors that have internal Minkowski geometry.
To build a physical model, the sectors can be realized
as pieces of paper or as boxes made from cardboard.

Using sector models
the three questions stated above
can be treated in a descriptive way.
The approach is suitable for 
undergraduate students
and can also be used with
advanced high school students.
Since this description of the basic ideas
is closely connected with the standard mathematical presentation,
the approach can also be used
as a complement to a standard textbook
in order to promote a geometric understanding
of curved spacetimes.

This paper addresses the first of the 
three fundamental questions
in more detail and 
shows how the properties of curved space
can be conveyed using sector models.
The Schwarzschild spacetime of a black hole 
will serve as an example for this introduction.
Section~\ref{sec.unterricht} outlines a workshop
introducing the notion of curved space.
The workshop has been held
several times 
for high-school 
and 
for undergraduate students.
The concept of the sector model
is introduced in this section in
a non-technical way.
Details on the computation
of the sector models 
and on their properties
are given in section~\ref{sec.modell},
a spacetime sector model is
presented in section~\ref{sec.raumzeit},
and section~\ref{sec.fazit} gives an outlook
on further applications of sector models.

\section{A do-it-yourself black hole}

\label{sec.unterricht}

This workshop introduces
the three-dimensional curved space
of a black hole. 
The concept of a curved space 
is presented in a descriptive way
in order to promote a qualitative understanding
of the properties of curved spaces.
The black hole is a con\-ven\-ient example for two reasons:
As an exotic object with frequent media coverage
it appeals to many people.
Also, close to a black hole
relativistic effects are sufficiently large
to be clearly visible in a true-to-scale model.

\subsection{Curved surfaces}

\label{sec.gekruemmte-flaechen}

For a start we introduce
the concept of curvature
using curved surfaces by way of example.
In doing so we distinguish 
positive, negative and zero curvature.
The sphere, the saddle and the plane
are introduced as prototypes of the three types
of curved surfaces.

We state the criterion that permits to
determine the type of curvature:
A small piece of the surface is cut out
and flattened on a plane.
If it tears when flattened, the curvature is positive,
if it buckles, the curvature is negative. 
A piece that can be flattened 
without tearing or buckling has zero curvature.
This criterion determines the sign of the
internal (Gaussian) curvature.
For practice it is applied to different
surfaces. Useful examples are in particular
the torus (negative curvature at the inner rim,
positive curvature at the outer rim) to
demonstrate that curvature can vary from point to point,
and the cylinder (zero curvature) to show
that curvature as defined here
is not in perfect agreement 
with the everyday use of the word.

\begin{figure}
  \centering
  \includegraphics[width=\db]{\bilder/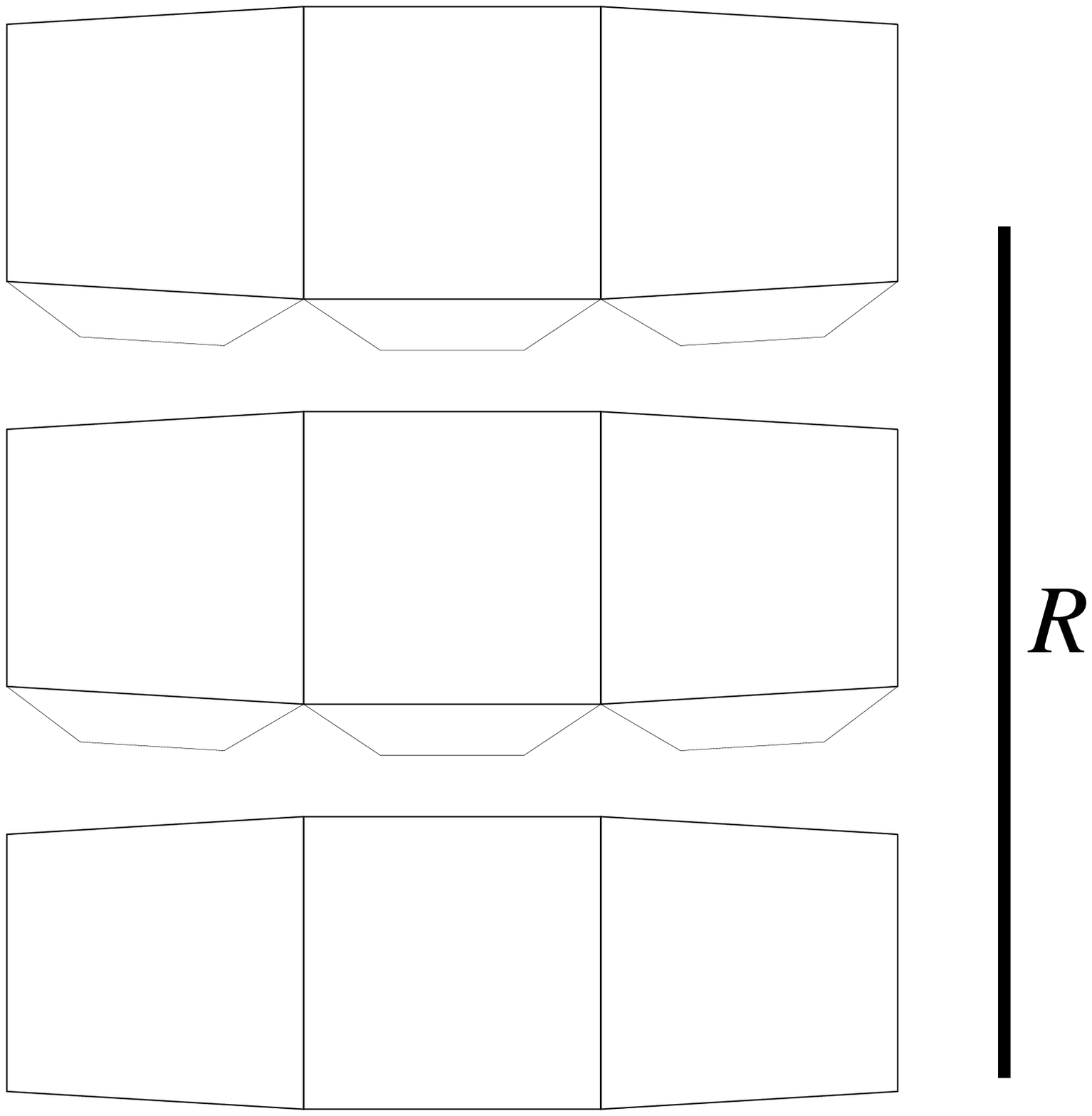}
  \hspace{1cm}
  \includegraphics[width=\db]{\bilder/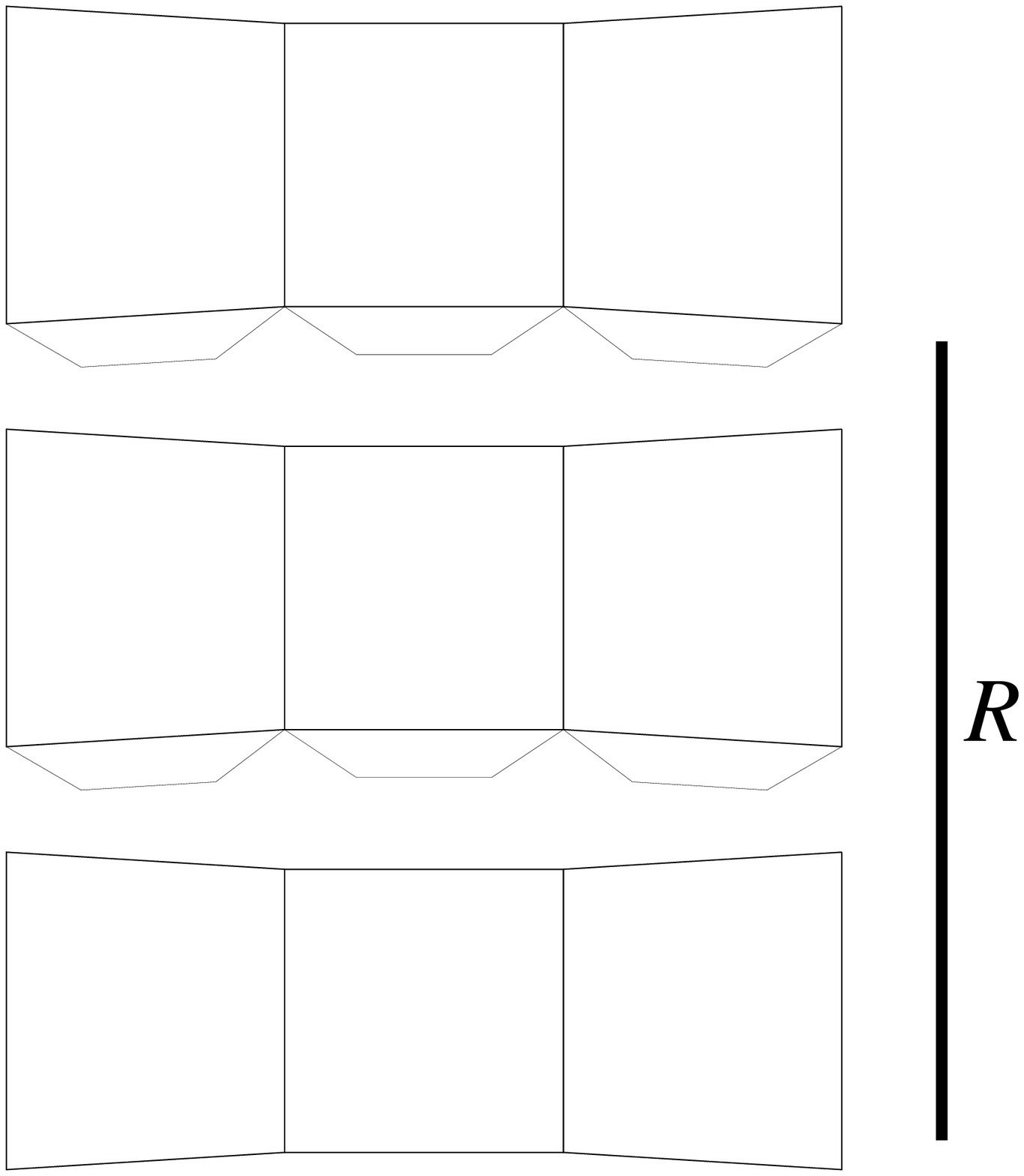}
  \caption{\label{fig.bb-flaeche}
Cut-out sheets for two curved surfaces.
Left: spherical cap with positive curvature
$K=1/R^2$, $R$ being the radius of the sphere.
Right: Saddle with constant negative curvature $K=-1/R^2$.
The length of the scale bar indicates the value of $R$.
}
\end{figure}

\begin{figure}
  \centering \subfigure[]{%
    \includegraphics[width=\hb]{\bilder/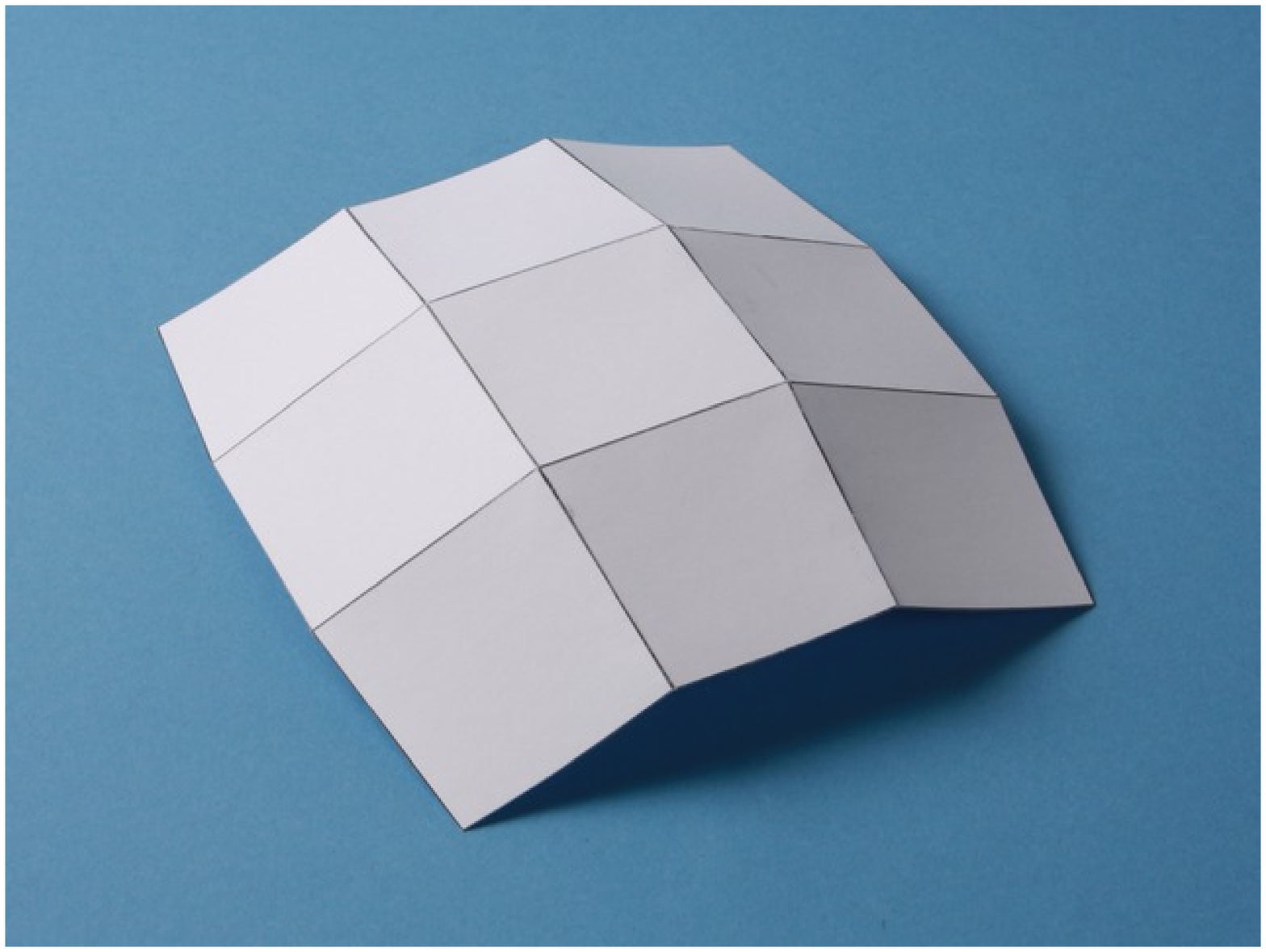}
    \label{fig.flaechepos}}\hfill%
  \subfigure[]{%
    \includegraphics[width=\hb]{\bilder/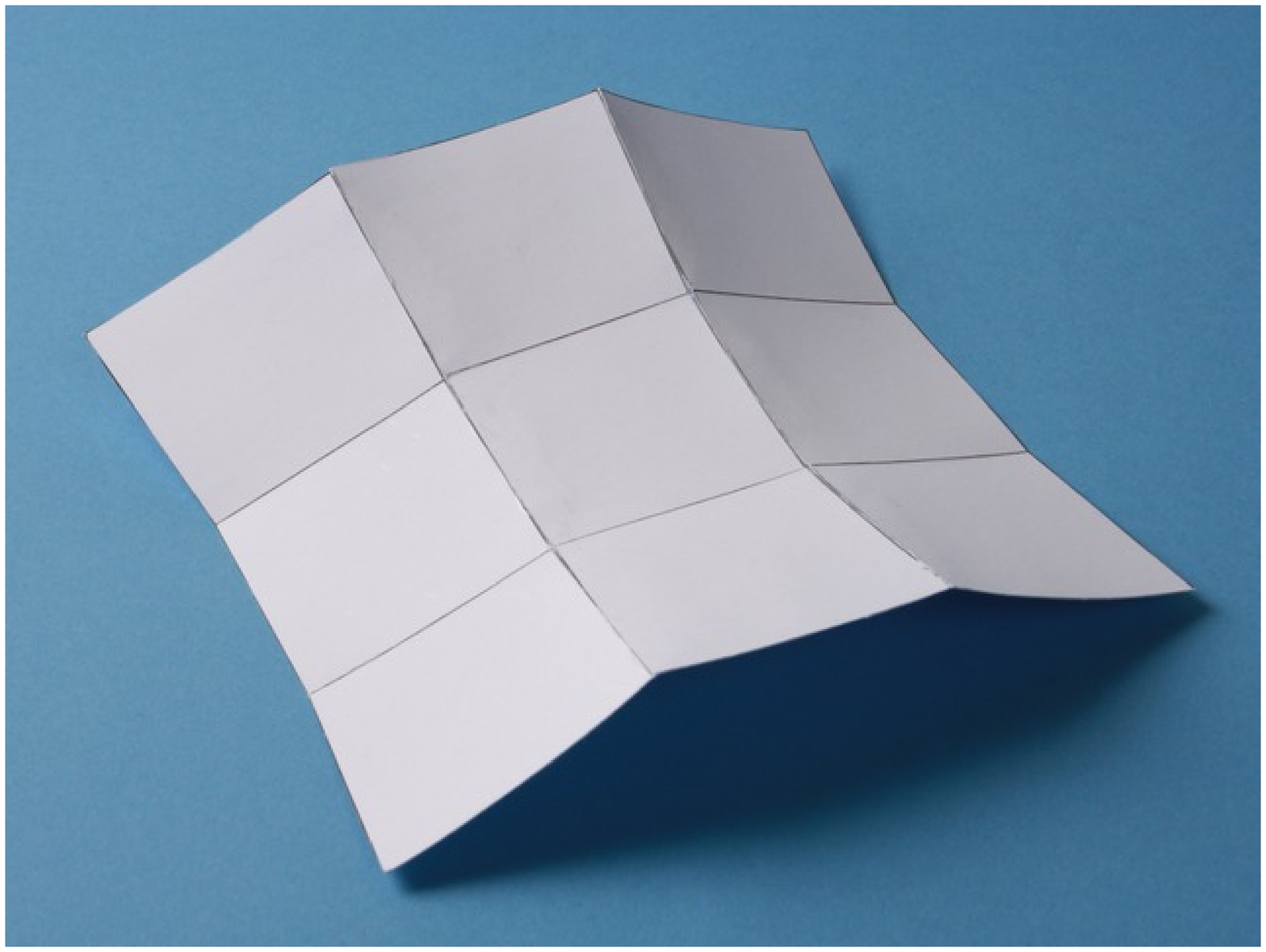}
    \label{fig.flaecheneg}}\\
  \subfigure[]{%
    \includegraphics[width=\hb]{\bilder/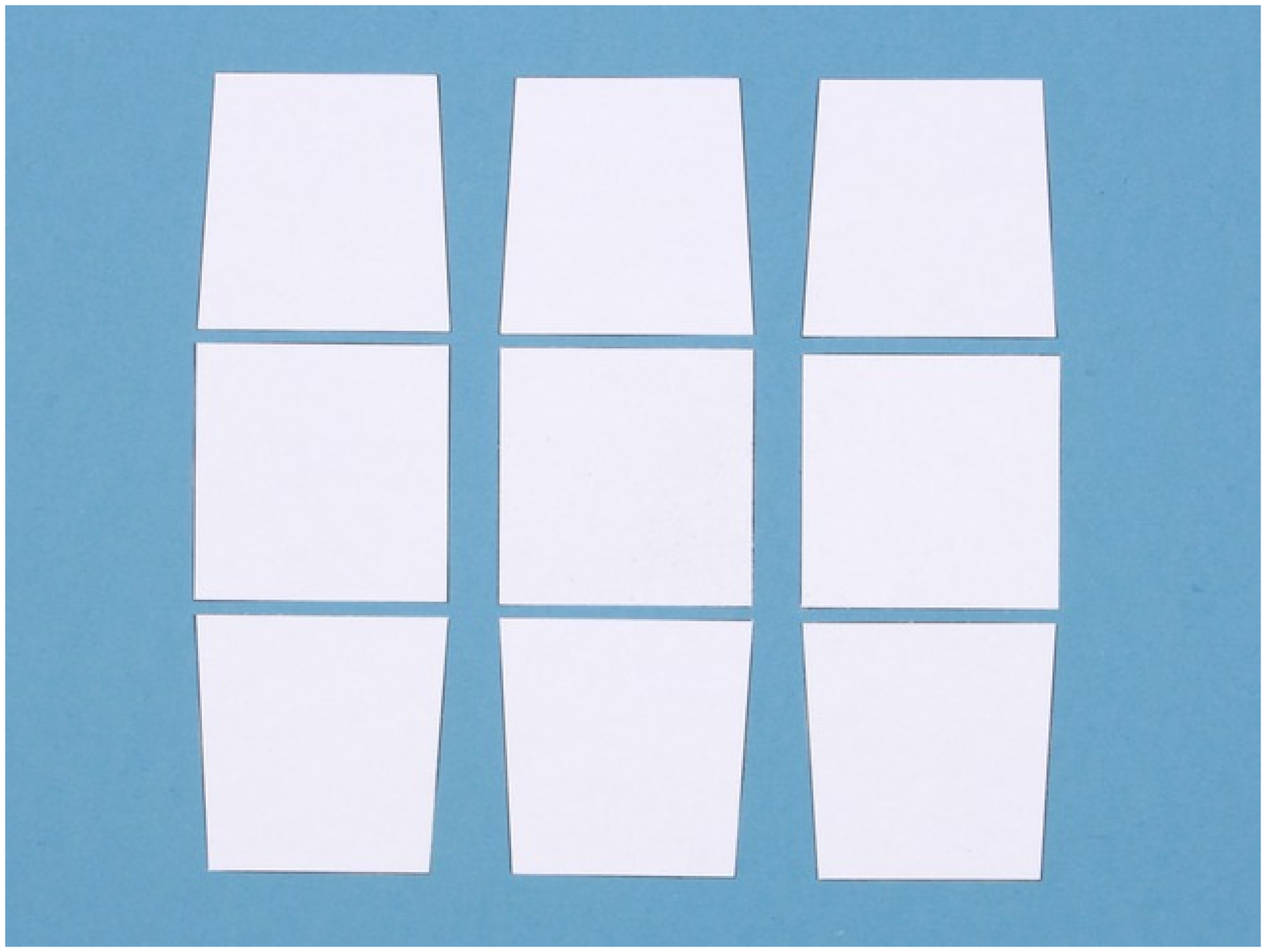}
    \label{fig.flaechepos-sk}}\hfill%
  \subfigure[]{%
    \includegraphics[width=\hb]{\bilder/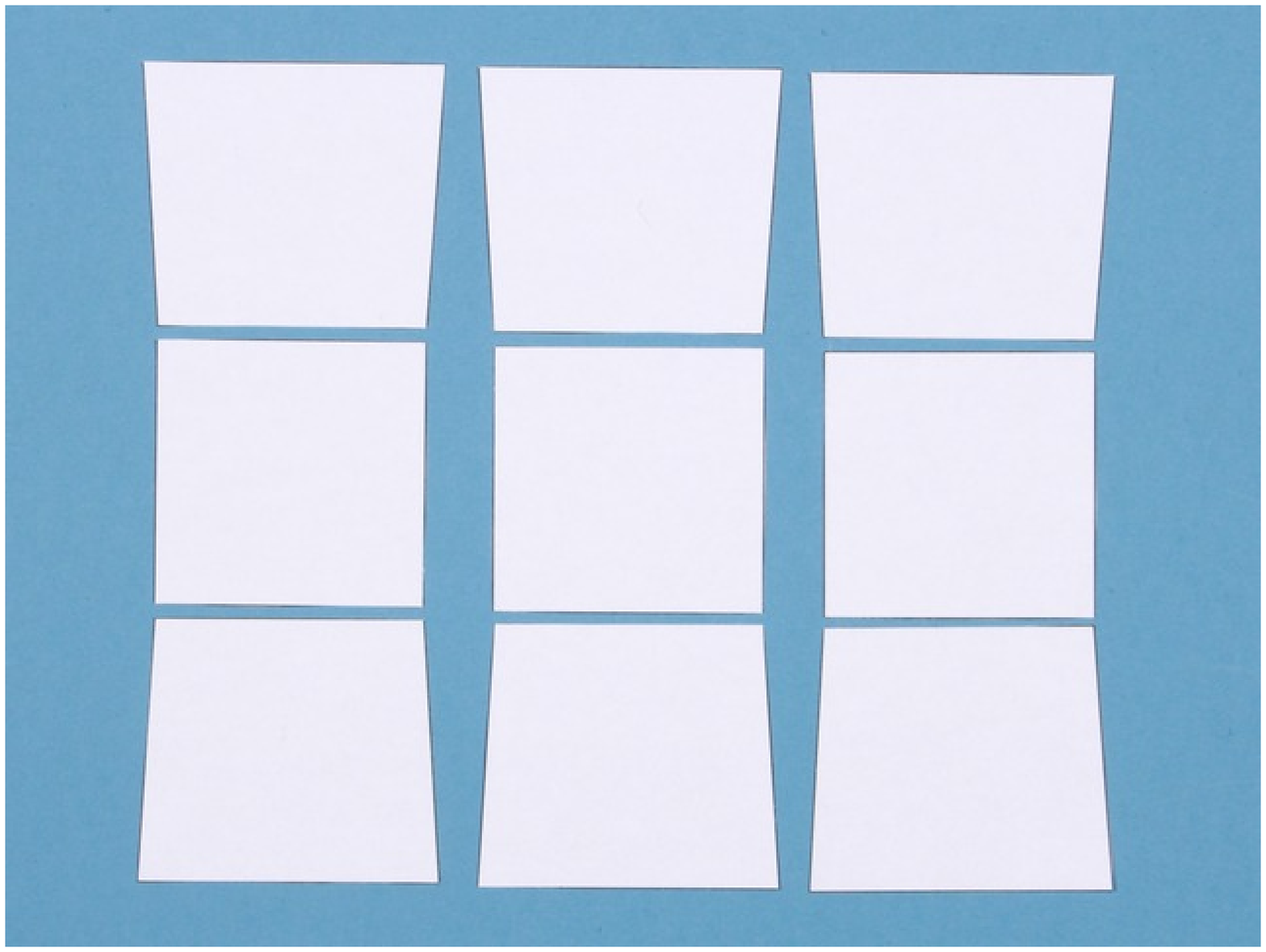}
    \label{fig.flaecheneg-sk}}\\
  \subfigure[]{%
    \includegraphics[width=\hb]{\bilder/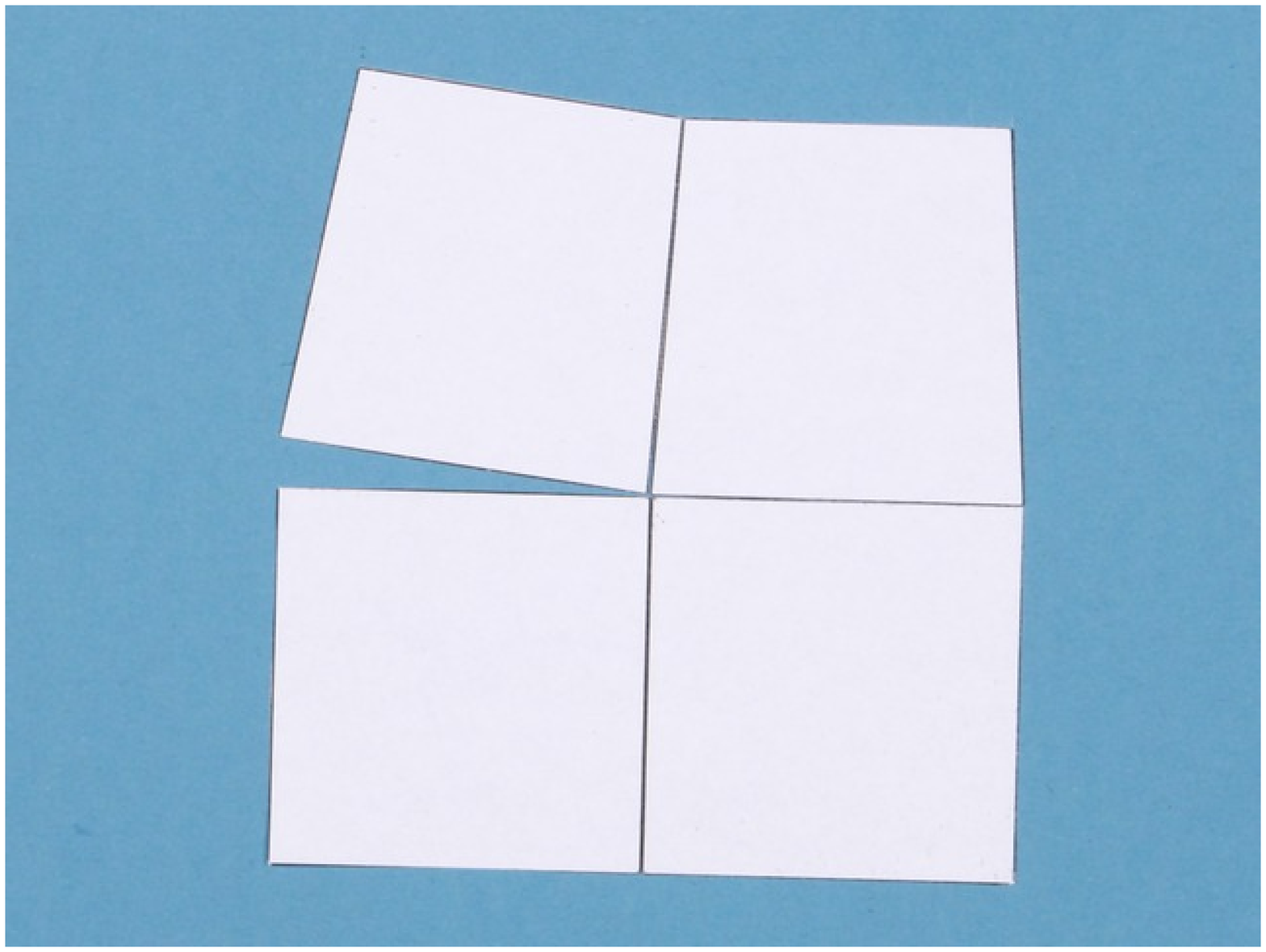}
    \label{fig.flaechepos-def}}\hfill%
  \subfigure[]{%
    \includegraphics[width=\hb]{\bilder/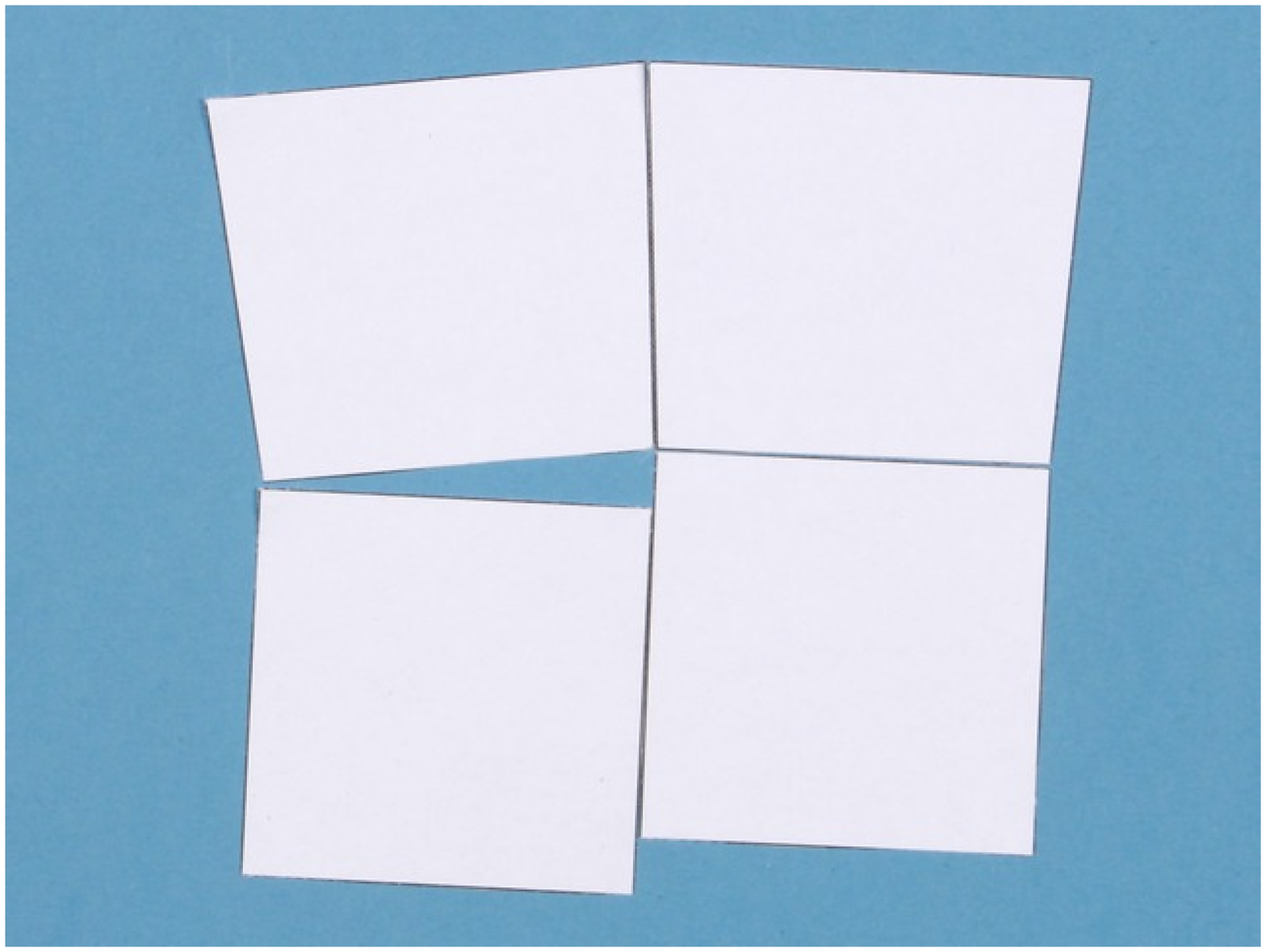}
    \label{fig.flaecheneg-def}}\\
  \caption{\label{fig.flaecheposneg} 
Surfaces with positive curvature (a) and negative curvature (b)
ap\-prox\-i\-mat\-ed by flat pieces.
Their sector models (c, d) indicate
the positive curvature by ``tearing'' (e) and
the negative curvature by ``buckling'' (f), respectively.
}
\end{figure}

The second step illustrates
how a curved surface
can be approximated by small flat pieces.
Using glue and the cut-out sheets 
shown in figure~\ref{fig.bb-flaeche}
one group is given the task to build two surfaces
(figure~\ref{fig.flaechepos}, \ref{fig.flaecheneg})
and to determine the sign of the curvature in each case.
A second group is instructed 
to just cut out the flat pieces without glueing them together,
to lay them out on the table 
(figure~\ref{fig.flaechepos-sk}, \ref{fig.flaecheneg-sk})
and 
to find out the sign of the curvature from this set-up.
The criterion ``tearing'' or ``buckling''
can easily be applied in this case also
(figure~\ref{fig.flaechepos-def}, \ref{fig.flaecheneg-def}).

The aim of the third step
is to accept
the representation of a curved surface 
by pieces laid out in the plane
(figure~\ref{fig.flaechepos-sk}, \ref{fig.flaecheneg-sk})
as a useful and indeed equivalent alternative
to the more familiar picture
of the surface bending into the third dimension
(figure~\ref{fig.flaechepos}, \ref{fig.flaecheneg}).
To prepare for this new point of view
we describe the world view of ``flatlanders'',
the inhabitants of ``Flatland'' in 
Edwin Abbott's tale of the same name \citep{abb1884}:

{\em I call our world Flatland (...).
Image a vast sheet of paper on which (...) figures
(...) move freely about,
but without the power of rising above
or sinking below it,
very much like shadows (...).}

The flatlanders move in two dimensions (forward -- backward,
right -- left), the third dimension (up -- down) 
is not only unaccessible to them,
but is beyond their imagination.
When we extend Abbot's flatland to curved surfaces,
the flatlanders still move only within the surface
forward -- backward and right -- left. 
Lacking the concept of up and down,
they cannot conceive of a surface
bending into an embedding three-dimensional space.
Nevertheless, they are able to study
the curvature of their world.
To do this, they fabricate a model 
similar to the ones shown in 
figures~\ref{fig.flaechepos-sk}, \ref{fig.flaecheneg-sk}:
A tract is subdivided into lots
that are small enough to be approximately flat.
The lengths of the edges of all the lots are measured,
the pieces reproduced on a reduced scale and laid out in the plane
(similar to the arrangement in figures~\ref{fig.flaechepos-sk}
and \ref{fig.flaecheneg-sk}). 
The pieces fit together without gaps
only if the region in question is flat.
Otherwise, the region has curvature
the sign of which can be found out
by applying the criterion ``tearing or buckling?''
at the vertices
(similar to the arrangement in figures~\ref{fig.flaechepos-def}
and \ref{fig.flaecheneg-def}).

\subsection{Curved space}

\label{sec.gekruemmte-raeume}

\begin{figure}
  \centering
  \includegraphics[width=\db]{\bilder/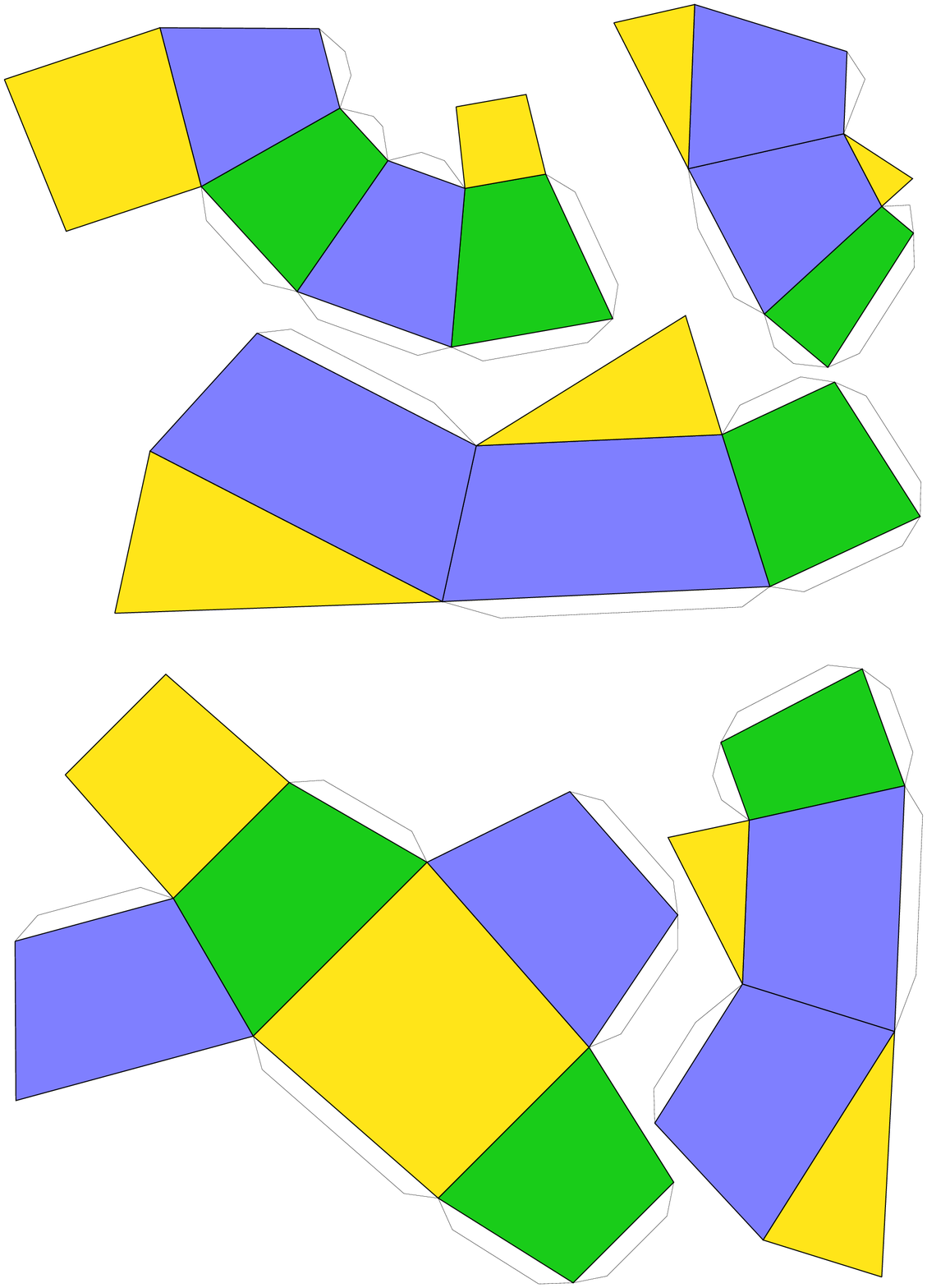}
  \hfill
  \includegraphics[width=\db]{\bilder/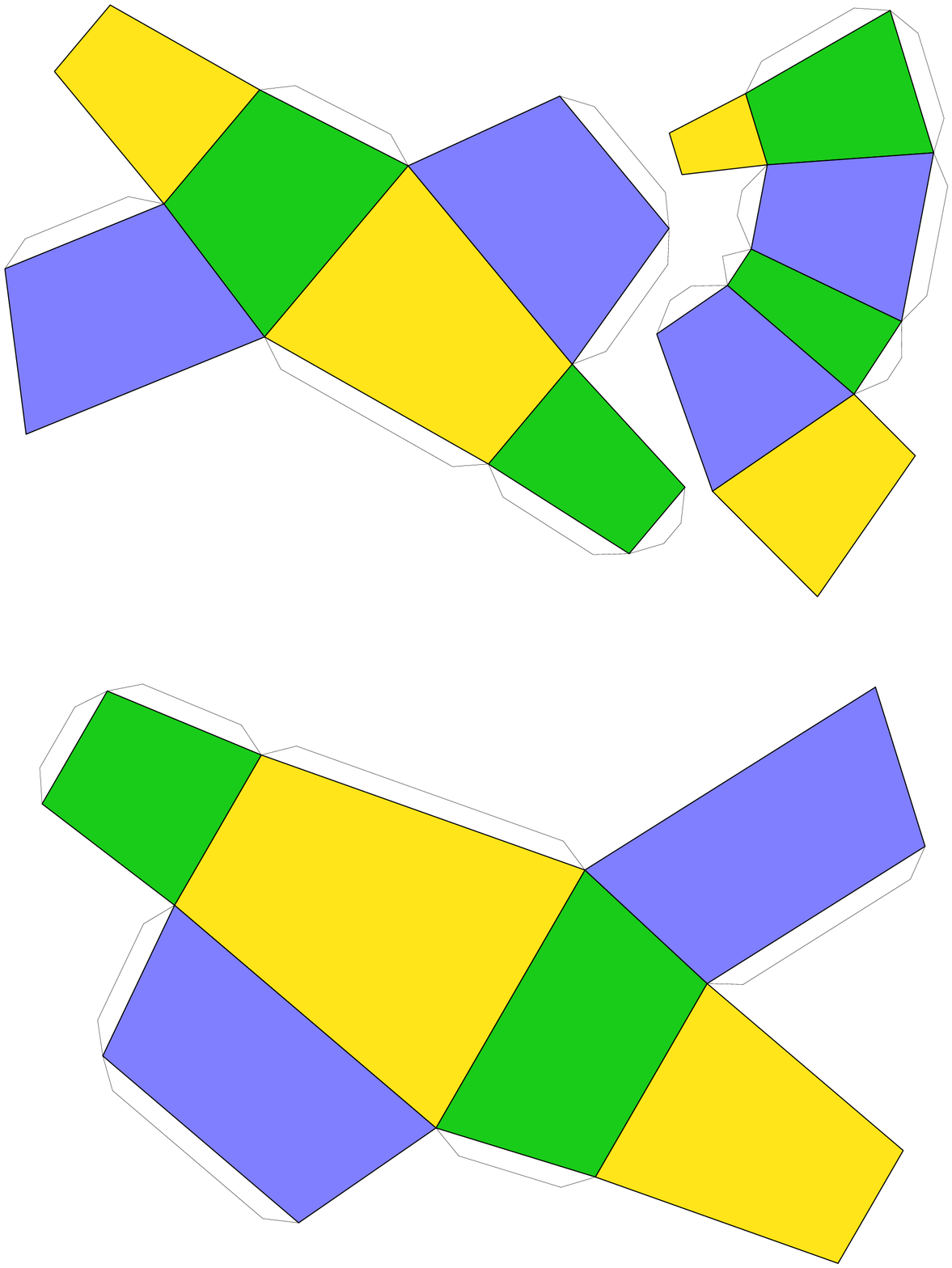}
  \hfill
  \includegraphics[width=\db]{\bilder/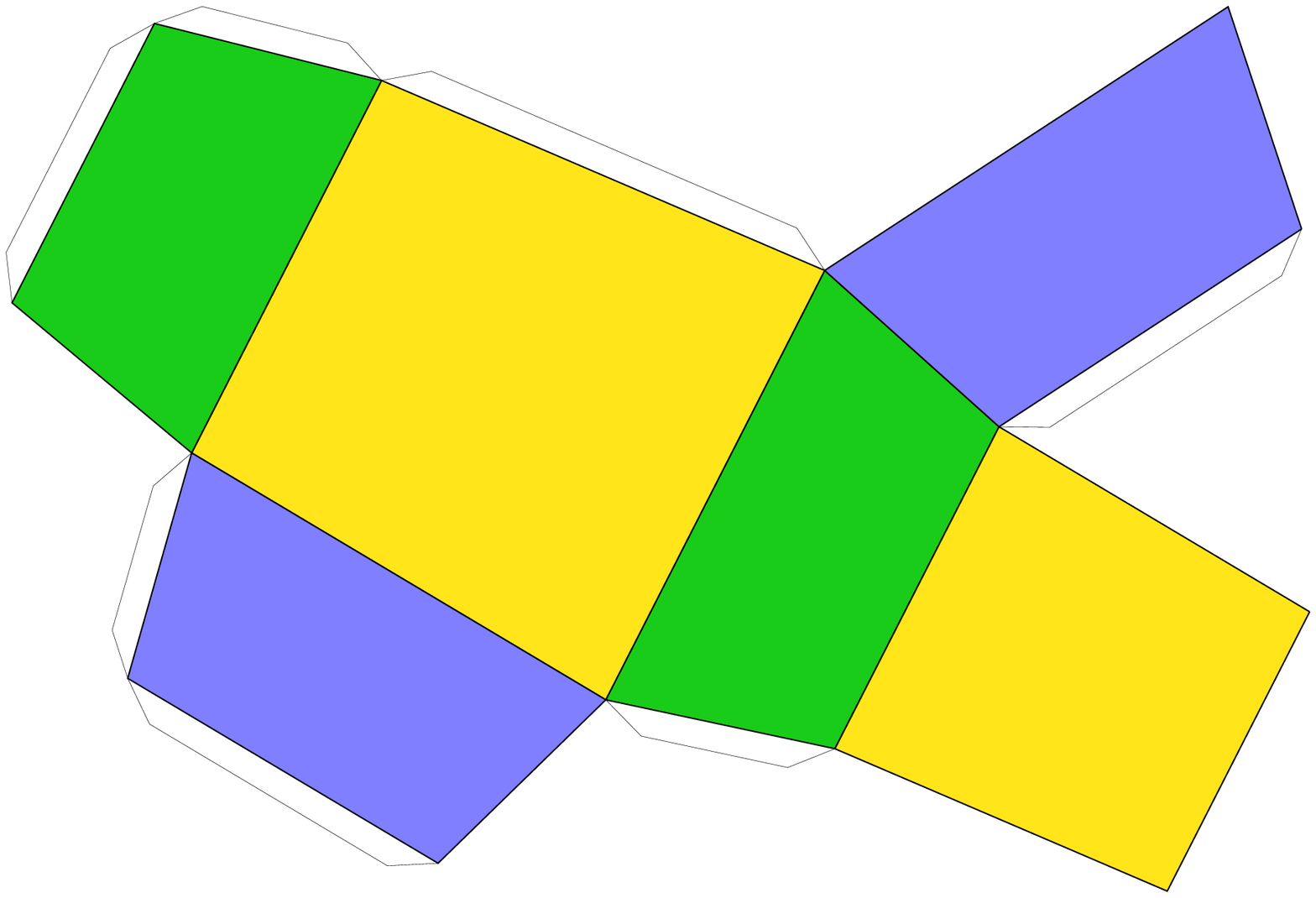}
\\
  \includegraphics[width=\db]{\bilder/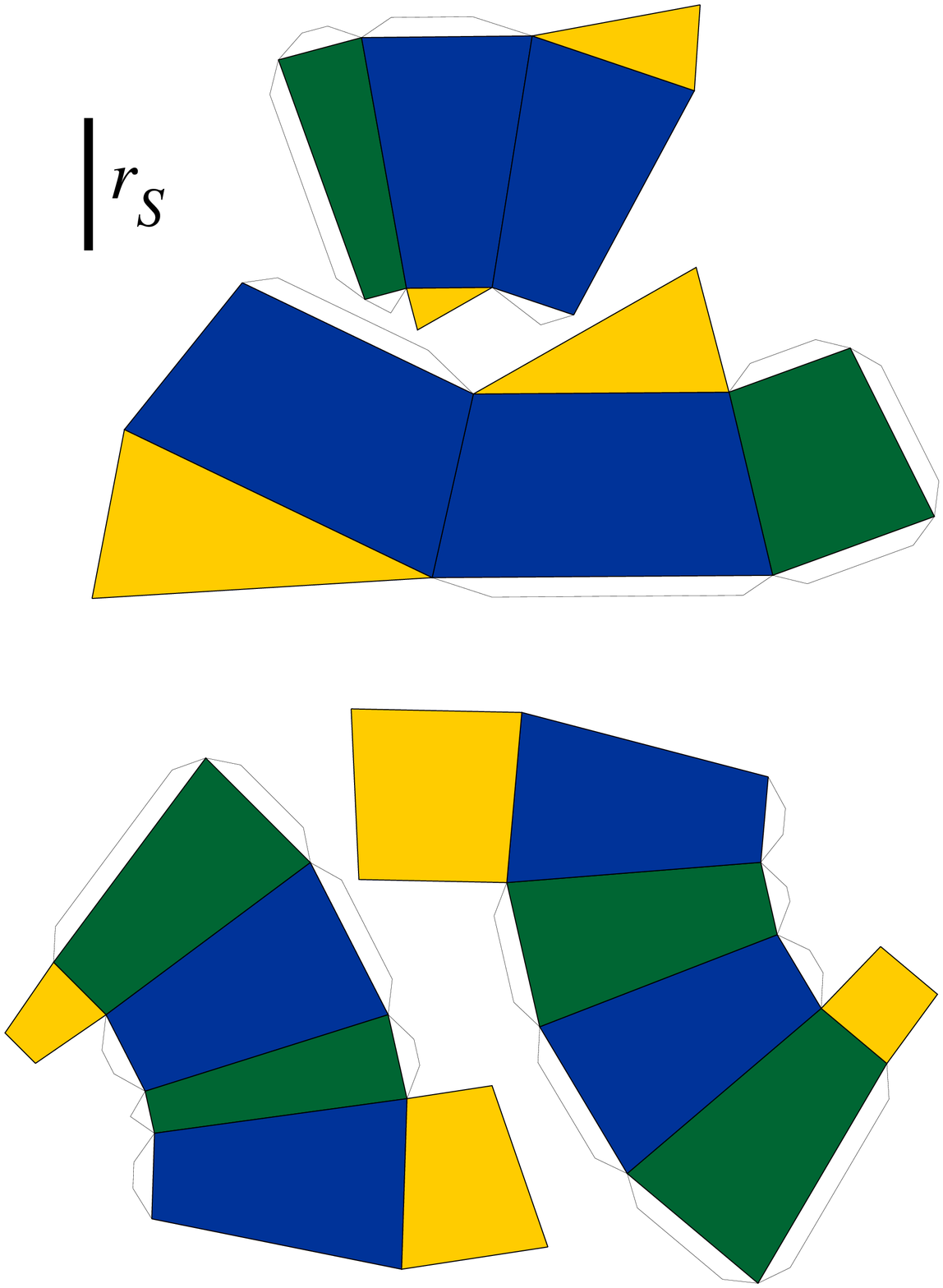}
  \hfill
  \includegraphics[width=\db]{\bilder/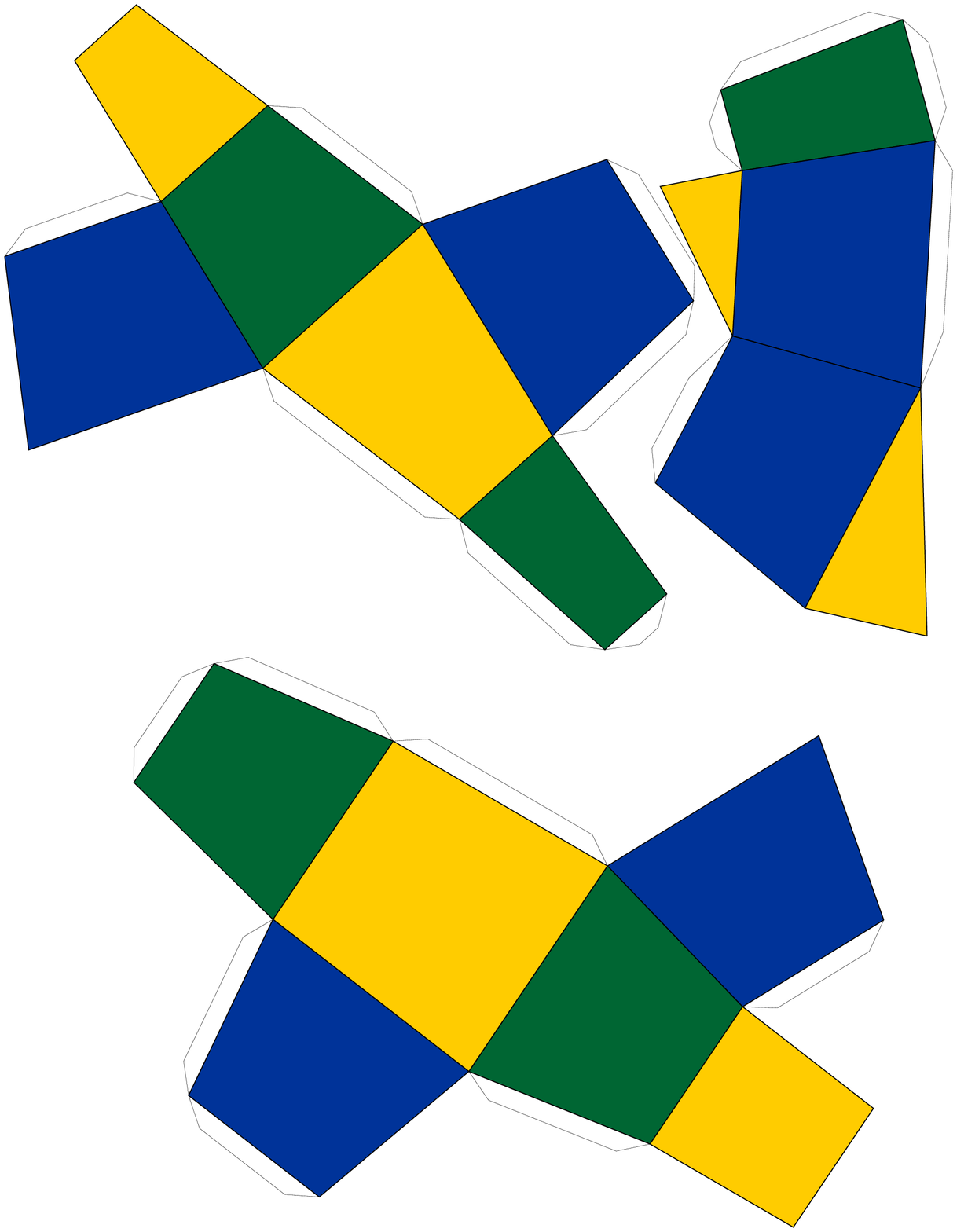}
  \hfill
  \includegraphics[width=\db]{\bilder/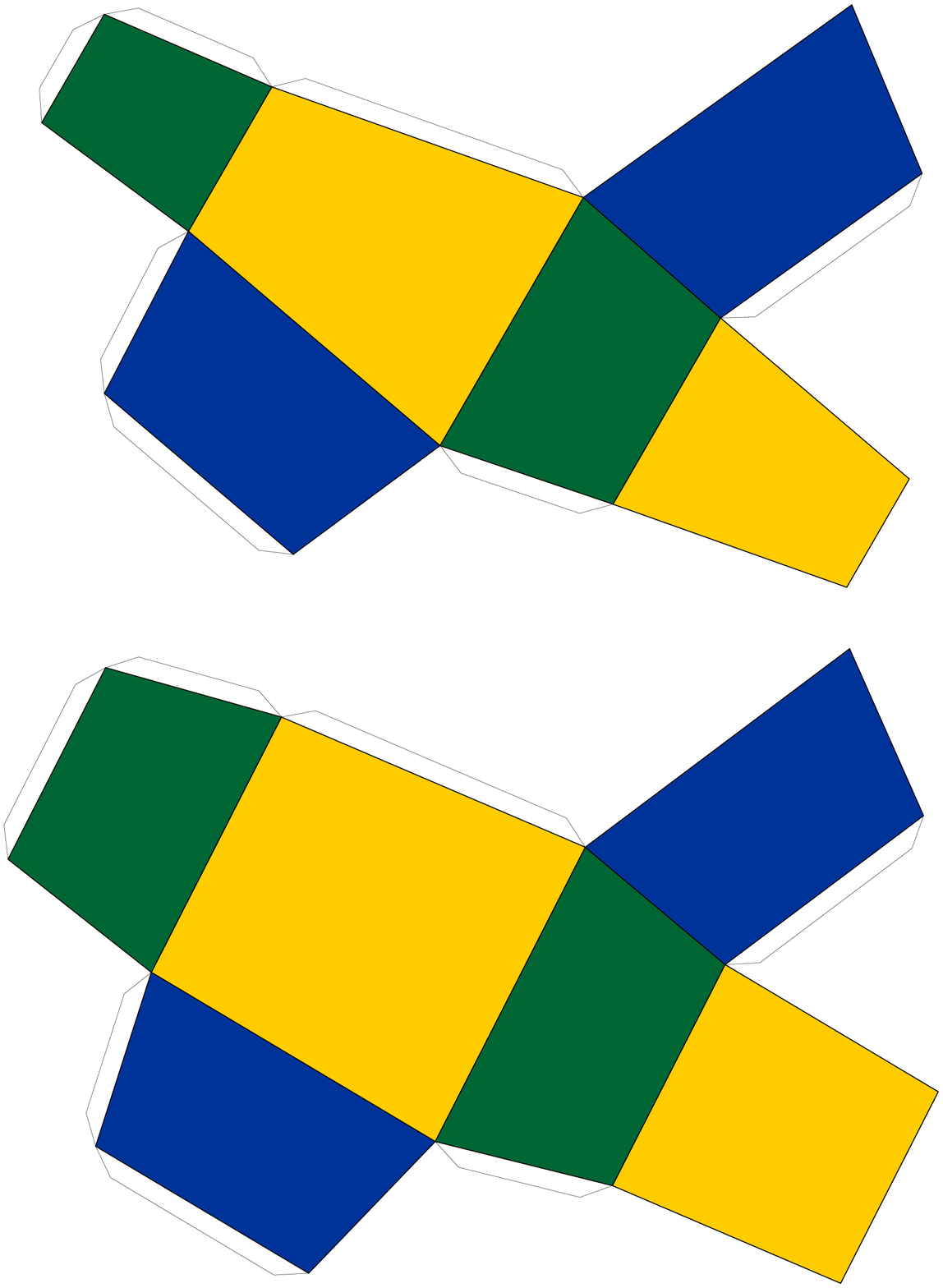}
\\
  \caption{\label{fig.bb-sl-3d}
Cut-out sheets for 
cardboard models 
depicting euclidean space (top) and 
a black hole (bottom), respectively.
The length of the scale bar indicates the Schwarzschild radius $\rs$
of the black hole.
}
\end{figure}

In step four we now study a three-dimensional curved space.
Being ``spacelanders'' familiar with three dimensions
but unable to conceive of a higher dimensional space,
we can examine the curvature of our three-dimensional space
the same way that flatlanders examine curved surfaces:
We subdivide a spatial region into pieces (three-dimensional blocks)
that are small enough to be approximately euclidean.
The lengths of all the edges are measured,
the blocks reproduced on a reduced scale
and laid out in euclidean space. 
If the spatial region has zero curvature,
the blocks can be assembled without gaps.
Otherwise, the model indicates the curvature
of the region.

\begin{figure}
\subfigure[]{%
\includegraphics[width=\hb]{\bilder/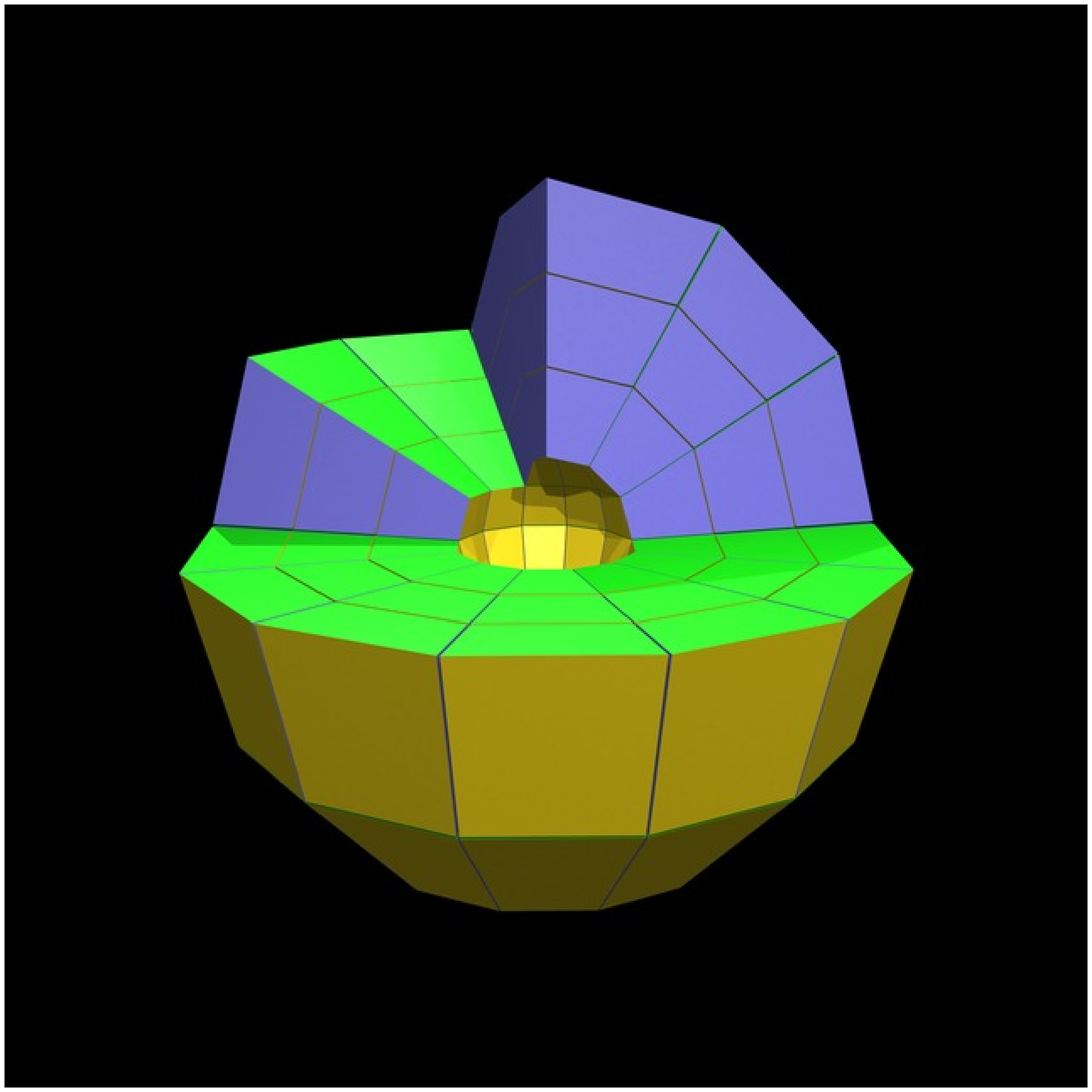}
}%
\hfill
\subfigure[]{%
\includegraphics[width=\hb]{\bilder/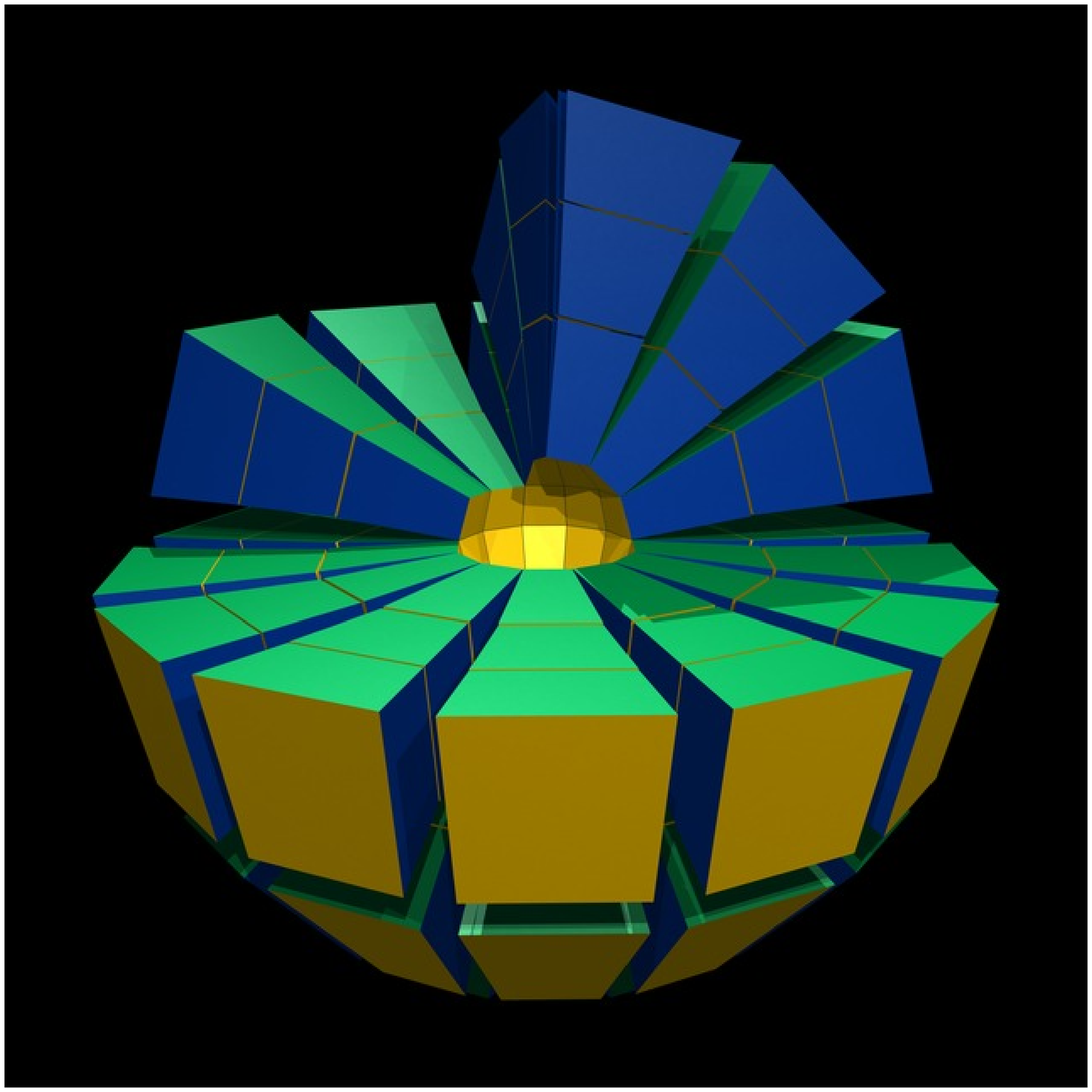}
}
  \caption{\label{fig.kloetze}
Computer generated pictures of the
sector models ``euclidean space'' (a) and
``black hole'' (b).
If a black hole of the appropriate
mass was placed in the centre of the
black hole model,
the blocks with the sizes and shapes as shown here
would fit without gaps.
    }
\end{figure}

Two such models -- one depicting euclidean space,
the other one depicting the curved space around a black hole --
are built by the participants.
\begin{figure}
  \centering
  \subfigure[]{\includegraphics[width=\hb]{\bilder/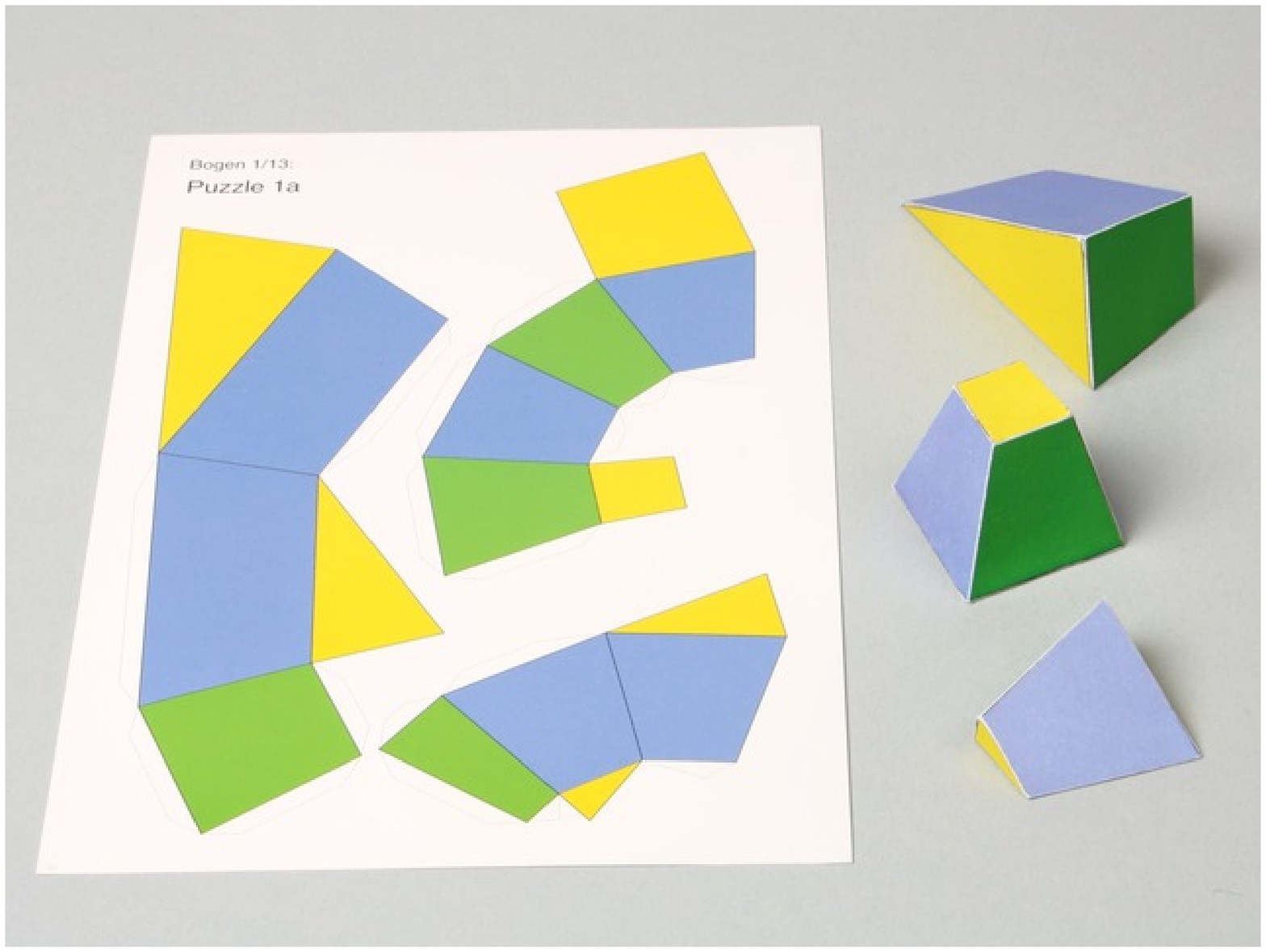}}\hfill
  \subfigure[]{\includegraphics[width=\hb]{\bilder/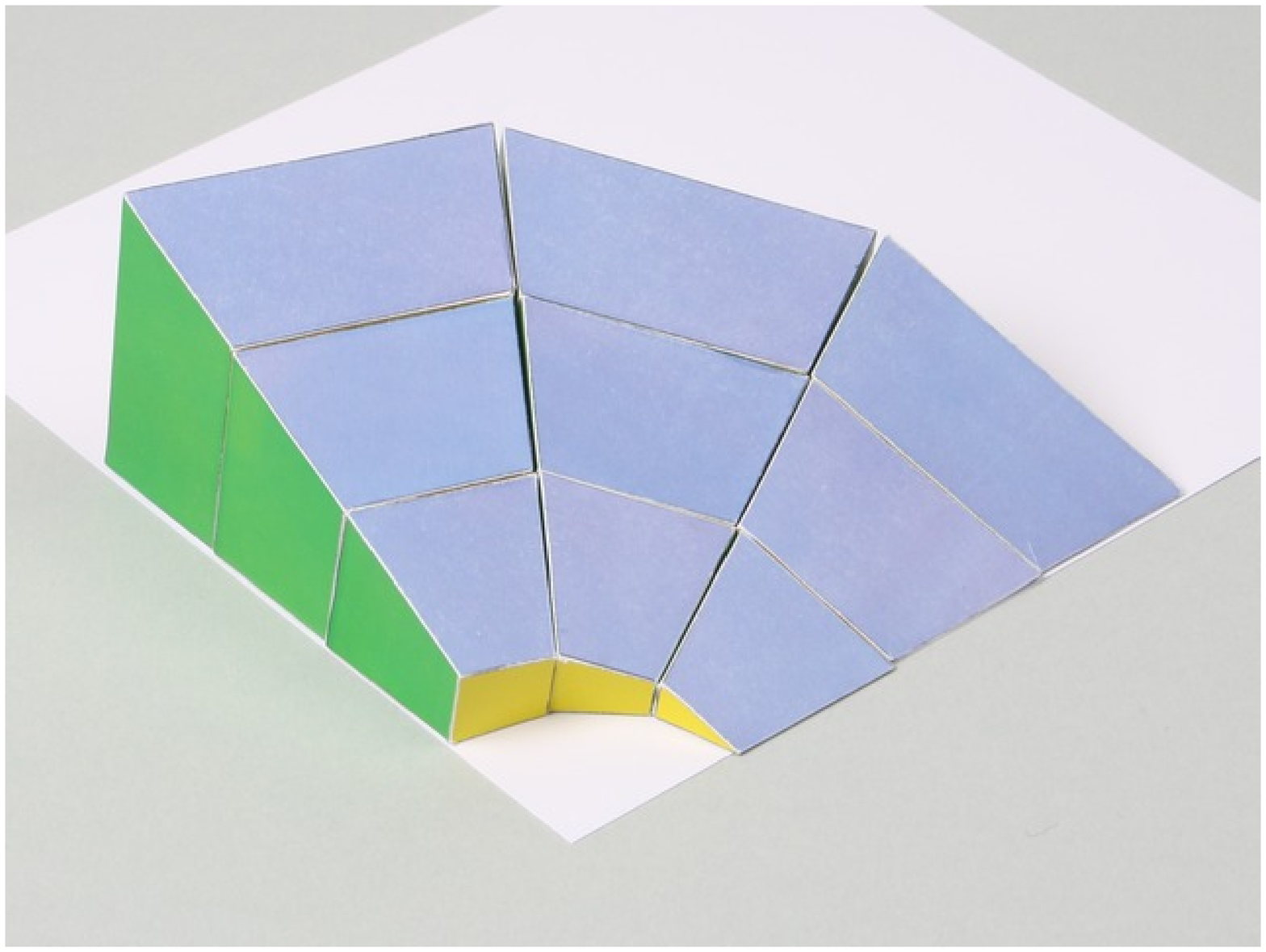}}\\
  \subfigure[]{\includegraphics[width=\hb]{\bilder/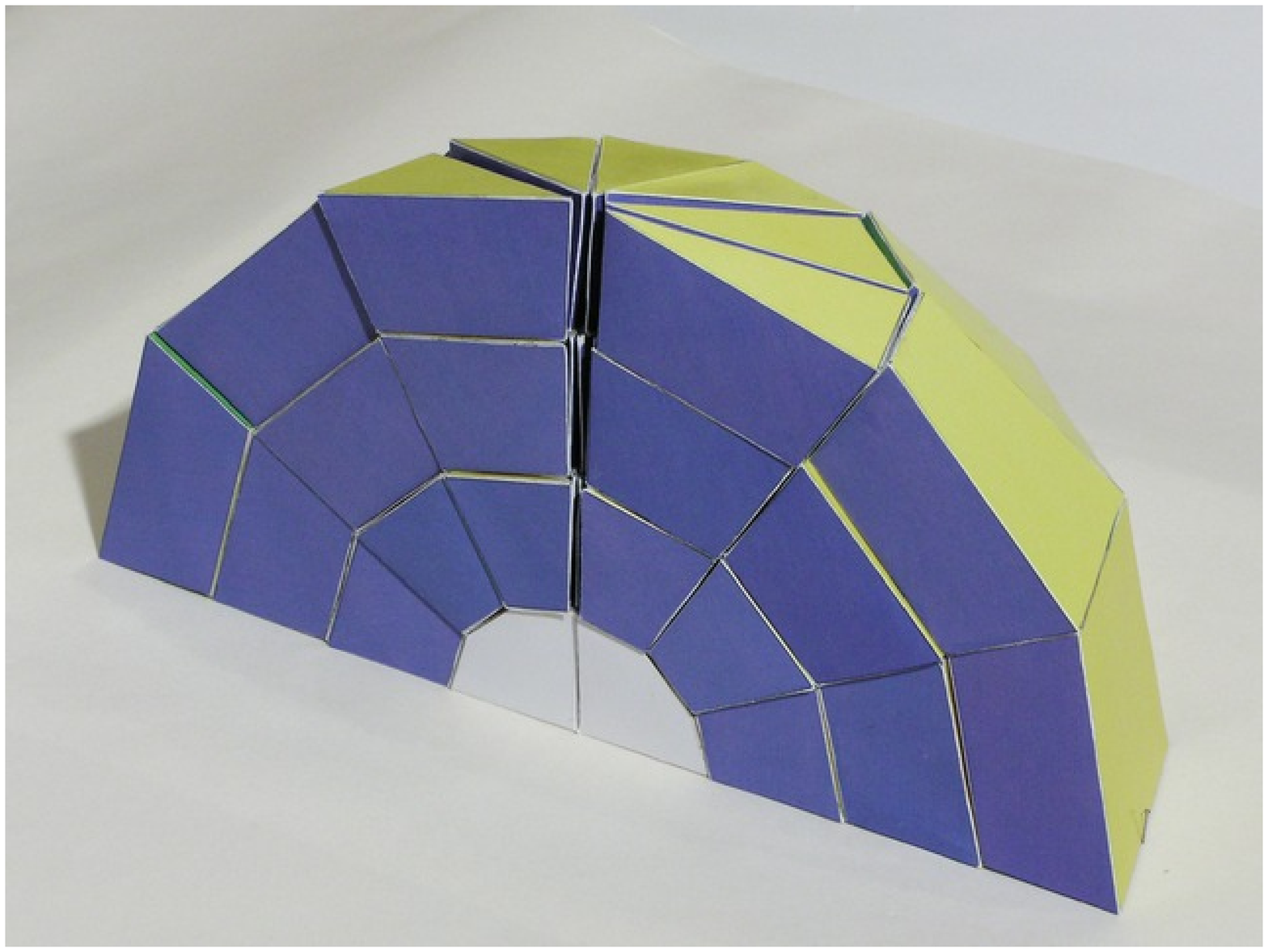}}\hfill
  \subfigure[]{\includegraphics[width=\hb]{\bilder/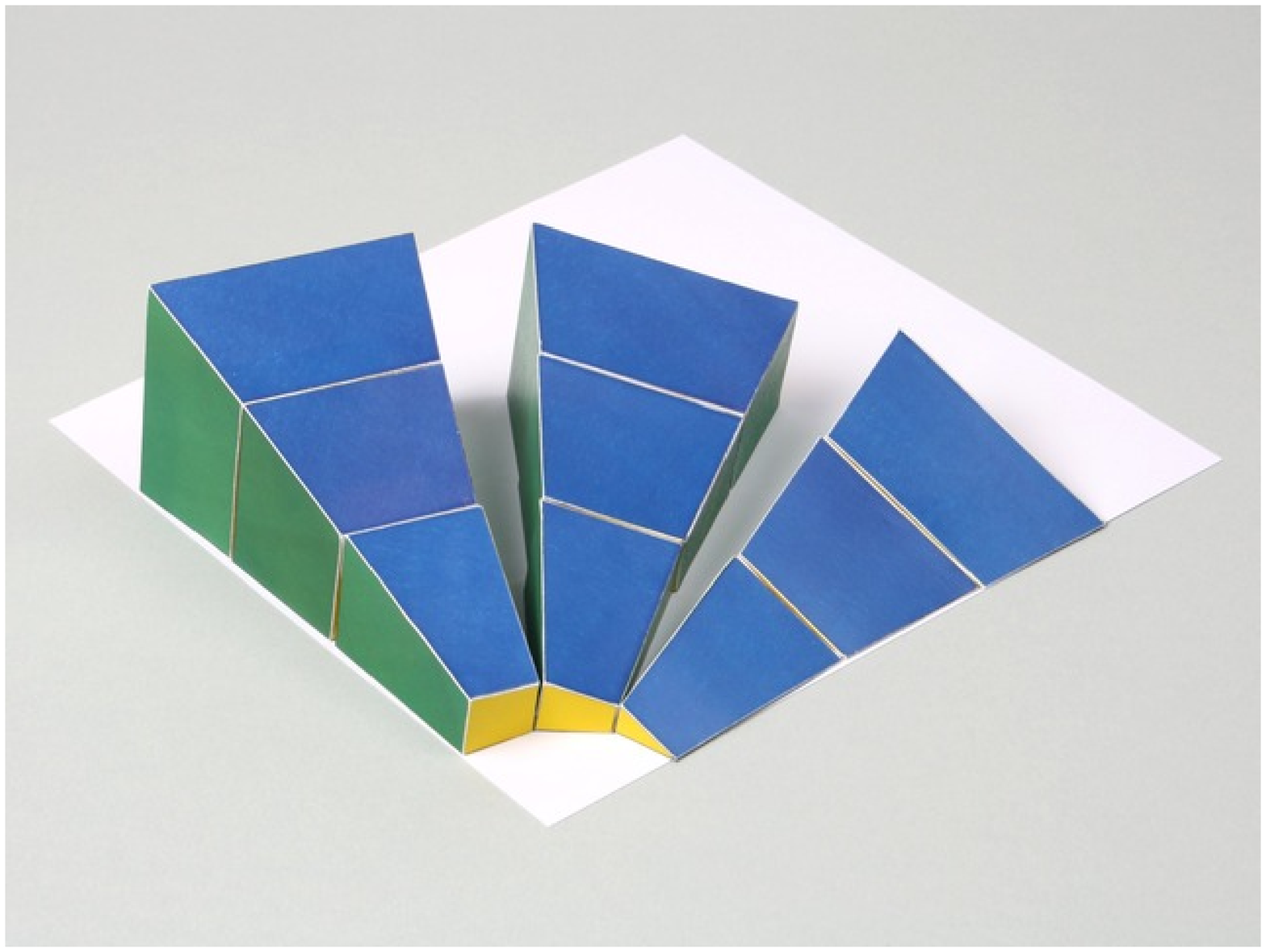}}
  \caption{\label{fig.schnitz}
  Sector models: individual sectors (a),
  single slice (b), and quarter sphere (c)
  of the model ``euclidean space'',
  single slice of the model ``black hole'' (d).
}
\end{figure}
The cut-out sheet for the euclidean model
(figure~\ref{fig.bb-sl-3d} top)
yields nine blocks with yellow, green, and blue sides
(figure~\ref{fig.schnitz}(a)). Assembled (yellow upon yellow
and green upon green) 
this part of
the model has the shape of
half an orange slice
(figure~\ref{fig.schnitz}(b)).
If possible, at least three such slices should be built.
Set up side by side
they form one eighth of a sphere
(figure~\ref{fig.schnitz}(c) shows one quarter of a sphere
made up of six slices)
with a small spherical cavity in the centre.
In figure~\ref{fig.schnitz}(c) an extra grey block 
fills the cavity to support the structure
(a cut-out sheet for the supporting block
is part of
the online resources
\citep{zah2014}).
Twenty-four slices add up to
the complete model of a hollow sphere.
The computer generated picture in
figure~\ref{fig.kloetze}(a)
shows the nearly complete sector model.
Obviously all the blocks 
fit together perfectly
as was to be expected 
for a euclidean space.

The second sector model
depicts a spatial region of the same shape:
A hollow sphere that has been subdivided 
into 24 slices of 9 blocks each,
following the same pattern as before.
This time, though, 
there is a black hole in the centre of the sphere
so that space in this region is strongly curved.
The reason for the spherical hollow
at the centre of the models 
becomes clear at this point:
The cavity is a little larger than the horizon of the black hole
and so completely contains the interior region
that cannot be represented by a rigid model.
Using the cut-out sheet ``black hole'' (figure~\ref{fig.bb-sl-3d} bottom)
nine blocks are built and combined to form a slice
(figure~\ref{fig.schnitz}(d)).
Figure~\ref{fig.kloetze}(b) shows a computer generated picture
of the nearly complete model.
The blocks obviously cannot be arranged to fill a sphere without gaps,
and this reveals that the space in question has non-vanishing
curvature.

Gaps appear whenever the pieces of a curved surface
are laid out in a plane or the blocks of a curved space
are assembled in euclidean space.
With a black hole of the appropriate mass in the centre of
the black hole model, though, the blocks \emph{the way they are} 
would fit without gaps. Just like the elements of area
would fit without gaps if laid out on a surface with the
appropriate curvature.

\begin{figure}
  \subfigure[]{%
    \includegraphics[width=\hb]{\bilder/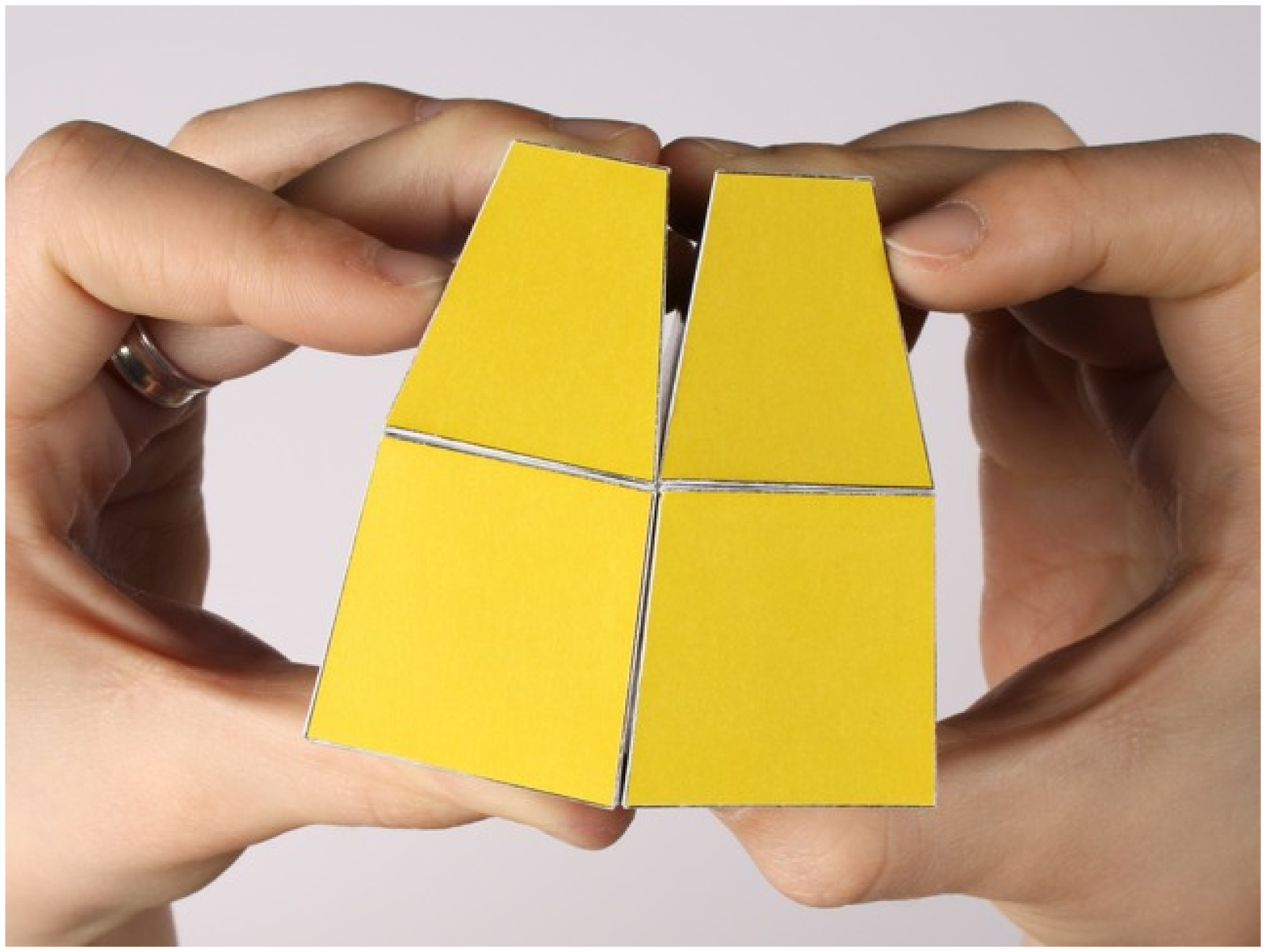}
    \label{fig.kaestchen-theta-phi}%
  }\hfill%
  \subfigure[]{%
    \includegraphics[width=\hb]{\bilder/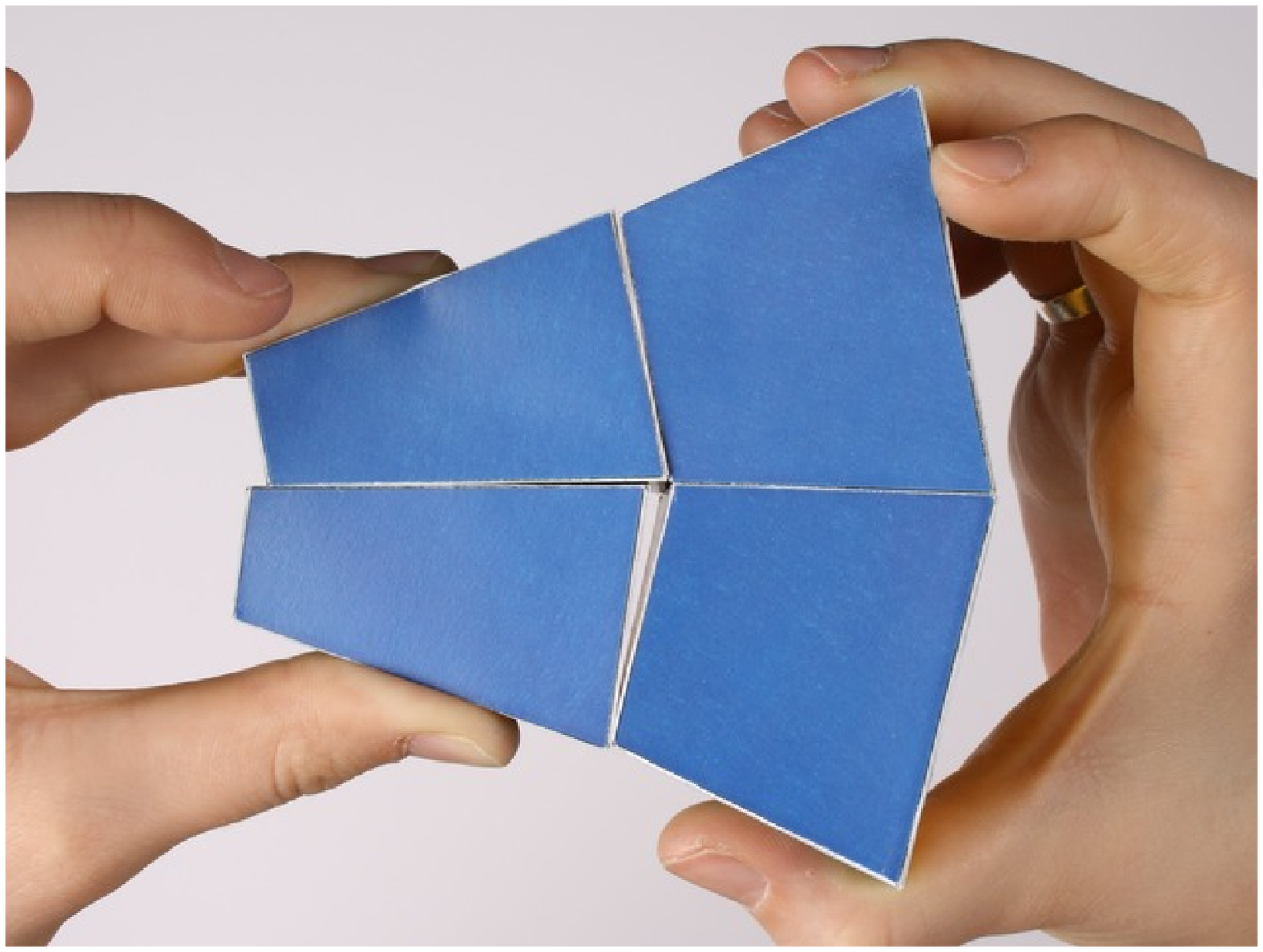}
    \label{fig.kaestchen-r-theta}%
  }\\%
  \subfigure[]{%
    \includegraphics[width=\hb]{\bilder/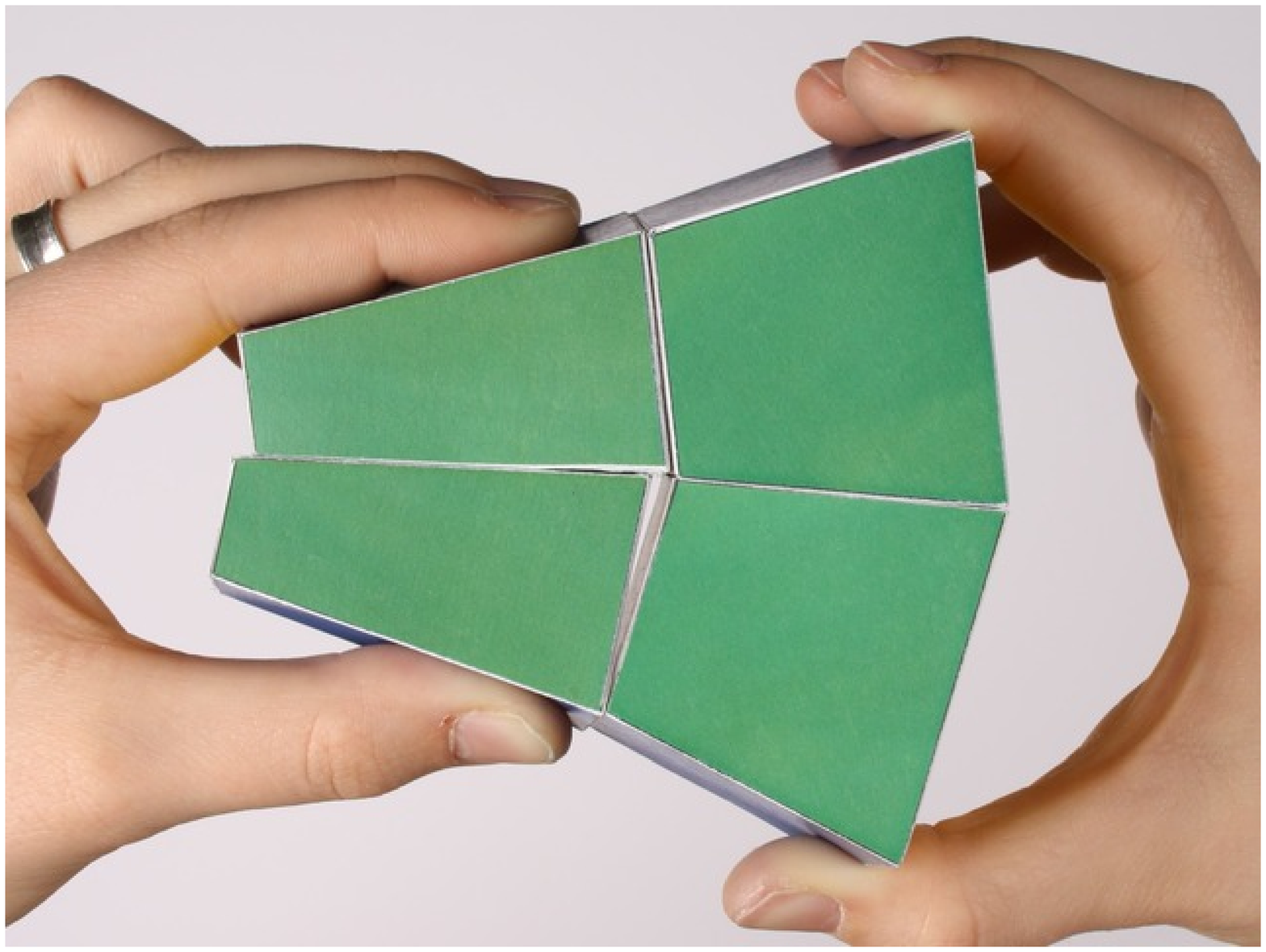}
    \label{fig.kaestchen-r-phi}%
  }\hfill%
  \subfigure[]{%
    \includegraphics[width=\hb]{\bilder/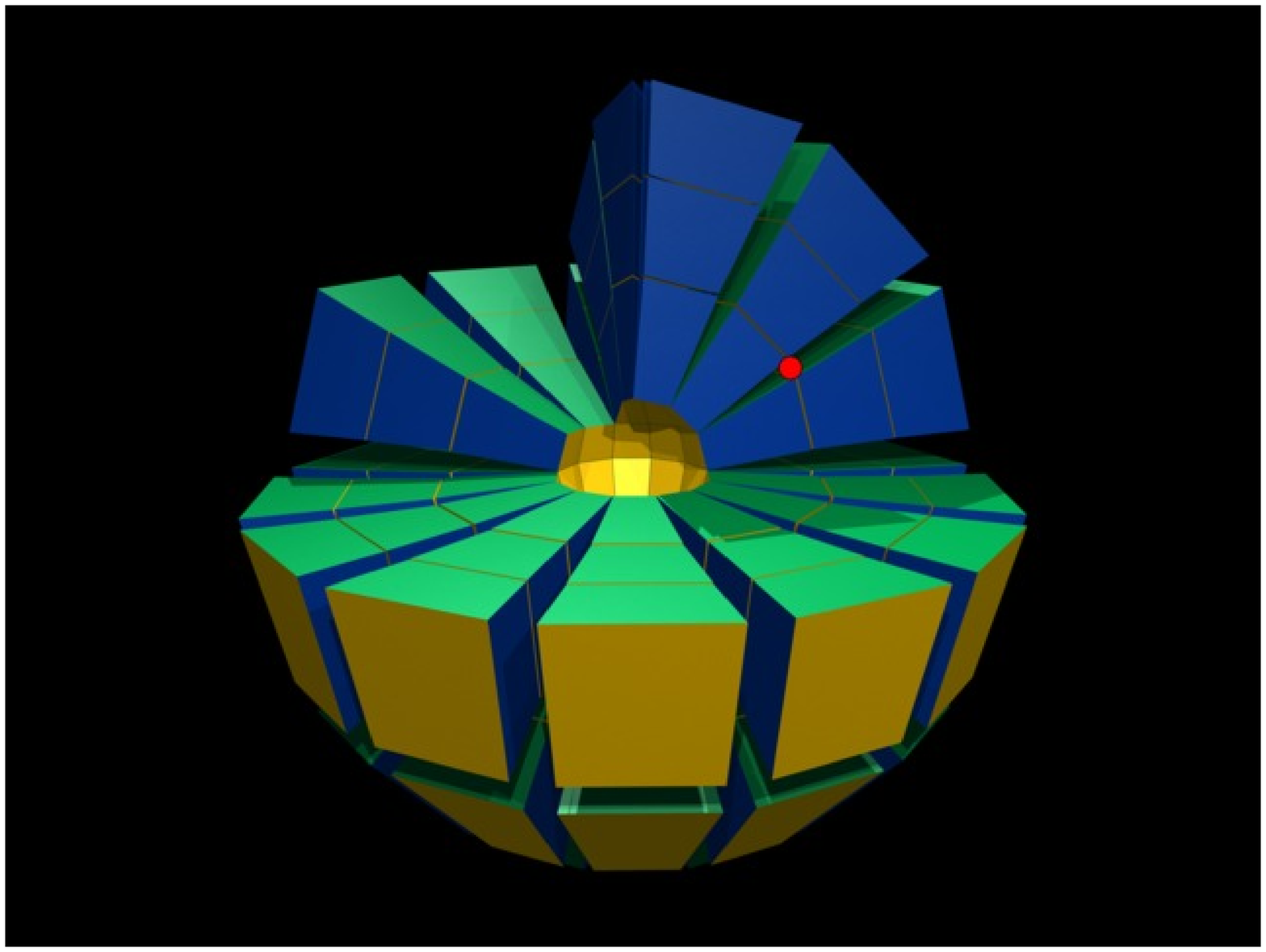}
    \label{fig.kaestchen-vertex}%
  }%
  \caption{\label{fig.kaestchen} 
Curvature in three-dimensional space.
The middle vertex in subimages (a), (b), and (c)
is the same in all cases,
it is marked in (d) with a red dot.
The curvature is positive with respect to the radial edge (a)
and negative with respect to both tangential edges (b, c).
}
\end{figure}

By means of the sector model, 
the curvature is then examined in more detail.
This is done along the lines of the two-dimensional case:
In figures~\ref{fig.flaechepos-def} and~\ref{fig.flaecheneg-def},
four elements of area each are joined at a common vertex
in order to resolve the question of ``tearing or buckling''.
In the spatial model, four blocks share not a vertex, but an edge.
When all four are joined at the common edge,
a gap may remain (``tearing'', indicating positive curvature)
as in figure~\ref{fig.kaestchen-theta-phi}
or the residual space after joining three blocks may be too small
to hold the fourth (``buckling'', indicating negative curvature)
as in figures~\ref{fig.kaestchen-r-theta}
and \ref{fig.kaestchen-r-phi}.
Note that the three subimages (a), (b), and (c) of figure~\ref{fig.kaestchen}
belong to three different edges 
{\em at the same place}.
I.e.\ the curvature at one and the same point
has a different value and indeed a different sign
depending on the orientation of the edge that is considered.
Hence, in more than two dimensions, ``curvature'' is not a number,
but a quantity with multiple components.
It is one of the strong points of sector models
that they clearly display this
fundamental property of curved spaces of more than two
dimensions~\citep{zah2008}.

By comparing the two sector models
one can moreover show that the laws of euclidean geometry
do not hold in a curved space.
For instance the relation between surface and volume
of an object: The outer spherical boundary of the models
has the same surface area in both cases.
This is easily seen
by placing the outer surfaces 
of corresponding blocks 
side by side. 
The same holds for the surface area of the inner
spherical boundary.
The volume enclosed within these two boundaries
is different, though:
In radial direction, 
each block of the black hole model is longer than
the corresponding block of the euclidean space model,
i.e.\ the surface with the same area encloses
a larger volume.
At this point one may recall the flatlanders
who in their two-dimensional world come to
quite similar conclusions:
When they measure circumference and area
of a circle 
on a hill, the
area is larger than it would be in the plane
given the same circumference.

\subsection{Visualization of sector models}

\begin{figure}
\subfigure[]{%
\includegraphics[width=\hb]{\bilder/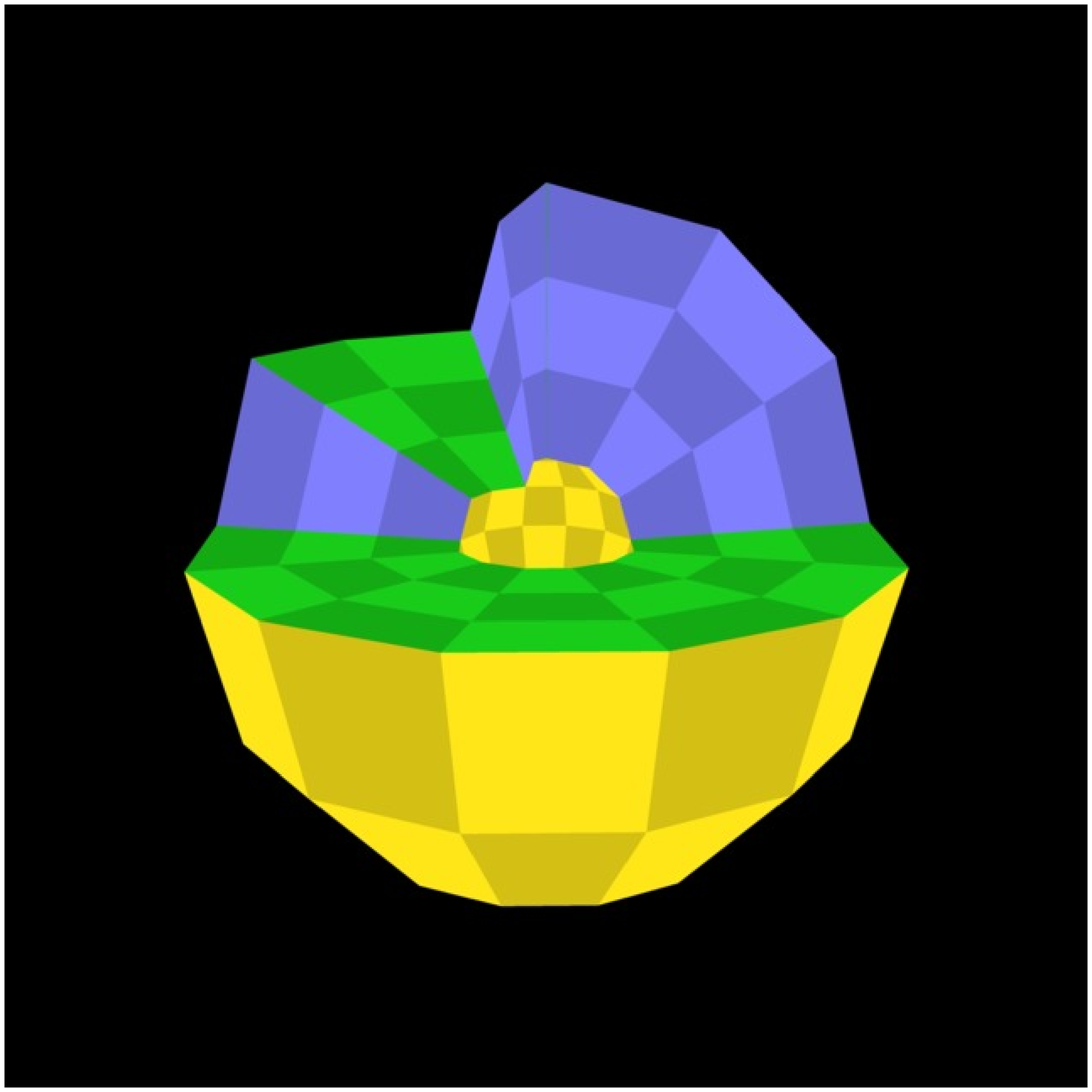}
}%
\hfill
\subfigure[]{%
\includegraphics[width=\hb]{\bilder/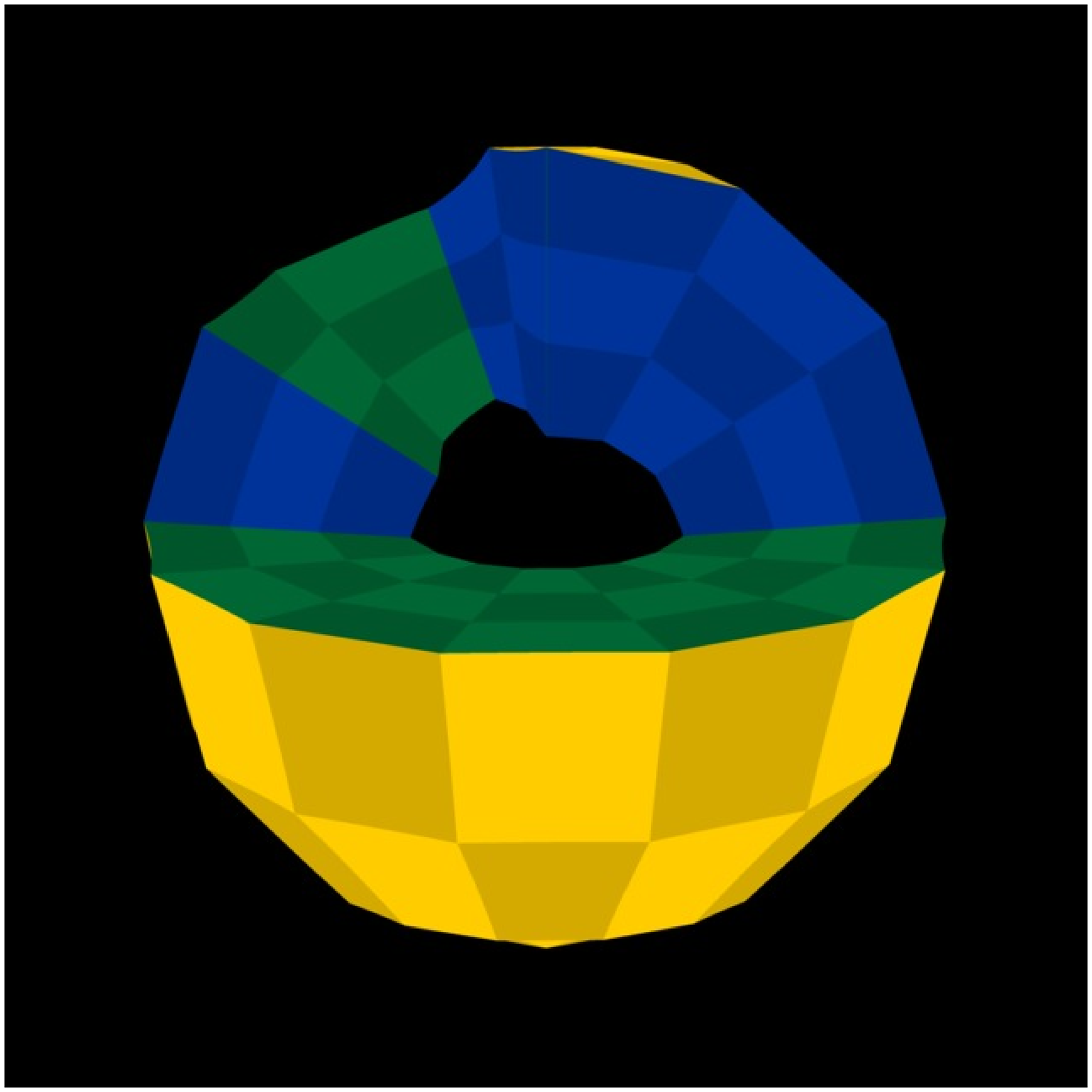}
}
\caption{\label{fig.graum1-pdnw}
View of the sector model ``at its place of origin'',
(a) euclidean space, (b) Schwarzschild space.
}
\end{figure}

By construction,
each sector model
fills space completely, i.e.\ without gaps,
at its place of origin.
This also holds for the sector model of a black hole
that in euclidean space cannot be assembled without gaps.
One may ask what this sector model would look like,
at its place of origin around a black hole,
when the shape of each block is exactly as shown in
figure~\ref{fig.kloetze}(b)
and these blocks join to fill a sphere.
Figure~\ref{fig.graum1-pdnw}(b) shows this view
with the set-up of
figure~\ref{fig.kloetze}(b).
The images in figure~\ref{fig.graum1-pdnw}
have been computed with the ray tracing method
of computer graphics by retracing, 
for each pixel, the incoming light ray
to its point of origin \citep{zah1991}.
As expected no gaps appear between the blocks%
\footnote{
In order to be able to distinguish
neighbouring blocks, 
they have been given slightly different shades of colour
in this image.}.
Comparing figure~\ref{fig.graum1-pdnw}(b) to
figure~\ref{fig.kloetze}(b),
one notices that the raytraced image ap\-pears
somewhat distorted and in particular does not show
the yellow inner spherical boun\-da\-ry.
These effects are due to gravitational light deflection
in the curved space\-time of the black hole.

\section{The sector models}

\label{sec.modell}

\subsection{Sector models of curved surfaces}
\label{sec.2dmodell}

The surface with positive curvature presented
in 
section~\ref{sec.gekruemmte-flaechen}
is a spherical cap with metric
\begin{equation}
\label{eq.pos}
\ud s^2 = R^2 \ud \theta^2 + R^2 \cos^2\theta \, \ud\phi^2
\end{equation}
and constant Gaussian curvature $K=1/R^2$,
$R$ being the radius of the sphere.
The surface with negative curvature
is a saddle with metric
\begin{equation}
\label{eq.neg}
\ud s^2 = R^2 \ud \theta^2 + R^2 \cosh^2\theta \, \ud\phi^2
\end{equation}
and constant Gaussian curvature $K=-1/R^2$.
In order to subdivide the surfaces,
(figures~\ref{fig.flaechepos-sk}, \ref{fig.flaecheneg-sk})
vertices were chosen at coordinates
\begin{eqnarray*}
\theta_i &= -\frac{\pi}{6} + i\cdot \frac{\pi}{9} \qquad & i=0\dots 3,\\
\phi_j   &= j\cdot \frac{\pi}{9} \qquad & j=0\dots 3.
\end{eqnarray*}
The lengths of the edges are the lengths of the
spacelike geodesics between neighbouring vertices.
They are computed by integrating the line element
along the geodesics.
The edge lengths together with the condition of
mirror symmetry uniquely determine the shapes of
the trapezoidal sectors.
\begin{figure}
  \centering
  \includegraphics[width=0.3\textwidth]{\bilder/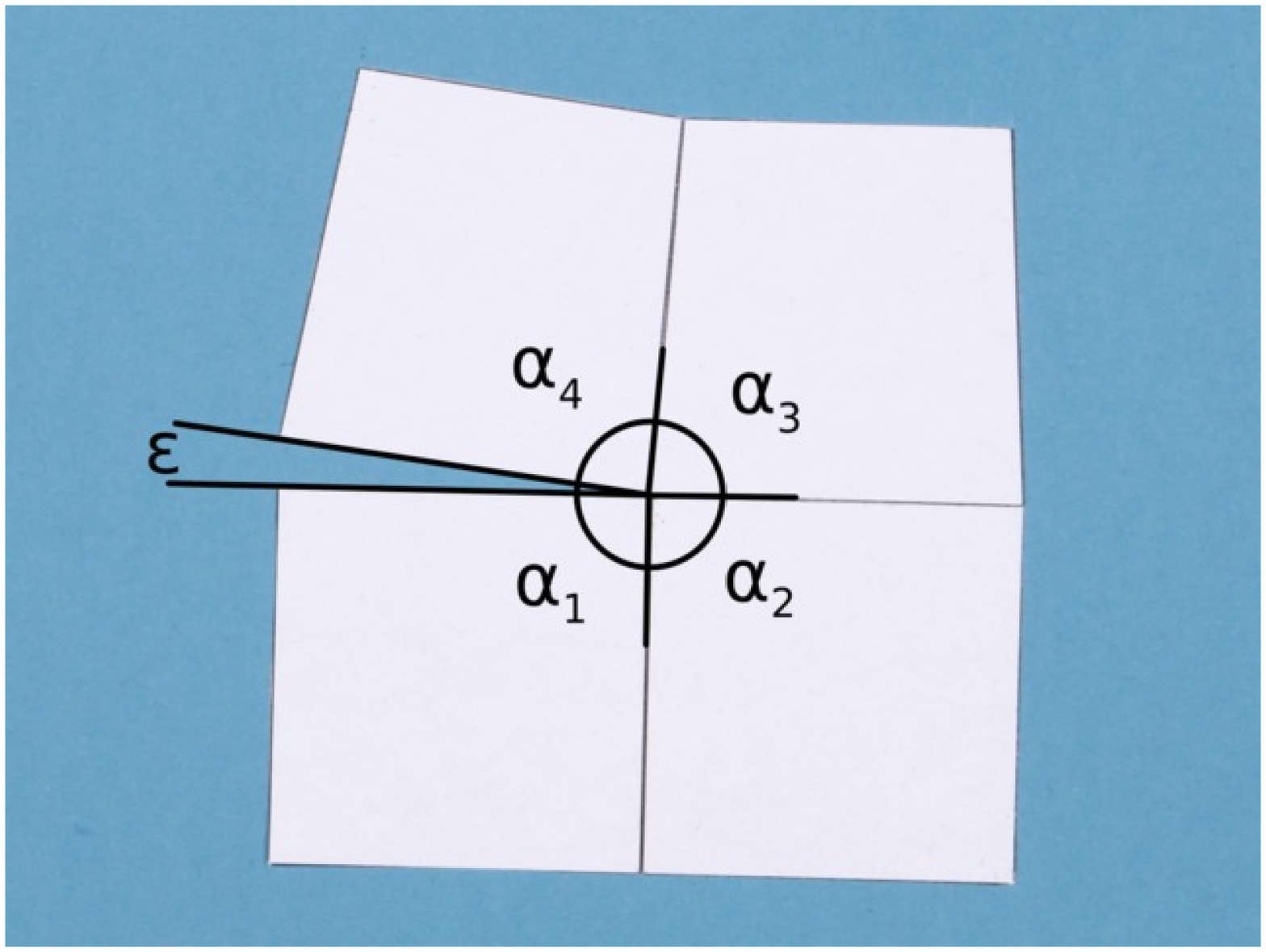}
  \caption{\label{fig.defizit}
Deficit angle $\epsilon$ of a vertex.
}
  \end{figure}

The deficit angle of a vertex
(figure~\ref{fig.defizit})
is defined as
\begin{equation}
\label{eq.def}
\epsilon = 2\pi - \sum_i \alpha_i,
\end{equation}
where the angles $\alpha_i$ are the interior angles
of all the sectors containing the vertex in question.
With increasing fineness of the subdivision into sectors
\begin{equation}
\label{eq.gauss}
K = \rho \epsilon
,
\end{equation}
where $\rho$ is the density of the vertices
and $\epsilon$ their deficit angle,
approximates 
the Gaussian curvature of the surface
\citep{reg1961}.

\subsection{Spatial sector models}
\label{sec.3dmodell}

In section~\ref{sec.gekruemmte-raeume}
sector models of euclidean space
and of the space around a black hole
were introduced.
The respective metrics are
\begin{equation}
\label{eq.eukl}
\ud s^2 = \ud r^2 + r^2(\ud \theta^2 + \sin^2\theta \, \ud \phi^2)
\end{equation}
for euclidean space and
\begin{equation}
\label{eq.ssmr-krumm}
\ud s^2 = \left( 1 - \frac{\rs}{r} \right)^{-1} \ud r^2 + 
r^2(\ud \theta^2 + \sin^2\theta \, \ud \phi^2) 
\end{equation}
for the three-dimensional spacelike hypersurface
of the Schwarzschild spacetime
that is defined by constant Schwarzschild time.
Here, $\rs = 2 G M / c^2$ is the Schwarzschild radius
of mass $M$, $G$ the gravitational constant,
and $c$ the speed of light in vacuum.
The sector model built 
from the cut-out sheets available in the online resources \citep{zah2014}
corresponds to a black hole mass of three earth masses
if the sheets are printed in A4 format.

For the partitioning of the spherical regions
(figure~\ref{fig.schnitz})
vertices were chosen with coordinates
\begin{eqnarray*}
r_i      &= i\cdot 1.25\, \rs \qquad & i=1\dots 4,\\
\theta_j &= j\cdot \frac{\pi}{6} \qquad & j=0\dots 6,\\
\phi_k   &= k\cdot \frac{\pi}{6} \qquad & k=0\dots 11.
\end{eqnarray*}
Partitioning is limited to the region outside $r=1.25\,\rs$.
Inside $\rs$ no static spacelike hypersurface can be defined.
The computation 
is performed as
in the case of
the curved surfaces: The lengths of the edges 
between neighbouring vertices are computed
by integrating the line element
along spacelike geodesics. 
In accord with the symmetric partitioning of the spherical
region, the sector faces are described as isosceles trapezia.
The shapes of the faces are uniquely defined by the edge lengths
together with this symmetry condition.
The complete 
spatial 
sector model
is built up of 216 sectors.
Since the partitioning consists
of 24 identical slices,
one slice 
contains the full information on the geometry.
The cut-out sheets 
comprise
the nine sectors of one slice,
corresponding to
$i=1\ldots 4$, $j=0\ldots 3$ and $k=0\ldots 1$.
The sector models of euclidean space and Schwarzschild space
are true-to-scale representations
of the respective metrics.
The models therefore display
the geometric properties of these
spaces
in a way that is quantitatively correct
(within the framework of the rather coarse discretization).
In particular
in the radial direction 
a sector of the Schwarzschild space model
is longer 
than the corresponding sector of the euclidean space model.
They differ by a factor
given by the integral of the metric coefficient
$\int \left( 1 - \rs/r \right)^{-1/2} \ud r$
along the edge.

The deficit angle with respect to some edge
is again defined by equation~(\ref{eq.def}).
In euclidean space it is zero,
in Schwarzschild space it depends on the orientation of the edge
(figure~\ref{fig.kaestchen}).
In a local orthonormal coordinate system
and for a partitioning into coordinate cubes,
the components $R^{\hat{j}}{}_{\hat{k}\hat{j}\hat{k}}$
of the Riemann curvature tensor are approximated as
\begin{equation}
  R^{\hat{j}}{}_{\hat{k}\hat{j}\hat{k}} = \rho \epsilon l
,
\end{equation}
where $\Ve_{\hat{i}}$, $\Ve_{\hat{j}}$, $\Ve_{\hat{k}}$ are the basis vectors,
$\rho$ is the density of the edges in $i$-direction,
$\epsilon$ their deficit angle and $l$ their length
\citep{reg1961, mis}.

\section{Spacetime sector models}

\label{sec.raumzeit}

\begin{figure}

\centering

\subfigure[]{
\begin{minipage}{\db}
\centering
\includegraphics[scale=0.4]{\bilder/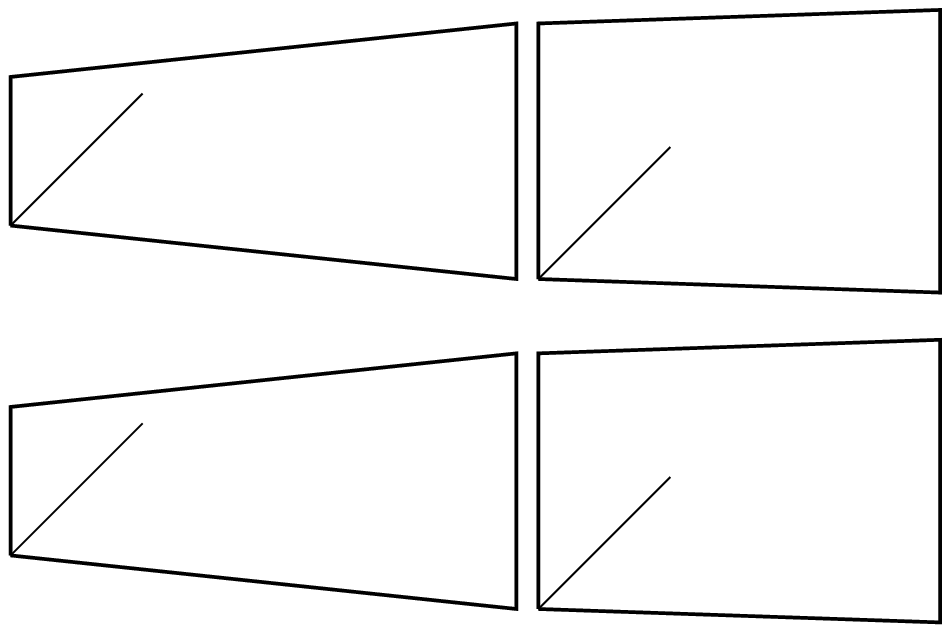}
\end{minipage}
\label{fig.rtsektoren-sekt-sym}
}%
\qquad
\subfigure[]{
\begin{minipage}{\db}
\centering
\includegraphics[scale=0.4]{\bilder/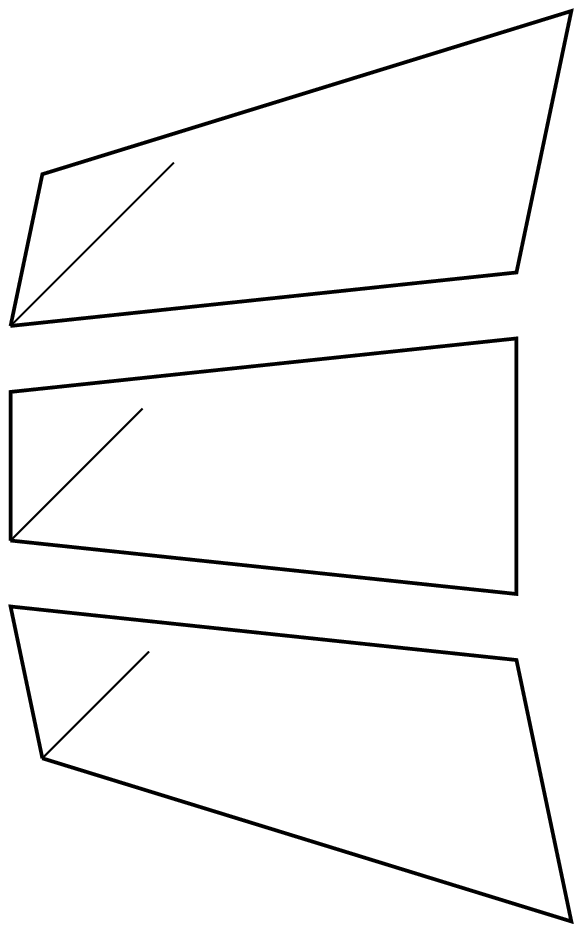}
\end{minipage}
\label{fig.rtsektoren-sektor-lt}
}

\subfigure[]{
\begin{minipage}{\db}
\centering
\includegraphics[scale=0.4]{\bilder/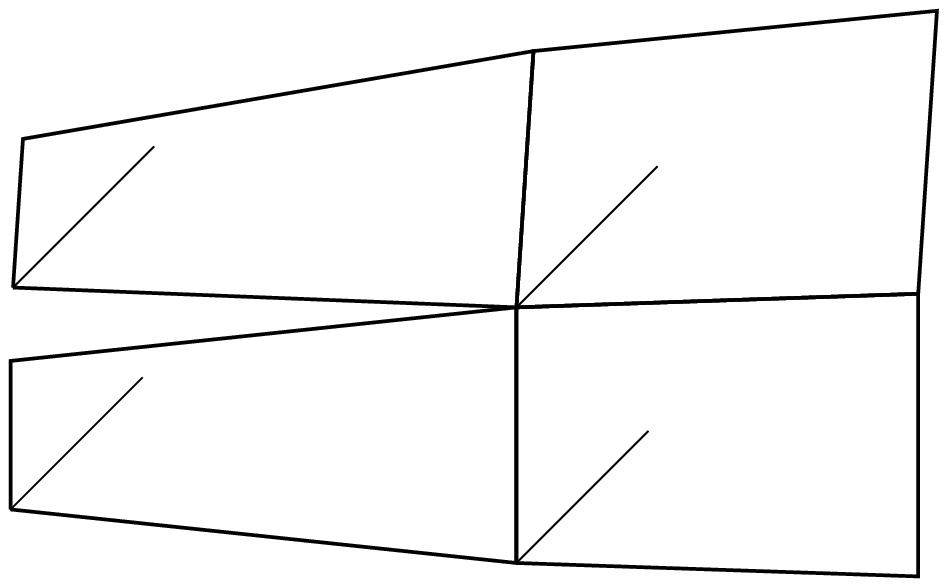}
\end{minipage}
\label{fig.rtsektoren-kruemm-3}
}%
\qquad
\subfigure[]{
\begin{minipage}{\db}
\centering
\includegraphics[scale=0.4]{\bilder/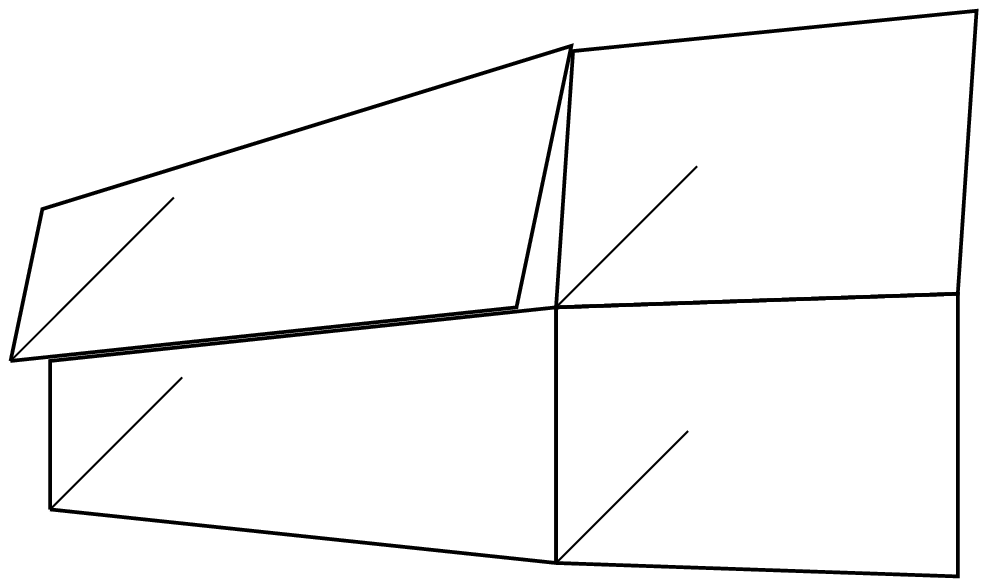}
\label{fig.rtsektoren-kruemm-2-l}
\end{minipage}
}

\caption{\label{fig.rtsektoren}
(a) 
Spacetime sector model of the $t$-$r$ subspace
of the Schwarzschild spacetime.
The spacelike coordinate axis is horizontal,
the timelike one vertical, and the diagonal lines
mark the light cones.
(b)
Different representations of the same spacetime sector.
Lorentz transformations convert them into each other.
(c) Deficit angle in spacelike direction: ``tearing''.
(d) Deficit angle in timelike direction: ``buckling''.
}
\end{figure}

The idea of approximating
a curved surface or a curved space
by small uncurved sectors
can also be applied to curved spacetimes.
In this case the uncurved sectors 
have Minkowski geometry.
Figure~\ref{fig.rtsektoren}
gives an example of a spacetime sector model.
The curved spacetime here
is the two-dimensional $t$-$r$ subspace of
the Schwarzschild spacetime
with the Schwarzschild radial coordinate $r$
as space coordinate and the Schwarzschild time $t$
as time coordinate.
The metric of this subspace is
\begin{equation}                                                                 
\label{eq.ssm-rt}                                                            
\ud s^2 = - \left (1 - \frac{\rs}{r} \right) \ud t^2 +
            \left (1 - \frac{\rs}{r} \right)^{-1} \ud r^2
\end{equation}
with the Schwarzschild radius $\rs$.
In this section geometric units are used
so that the
speed of light in vacuum $c$ is equal to unity.

The vertices for the subdivision into sectors
are at
\begin{eqnarray*}
t_i  &= i\cdot 1.25\, \rs \qquad & i=0\dots 2,\\
r_j  &= j\cdot 1.25\, \rs \qquad & j=1\dots 3.
\end{eqnarray*}
The geodesics that connect neighbouring vertices
are partly spacelike and partly timelike.
In both cases the length of the edges
is determined by integrating the line element
along the geodesics. 
The metric coefficients do not depend on the $t$ coordinate,
therefore identical sectors are lined up in $t$ direction.
Each sector is described as an isosceles trapezium with
the timelike edges as bases.
This symmetry condition together with the edge lengths
uniquely determines the shapes of the sectors.
Fig.~\ref{fig.rtsektoren-sekt-sym} shows the representation
of the sectors in a Minkowski diagram.

The geometry in the interior of each sector
is Minkowskian
and the statements of the special theory of relativity
about Minkowski spacetime
apply.
In particular this includes
the free choice of the inertial frame
in which events are described
and the use of the Lorentz transformation
in order to change from one inertial frame
to another.
When two inertial frames
connected by a Lorentz transformation
are visualized in a Minkowski diagram,
their time axes form an angle,
the space axes likewise but in the opposite direction,
and the light cone is common to both frames.
Thus, the representation of the sectors in a Minkowski diagram
as shown in figure~\ref{fig.rtsektoren-sekt-sym}
corresponds to the choice of a certain inertial frame,
and a boost transformation into a different inertial frame
changes the representation of the sectors 
(figure~\ref{fig.rtsektoren-sektor-lt}).
In the spatiotemporal sense, though, the sectors have the same
shape and symmetry in all inertial systems,
since these are described in terms of scalar products
that are invariant with respect to Lorentz transformations.
In order to illustrate the curvature
in this subspace,
sectors must again be assembled around a vertex.
In a spatial sector model
a sector is rotated
in order to lay it alongside the neighbouring sector.
In the spatiotemporal case
the rotation is replaced by a Lorentz transformation.
One can see
that joining neighbouring sectors by means of rotation
in the $t$-$r$ plane 
is not possible
by considering the light cones
that must be identical in the two sectors that
are joined. 
Since a change of inertial frame
entails a change in the representation of a sector as described above,
the common edge of neighbouring sectors can be made to
coincide in the Minkowski diagram by a Lorentz transformation
(figure~\ref{fig.rtsektoren-kruemm-3}).

In case of a global Minkowski spacetime,
all the sectors can be combined in this way
to cover without gaps a section of the spacetime.
In the $t$-$r$ subspace of the Schwarzschild spacetime, however,
there is a deficit angle at the vertex
(figure~\ref{fig.rtsektoren-kruemm-3}),
indicating there is a non-zero spatiotemporal curvature.
If a gap appears between two spacelike edges
(as in this example),
the spatiotemporal curvature is positive,
in case of an overlap
it would be negative\footnote{%
Note that
when the signature is chosen to be ($+$ $-$ $-$ $-$),
the spatiotemporal curvature has the opposite sign:
It is negative (positive),
if a gap (an overlap) appears between two spacelike edges.
}.
If the sectors are arranged
in such a way
that the deficit angle appears between two timelike edges,
an overlap results in the example shown here
(figure~\ref{fig.rtsektoren-kruemm-2-l}).
The gap between the spacelike edges
and the overlap of the timelike edges
specify the same Lorentz transformation
that would make the edges on both sides of the deficit angle meet.

Similar to the spatial case, the deficit angle
is related to the respective component of the
curvature tensor:
In a local orthonormal coordinate system,
and for a partitioning into coordinate cubes,
the component $R^{\hat{t}}{}_{\hat{r}\hat{t}\hat{r}}$
is approximated as
\begin{equation}
  R^{\hat{t}}{}_{\hat{r}\hat{t}\hat{r}} = \rho \alpha
,
\end{equation}
where $\Ve_{\hat{t}}$ and $\Ve_{\hat{r}}$
are the timelike and spacelike basis vectors, respectively,
$\rho$ is the local 
density of vertices 
and $\alpha$ the rapidity 
of the Lorentz transformation
specified by the local deficit angle
(the velocity being $\beta=\tanh{\alpha}$).

\section{Conclusions and outlook}

\label{sec.fazit}

Sector models are physical models
that represent two- and three-dimensional curved spaces or spacetimes
true to scale.
This description of curved spaces does not use coordinates,
it therefore provides physical insight in a direct and intuitive way.
We have shown how the curvature of a three-dimensional space
and of a 1+1-dimensional spacetime can be described
by means of such models.
This is the answer that we propose to give to the first
of the three questions raised in the introduction,
``What is a curved spacetime?''.
The work with the sector models can stand on its own,
or it can supplement the usual mathematical introduction of the
Riemann curvature tensor by providing a visualization.

The workshop on curved spaces
outlined in section~\ref{sec.unterricht}
has been held a number of times
with undergraduate students, high school students
and interested adults
\citep{zah2004,kra2005,zah2010,zah2012}.
The programme
and the materials 
have been developed in 
several cycles of testing and revision.
For several years,
this workshop has been held 
at Hildesheim university
for students 
who intend to become physics teachers.

In comparison with the introduction of curvature
via prototypical curved surfaces
as described in the introduction,
the use of sector models
has a considerably wider scope.
Curved two-dimensional surfaces
when depicted in the usual way 
are shown embedded in three-dimensional space.
This picture
cannot be carried over
to curved spaces
because the necessary higher dimensional embedding space
cannot be visualized. 
The sector models, in contrast, need no extra dimension.
The two-dimensional case (flatland)
is carried over directly to 
the three-dimensional case and
so makes the curved three-dimensional case
accessible to ``spacelanders''.
In particular, sector models illustrate
the tensorial character of curvature
in spaces of more than two dimensions.
This can be accomplished neither with prototypical curved surfaces
nor with embedding diagrams.

The usefulness of sector models
extends well beyond the visualization of curvature
as will be shown in a sequel to this paper.
Sector models permit to introduce other geometric concepts
(e.g.\ parallel transport, geodesics)
in an intuitive way. 
They can be used to study relativistic phenomena
by construction instead of computation
(e.g.\ particle paths, redshift).
Other spacetimes that e.g.\ contain matter
or are not static
can be described in the same
way as the Schwarzschild spacetime
considered here.
Thus, the second and third of the fundamental questions
stated in the introduction,
namely the motion of particles 
and the connection between curvature and matter
can also be addressed within this framework.

\end{document}


\noindent\today
\bigskip

\title[Sektormodelle -- Teil 1: Gekrümmte Räume und Raumzeiten]{%
Sektormodelle -- Ein Werkzeugkasten zur Vermittlung der
               Allgemeinen Relativitätstheorie.\\
Teil 1: Gekrümmte Räume und Raumzeiten
}

\author{C Zahn und U Kraus}

\address{Institut für Physik, Universität Hildesheim, Marienburger
  Platz 22, 31141 Hildesheim, Germany}

\eads{\mailto{corvin.zahn@uni-hildesheim.de},
  \mailto{ute.kraus@uni-hildesheim.de}}

\begin{abstract}

\parindent=0pt

Für die Vermittlung der Allgemeinen Relativitätstheorie
in der Schule, im Grund- oder im Nebenfachstudium 
besteht das Anliegen, eine fachlich befriedigende
Darstellung zu geben, die nicht mehr als Schulmathematik
voraussetzt.
Wir stellen in diesem Beitrag einen solchen Zugang vor.
Das zentrale Werkzeug unserer Einführung
sind sogenannte Sektormodelle,
die gekrümmte Räume
im Sinne des Regge-Kalküls
durch eine Zer\-le\-gung in kleine, ungekrümmte Sektoren
beschreiben,
ähnlich der Triangulierung einer ge\-krümm\-ten Fläche.
Wir schildern
einen Workshop für Schüler/innen und
Studierende,
in dem gekrümmte
Räume anhand von Sektormodellen Schwar\-zer Lö\-cher
eingeführt werden.
Wir beschreiben ferner
die Erweiterung 
auf Sek\-tor\-mo\-del\-le
ge\-krümm\-ter Raumzeiten.
Raumzeitliche Sektormodelle
setzen Grundkenntnisse
der Speziellen Relativitätstheorie voraus.
Die vorgestellten Materialien stehen
online unter 
\url{http://www.tempolimit-lichtgeschwindigkeit.de}
für den Unterricht zur Verfügung.

\end{abstract}

\maketitle

\section{Einleitung}

Die Allgemeine Relativitätstheorie ist
neben der Quantentheorie 
eine der beiden großen 
Weiterentwicklungen der Physik im 20. Jahrhundert.
Durch Experimente im Sonnensystem 
mit hoher Genauigkeit getestet,
ist sie eine gut etablierte Theorie.
Sie gehört zu den Grundlagen des heutigen physikalischen Weltbildes;
von zentraler Bedeutung ist sie
für das Verständnis vieler astrophysikalischer Phänomene.
Die Relativitätstheorie ist auch von hohem Interesse
für die allgemeine Bevölkerung, 
nicht zuletzt wegen ihrer Bedeutung 
für das Verständnis exotischer Objekte wie Schwarzer Löcher 
sowie für Fragen nach Anfang und Ende der Welt.

Die Vermittlung dieser wichtigen und für Studierende und
Schüler/innen
in der Regel faszinierenden Theorie
steht jedoch vor einem grundlegenden Problem.
Die Stan\-dard\-dar\-stel\-lung beginnt mit den benötigten mathematischen
Werkzeugen, begründet dann die Einsteinschen Feldgleichungen, leitet
Lösungen der Feldgleichungen her und untersucht schließlich deren
Eigenschaften, insbesondere die Bahnen von Teilchen und Licht.
Die mathematischen Werkzeuge, auf denen dieses Programm aufbaut, 
sind umfangreich und gehen über Schulmathematik weit hinaus.
Für Schüler/innen und Studienanfänger/innen sind die mathematischen
Anforderungen also zu hoch. Ebenso im Studium von Physik als
Nebenfach oder als einem von zwei Hauptfächern (z.\,B.\ Lehramtsstudium),
da hier nicht genügend Zeit zur Verfügung steht, um 
die mathematischen Grundlagen zu entwickeln und dann den Bogen
bis zu den aktuellen Anwendungen zu schlagen.

Daraus folgt das Anliegen, 
die Allgemeine Relativitätstheorie 
ohne ihren ma\-the\-ma\-ti\-schen Apparat zu vermitteln,
d.\,h. eine fachlich befriedigende und an\-schluss\-fä\-hi\-ge Darstellung
zu geben, die 
nicht mehr als Schulmathematik verwendet.
Dieses Anliegen, bereits von Einstein 1916
formuliert \citep{ein1},
ist seitdem in zahlreichen Unterrichtsvorschlägen
sowie in populärwissenschaftlichen Darstellungen
weiterverfolgt worden.
Die folgenden vier Herangehensweisen haben eine größere
Verbreitung gefunden:

\smallskip

1. Folgerungen aus dem Äquivalenzprinzip:
Anhand von Gedankenexperimenten werden
die gravitative Licht\-ab\-len\-kung
und die gravitative Zeitdilatation thematisiert
\citep[z.\,B.\ in][]{ein1,gam,sar,tip,stan}.

2. Beschreibung der Geometrie gekrümmter Flächen:
Am Beispiel einfacher ge\-krümm\-ter Flächen wie
z.\,B.\ der Kugeloberfläche werden
die geometrischen Begriffe eingeführt, die im Zusammenhang
mit gekrümmten Raumzeiten eine wichtige Rolle spielen, 
beispielsweise
Metrik, Geodäten 
und Krümmung
\citep[z.\,B.\ in][]{sar, har, stan, nat}.

3. Rechnungen auf der Grundlage der Newtonschen Dynamik:
Zu Phänomenen, die eigentlich mit der Relativitätstheorie
zu beschreiben wären,
beispielsweise Licht\-ab\-len\-kung, Schwarze Löcher 
oder die kosmologische Expansion,
werden Rechnungen im Rahmen der Newtonschen Theorie
durchgeführt \citep[z.\,B.\ in][]{ehl,lot1}.
Dahinter steht der didaktische Ansatz, dass
mit vertrauten Begriffen wie Kraft und Energie gearbeitet wird
um einen direkten Zugang zu den Phänomenen 
(z.\,B.\ Licht\-ab\-len\-kung, beschrieben als Ablenkung
eines klassischen Teilchenstroms)
und damit zur relativistischen Astrophysik
(z.\,B.\ Gravitationslinsen) zu schaffen.
Die Frage nach den Grundkonzepten der 
Allgemeinen Relativitätstheorie
und dem Konzeptwechsel gegenüber der
Newtonschen
Theorie wird 
bei diesem Zugang nicht gestellt.

4. Analogien:
Im populärwissenschaftlichen Bereich
ist der Einsatz von Analogien vorherrschend.
Analogien werden sowohl zur Einführung von Grundkonzepten 
(Beispiel: ein von einer Kugel eingedelltes Gummituch illustriert das
Konzept \glqq Masse krümmt den Raum\grqq) als
auch in Bezug auf relativistische Phänomene
(Beispiel: der Fuß eines Weinglases als \glqq Gravitationslinse\grqq)
eingesetzt \citep[z.\,B.\ in][]{pri,lot2}.
Der Zugang über Analogien birgt ein nicht unerhebliches Risiko
für die Entstehung von Fehlvorstellungen \citep{zah2010}.

\smallskip

Ein weiterer, wenn auch anspruchsvollerer Ansatz
zur bildlichen Darstellung
ge\-krümm\-ter Räume und Raumzeiten
nutzt Einbettungsdiagramme.
Einbettungen werden für rein räumliche zweidimensionale Unterräume
\citep{fla1916, eps1994,jon2001,jon2005}
sowie für raumzeitliche Unterräume
\citep{mar1998}
beschrieben.
Einbettungen gehen über Analogien wie die oben
unter Punkt~2 und~4 erwähnten hinaus,
da sie physikalisch bedeutsame
Unterräume konkreter Raumzeiten darstellen, 
so dass ihre
geometrischen Eigenschaften physikalische Bedeutung haben. 
Sie sind,
wie die oben erwähnten Analogien, 
auf zweidimensionale Unterräume beschränkt.

Dieser Beitrag stellt einen neuartigen Zugang 
zur Allgemeinen Relativitätstheorie vor,
der wie die oben genannten das Anliegen hat mit Schulmathematik
auszukommen. Er zielt auf die Grundkonzepte der Allgemeinen Relativitätstheorie
ab, wie sie in knappster Form in dem bekannten Ausspruch von John Wheeler
zusammengefasst sind \citep{whe}:

\begin{quote}
{\em Spacetime tells matter how to move.\par
Matter tells spacetime how to curve.}%
\footnote{%
Die Raumzeit sagt der Materie, wie sie sich bewegen soll.
Die Materie sagt der Raumzeit, wie sie sich krümmen soll.
}
\end{quote}

Der Ausspruch verdeutlicht, dass
drei grundlegende Fragen zu beantworten sind:

\begin{enumerate}
\item Was ist eine gekrümmte Raumzeit?
\item Wie bewegt sich Materie in einer gekrümmten Raumzeit?
\item Welcher Zusammenhang besteht zwischen der Materieverteilung und
   der Krüm\-mung der Raumzeit?
\end{enumerate}

Zentrales Werkzeug des hier beschriebenen Zugangs sind sogenannte
\emph{Sektormodelle}. Es handelt sich um  gegenständliche Modelle, 
die zwei- oder dreidimensionale
Un\-ter\-räu\-me gekrümmter Raumzeiten maßstabsgetreu darstellen.
Die Modelle sind eine Umsetzung der
im Regge-Kalkül verwendeten
koordinatenfreien Beschreibung einer gekrümmten Raumzeit
durch Zerlegung in kleine, ungekrümmte Sektoren
\citep{reg1961}.
Im Falle eines zweidimensionalen Raums
sind die Sektoren ebene Flächenstücke,
im Falle eines dreidimensionalen Raums Klötzchen,
in deren Innerem die Geometrie euklidisch ist.
Raumzeitliche Unterräume bestehen aus Sektoren mit
Minkowskigeometrie.
Beim Bau des Anschauungsmodells werden die Sektoren
beispielsweise als Papierstücke oder Pappklötzchen ausgeführt.

Anhand von Sektormodellen lassen sich 
die oben genannten drei Grundfragen
auf anschauliche Weise behandeln.
Der Zugang ist für das nicht vertiefte Physikstudium geeignet
und kann auch 
mit Studienanfänger/innen und fortgeschrittenen Schüler/innen
eingesetzt werden.
Da die Darstellung bruchlos an die Standardformulierung
anschließt, kann der Zugang auch als Ergänzung 
zur Standarddarstellung genutzt werden, um 
eine geometrische Anschauung für
gekrümmte Raumzeiten zu vermitteln.

Dieser Beitrag geht auf die erste 
der oben genannten drei Grundfragen näher ein und zeigt,
wie man anhand von Sektormodellen die
Eigenschaften eines gekrümmten Raums 
verdeutlichen kann.
Als Beispiel dient die Schwarzschildraumzeit eines Schwarzen Lochs.
Der folgende Abschnitt~\ref{sec.unterricht} schildert
zunächst
beispielhaft den Ablauf eines
Workshops zum Begriff des gekrümmten Raums.
Er wurde in dieser Weise 
schon mehrfach mit Schü\-ler\-grup\-pen sowie
mit Lehramtsstudierenden
durchgeführt.
Das Konzept des Sek\-tor\-mo\-dells wird darin auf
nichttechnische Weise eingeführt.
Hintergründe zu Berechnung 
und Eigenschaften der Sektormodelle
folgen in Abschnitt~\ref{sec.modell},
ein raumzeitliches Sektormodell wird 
in Abschnitt~\ref{sec.raumzeit} vorgestellt,
einen Ausblick auf
weitere mögliche Anwendungen von Sektormodellen
gibt Abschnitt~\ref{sec.fazit}.

\section{\glqq Wir basteln ein Schwarzes Loch\grqq}

\label{sec.unterricht}

In diesem Workshop wird der dreidimensionale gekrümmte
Raum um ein Schwarzes Loch veranschaulicht. Mit Veranschaulichung ist
gemeint, dass eine geometrische Anschauung entwickelt werden soll, die
es erlaubt, die geometrischen Eigenschaften dieses gekrümmten Raums
auf qualitative Weise zu erfassen.

Das Schwarze Loch als Beispielobjekt hat hierbei zwei Vorzüge: Es
weckt in der Regel großes Interesse, da Schwarze Löcher aus den
Medien als exotische Objekte bekannt sind. Und relativistische Effekte
sind in der Nähe eines Schwarzen Lochs so groß, dass sie in einem
maßstabsgetreuen Modell problemlos zu erkennen sind.

\subsection{Gekrümmte Flächen}

\label{sec.gekruemmte-flaechen}

Wir beginnen den Workshop damit, dass der Begriff der
Krümmung anhand von Flächen im Raum eingeführt wird. Dabei
unterscheiden wir positive, negative und verschwindende Krümmung;
als Prototypen werden die Sphäre, die Sattelfläche 
sowie die Ebene vorgestellt.

Als Kriterium zur Ermittlung der Krümmung geben wir an: Ein kleines
Stück der Fläche wird ausgeschnitten und  flachgedrückt.
Reißt
es dabei ein, ist die Krümmung positiv, wirft es Falten, ist die
Krümmung negativ. Lässt es sich ohne Einreißen oder Faltenwerfen
flach ausbreiten, dann ist die Krümmung null. Dieses Kriterium
stellt das Vorzeichen der inneren (Gaußschen) Krümmung fest. An
einigen Beispielen wird das Zuordnen von Krümmungen
geübt; nützliche Beispiele
sind insbesondere der Torus (Krümmung am
Innenrand negativ, am Außenrand positiv), der zeigt, dass die
Krümmung über die Fläche variieren kann und der Zylinder
(Krümmung null), der aufzeigt, dass die hier definierte innere Krümmung
mit der Alltagsvorstellung einer gekrümmten Fläche nicht
vollständig übereinstimmt.

\begin{figure}
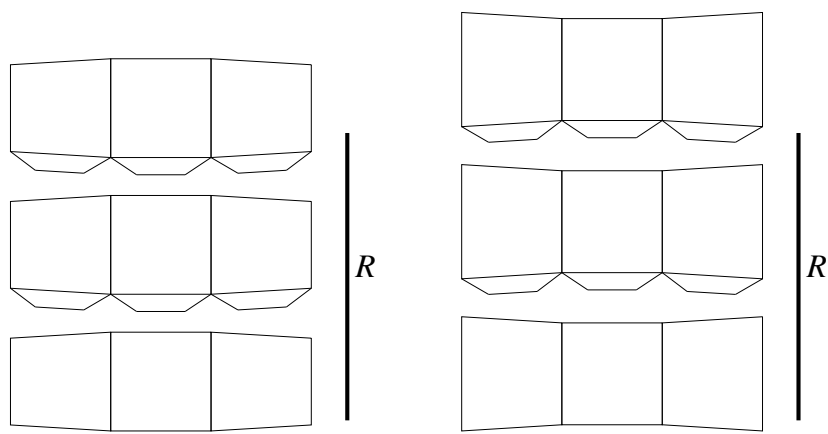

  \centering
  \includegraphics[width=\db]{\bilder/figure1a.eps}
  \hspace{1cm}
  \includegraphics[width=\db]{\bilder/figure1b.eps}
  \caption{\label{fig.bb-flaeche}
Bastelbögen gekrümmter Flächen. Links:
Kugelkalotte
zum Kugelradius $R$ mit konstanter positiver Krümmung $K=1/R^2$.
Rechts:
Sattelfläche mit konstanter negativer Krümmung $K=-1/R^2$.
Eine Strecke der Länge $R$ ist als Balken markiert. 
}
\end{figure}

\begin{figure}
  \centering \subfigure[]{%
    \includegraphics[width=\hb]{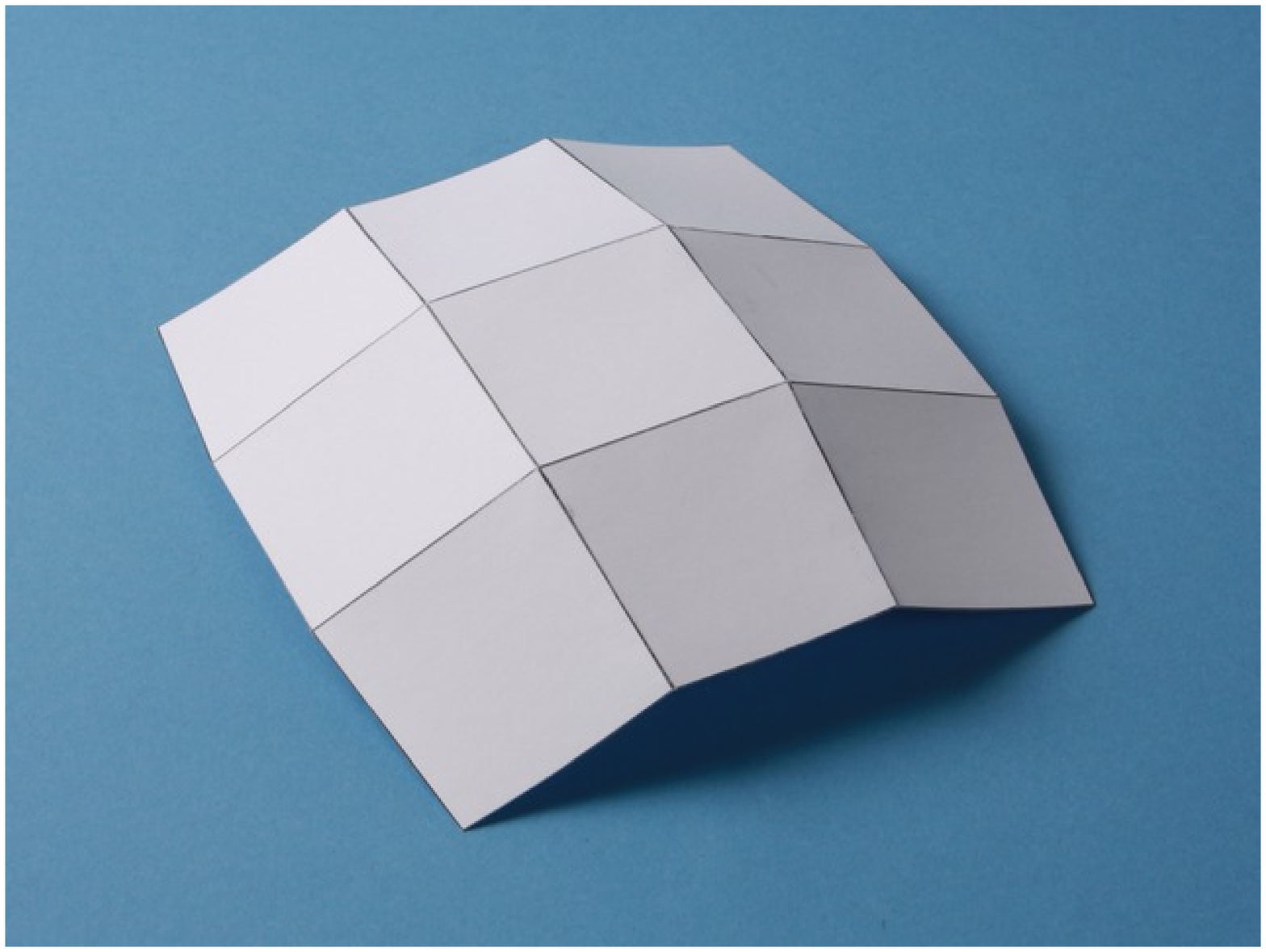}
    \label{fig.flaechepos}}\hfill%
  \subfigure[]{%
    \includegraphics[width=\hb]{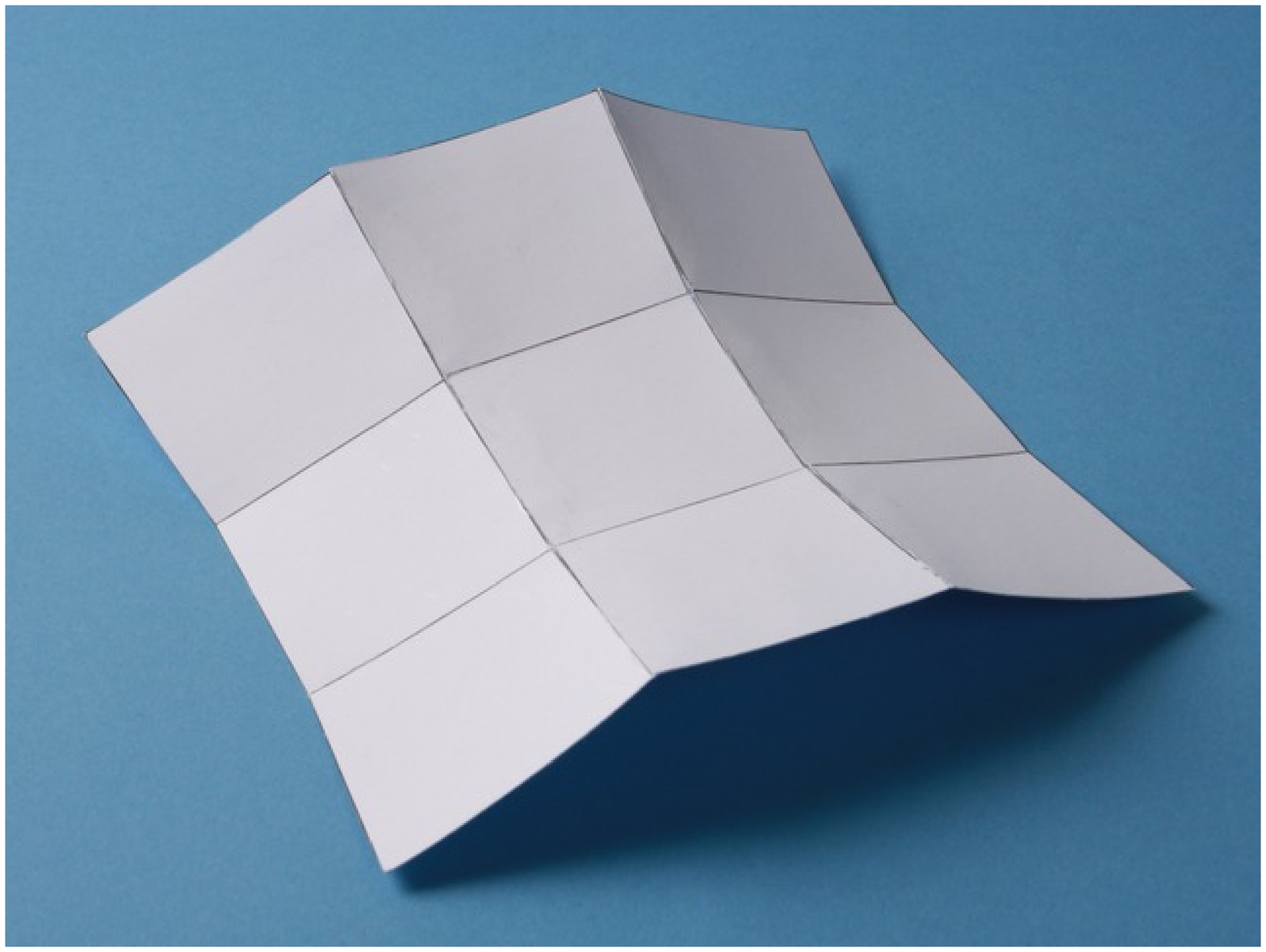}
    \label{fig.flaecheneg}}\\
  \subfigure[]{%
    \includegraphics[width=\hb]{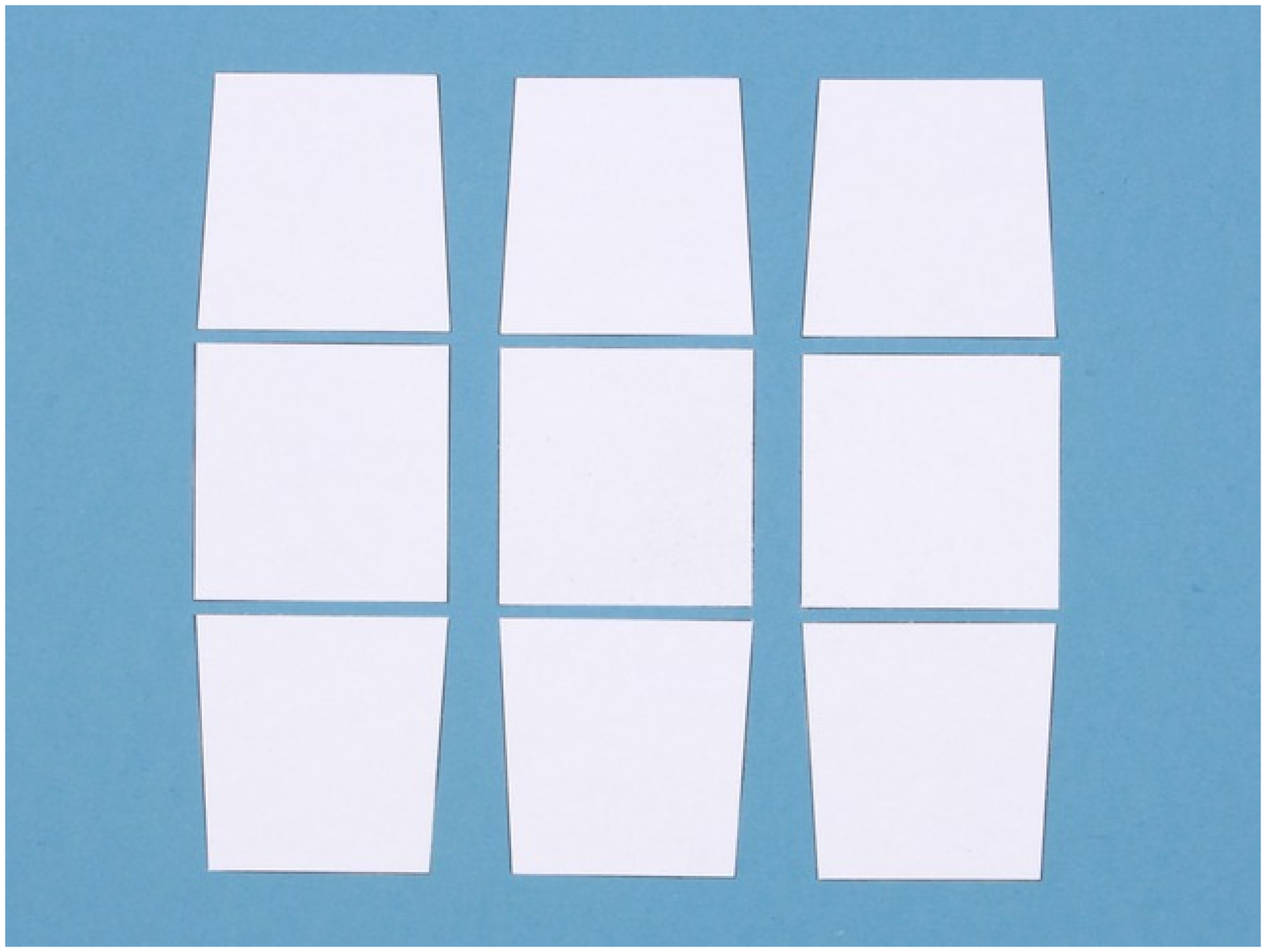}
    \label{fig.flaechepos-sk}}\hfill%
  \subfigure[]{%
    \includegraphics[width=\hb]{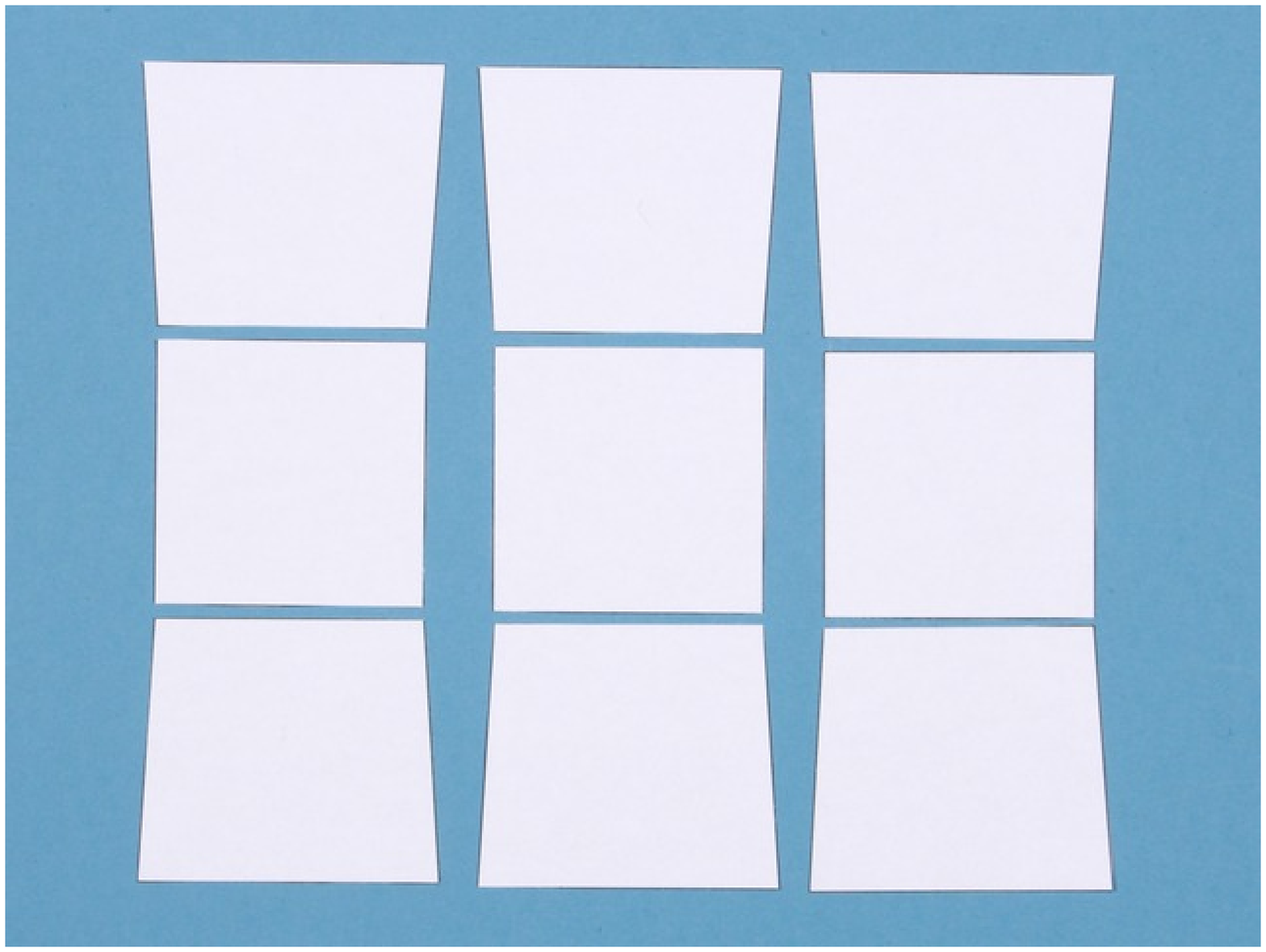}
    \label{fig.flaecheneg-sk}}\\
  \subfigure[]{%
    \includegraphics[width=\hb]{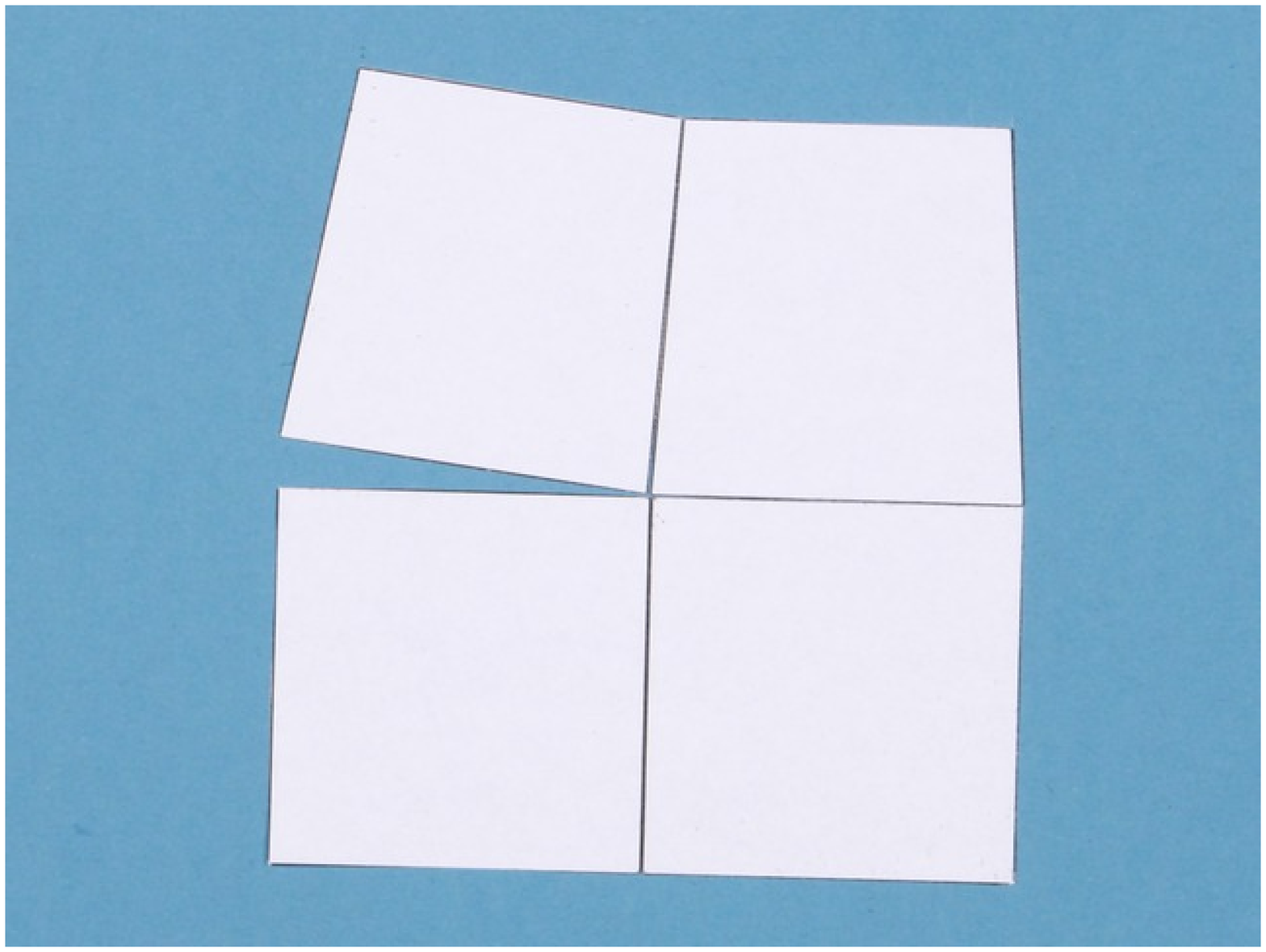}
    \label{fig.flaechepos-def}}\hfill%
  \subfigure[]{%
    \includegraphics[width=\hb]{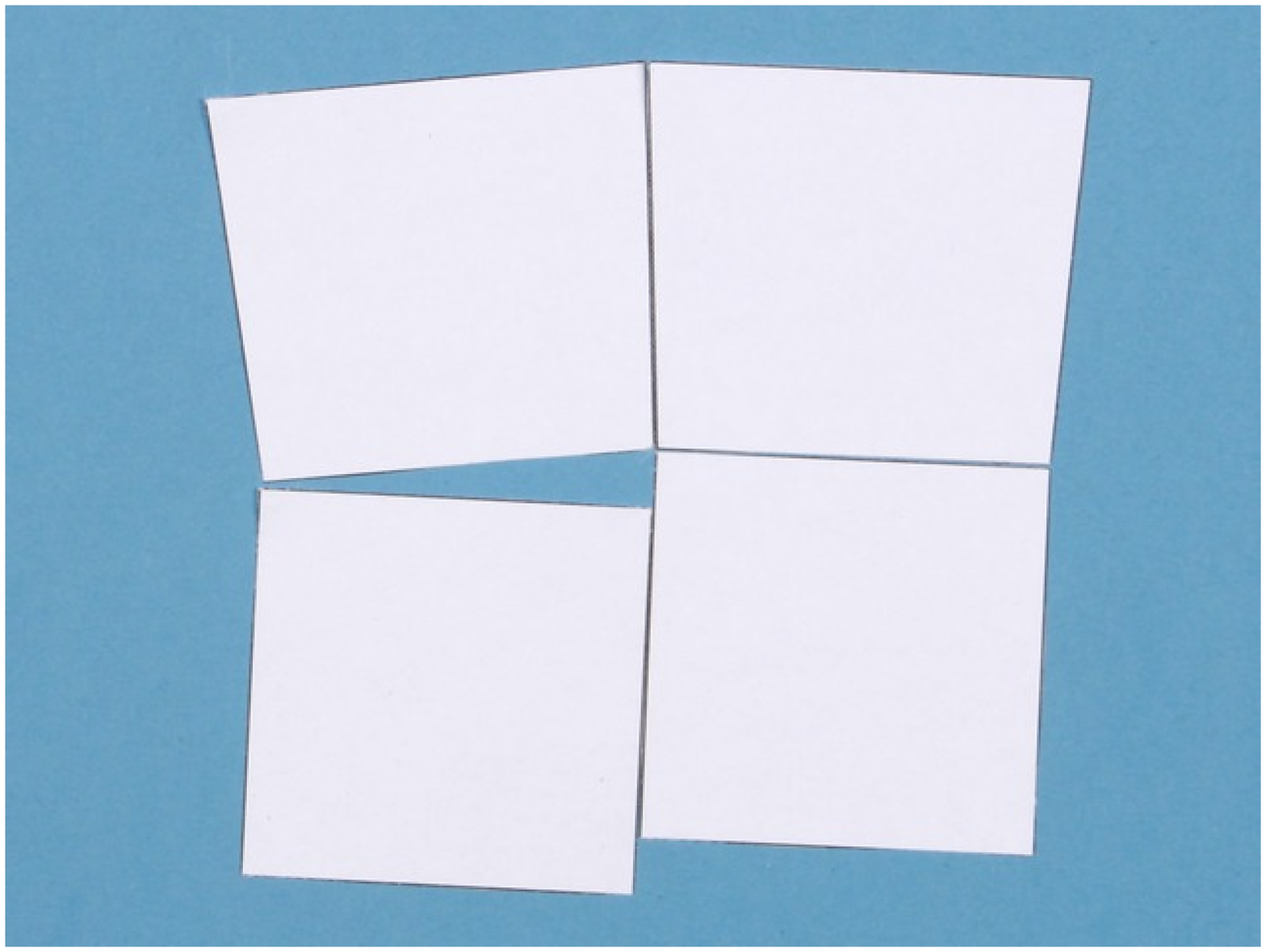}
    \label{fig.flaecheneg-def}}\\
  \caption{\label{fig.flaecheposneg} 
    Durch ebene Flächenstücke
    angenäherte gekrümmte Flächen mit positiver Krümmung (a) bzw.
    negativer Krümmung (b).
    Die zugehörigen Sektormodelle (c, d) zeigen die
    positive Krümmung durch
    \glqq Aufreißen\grqq\ (e) und die negative Krümmung durch
    \glqq Faltenwerfen\grqq\ (f) an einem Vertex an.
}
\end{figure}

Der zweite Schritt illustriert, wie eine gekrümmte Fläche durch
kleine ebene Flä\-chen\-stücke angenähert werden kann. Dazu dienen
die beiden Bastelbögen in Abb.~\ref{fig.bb-flaeche}. Eine Gruppe
erhält die Aufgabe, aus den Bastelbögen Flächen zusammenzukleben
(Abb.~\ref{fig.flaechepos}, \ref{fig.flaecheneg}) und jeweils das Vorzeichen der Krümmung zu
bestimmen. Eine zweite Gruppe wird beauftragt, die Flächenstücke
zwar auszuschneiden, aber nicht zusammenzukleben, sondern auf dem
Tisch auszulegen (Abb.~\ref{fig.flaechepos-sk}, \ref{fig.flaecheneg-sk}) und das Vorzeichen der
Krümmung daraus zu bestimmen. Das Kriterium \glqq Aufreißen oder
Faltenwerfen\grqq\ ist auch in dieser Darstellung leicht anwendbar
(Abb.~\ref{fig.flaechepos-def}, \ref{fig.flaecheneg-def}).

Im dritten Schritt geht es darum, sich auf die Darstellung einer
gekrümmten Fläche durch ausgelegte Flächenstücke
(Abb.~\ref{fig.flaechepos-sk}, \ref{fig.flaecheneg-sk}) einzulassen und zu erkennen, dass man auf die
vertraute Darstellung der Fläche im umgebenden dreidimensionalen
Raum (Abb.~\ref{fig.flaechepos}, \ref{fig.flaecheneg}) sogar ganz verzichten kann.
Um auf diese neue Sichtweise einzustimmen, schildern wir die Weltsicht
der \glqq Flächenwesen\grqq, die Edwin Abbott 
in seinem Roman \glqq Flächenland\grqq\ \citep{abb1884}
beschreibt:

{\em Ich nenne unsere Welt Flächenland (...).
Stellt euch ein weitausgedehntes Blatt Papier vor,
auf dem sich (...) Figuren (...) frei hin- und herbewegen,
jedoch ohne das Vermögen, sich darüber hinaus zu
erheben oder darunter zu sinken, etwa wie Schatten (...)
}  \citep[deutsche Übersetzung:][]{abb1982}.

Die Flächenwesen bewegen sich in zwei Dimensionen (vor -- zurück,
rechts -- links), die dritte (oben -- unten) ist ihnen nicht nur
unzugänglich, sondern liegt jenseits ihres Vorstellungsvermögens.
Wenn wir die Abbott'sche Flächenwelt auf eine gekrümmte Fläche
mit Hügeln und Tälern erweitern, bewegen sich auch hier die
Flächenwesen nur in der Fläche, vor -- zurück und rechts --
links. Ohne einen Begriff für oben und unten ist die Wölbung der
Fläche in den umgebenden dreidimensionalen Raum für die
Flächenwesen nicht vorstellbar. Dennoch können sie die Krümmung
ihrer Welt untersuchen. Dazu fertigen sie eine Darstellung wie in
Abb.~\ref{fig.flaechepos-sk}, \ref{fig.flaecheneg-sk} an:
Ein Gebiet wird in Stücke eingeteilt, die so klein sind,
dass sie näherungsweise eben sind. Die Kantenlängen der Stücke
werden ausgemessen; ein verkleinertes Modell wird zugeschnitten und 
als Sektormodell 
in der Ebene ausgelegt 
(entsprechend Abb.~\ref{fig.flaechepos-sk}, \ref{fig.flaecheneg-sk}). 
Nur wenn das
vermessene Gebiet eben war, fügen sich die Stücke lückenlos
zusammen.  Andernfalls kann man durch Zusammenschieben benachbarter Stücke
um ihren gemeinsamen Eckpunkt 
(wie in Abb.~\ref{fig.flaechepos-def}, \ref{fig.flaecheneg-def})
herausfinden, ob die Fläche dort positiv oder negativ gekrümmt ist.

\subsection{Gekrümmte Räume}

\label{sec.gekruemmte-raeume}

\begin{figure}
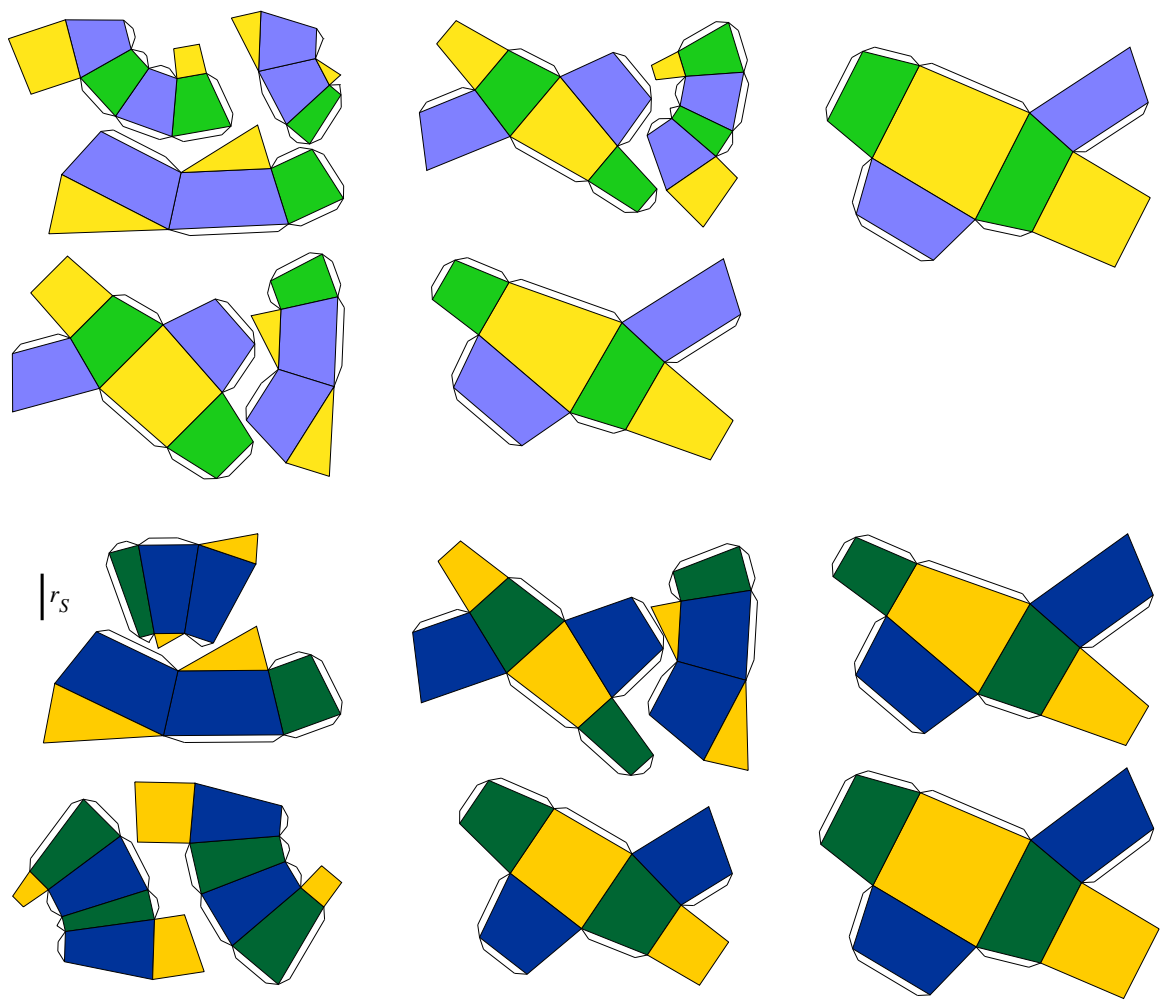

  \centering
  \includegraphics[width=\db]{\bilder/figure3a.eps}
  \hfill
  \includegraphics[width=\db]{\bilder/figure3b.eps}
  \hfill
  \includegraphics[width=\db]{\bilder/figure3c.eps}
\\
  \includegraphics[width=\db]{\bilder/figure3d.eps}
  \hfill
  \includegraphics[width=\db]{\bilder/figure3e.eps}
  \hfill
  \includegraphics[width=\db]{\bilder/figure3f.eps}
\\
  \caption{\label{fig.bb-sl-3d}
  Bastelbögen f\"ur die Pappmodelle \glqq ungekrümmter Raum\grqq\ (oben) und
           \glqq Schwarzes Loch\grqq\ (unten).
   Der Schwarzschildradius $\rs$ des Schwarzen Lochs 
   ist als Balken markiert.
}
\end{figure}

Im vierten Schritt untersuchen wir nun einen dreidimensionalen
gekrümmten Raum. Als \glqq Raumwesen\grqq, die drei Dimensionen
kennen, von höherdimensionalen Räumen aber keine Vorstellung
haben, können wir die Krümmung unseres dreidimensionalen Raums auf
die gleiche Weise untersuchen wie die Flächenwesen die Krümmung
ihrer Flächenwelt: 
Wir teilen ein Gebiet in Stücke ein (dreidimensional: Klötzchen),
die so klein sind, dass der Raum in ihrem Inneren näherungsweise
ungekrümmt ist. Die Kantenlängen werden ausgemessen, die
Klötzchen als verkleinertes Modell nachgebaut und im ungekrümmten
Raum zusammengesetzt. Nur wenn 
die Krümmung des vermessenen Gebiets null ist,
passen die Klötzchen lückenlos zusammen. Ansonsten 
zeigt das Modell die Krümmung des Raums an.

\begin{figure}
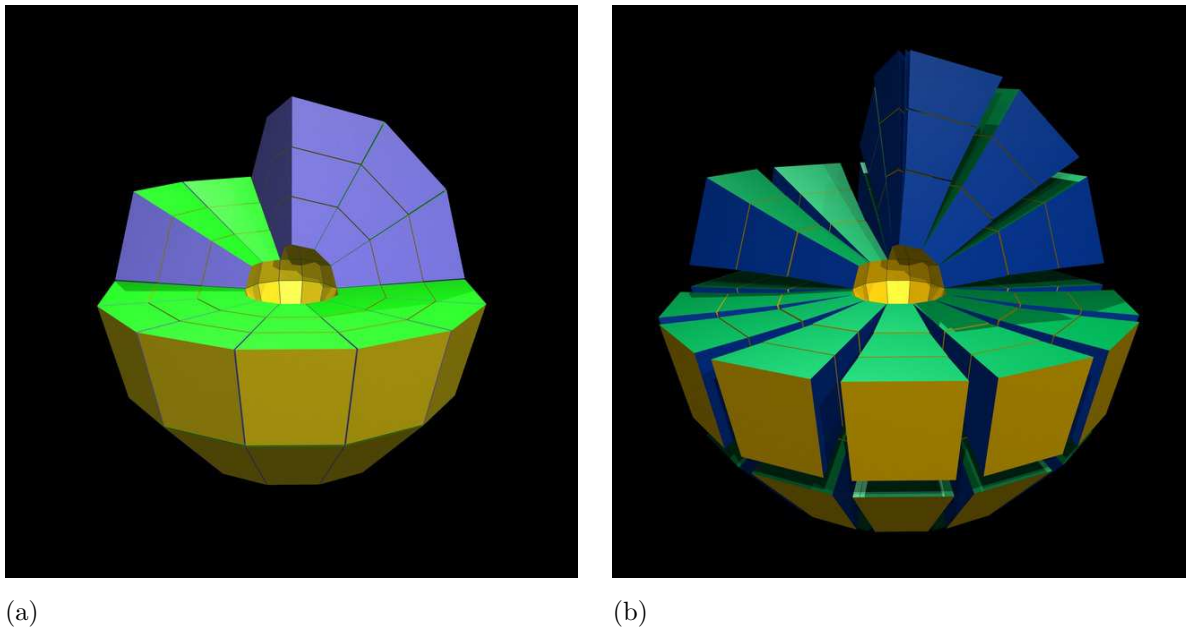

\subfigure[]{%
\includegraphics[width=\hb]{\bilder/figure5a.eps}
}%
\hfill
\subfigure[]{%
\includegraphics[width=\hb]{\bilder/figure5b.eps}
}
  \caption{\label{fig.kloetze}
    Sektormodelle \glqq ungekrümmter Raum\grqq\ (a) und
    \glqq Schwarzes Loch\grqq\ (b) als Computergrafiken.
    Angeordnet um ein Schwarzes Loch der 
    entsprechenden Masse würden 
    die Klötzchen des Modells \glqq Schwarzes Loch\grqq,
    so wie sie sind, lückenlos zusammenpassen.
    }
\end{figure}

Zwei solche Klötzchenmodelle -- ein Modell des ungekrümmten Raums
und ein Modell des gekrümmten Raums um ein Schwarzes Loch -- werden
in der Folge aus Bastelbögen von den Teilnehmern selbst gebaut und
untersucht. 
\begin{figure}
  \centering
  \subfigure[]{\includegraphics[width=\hb]{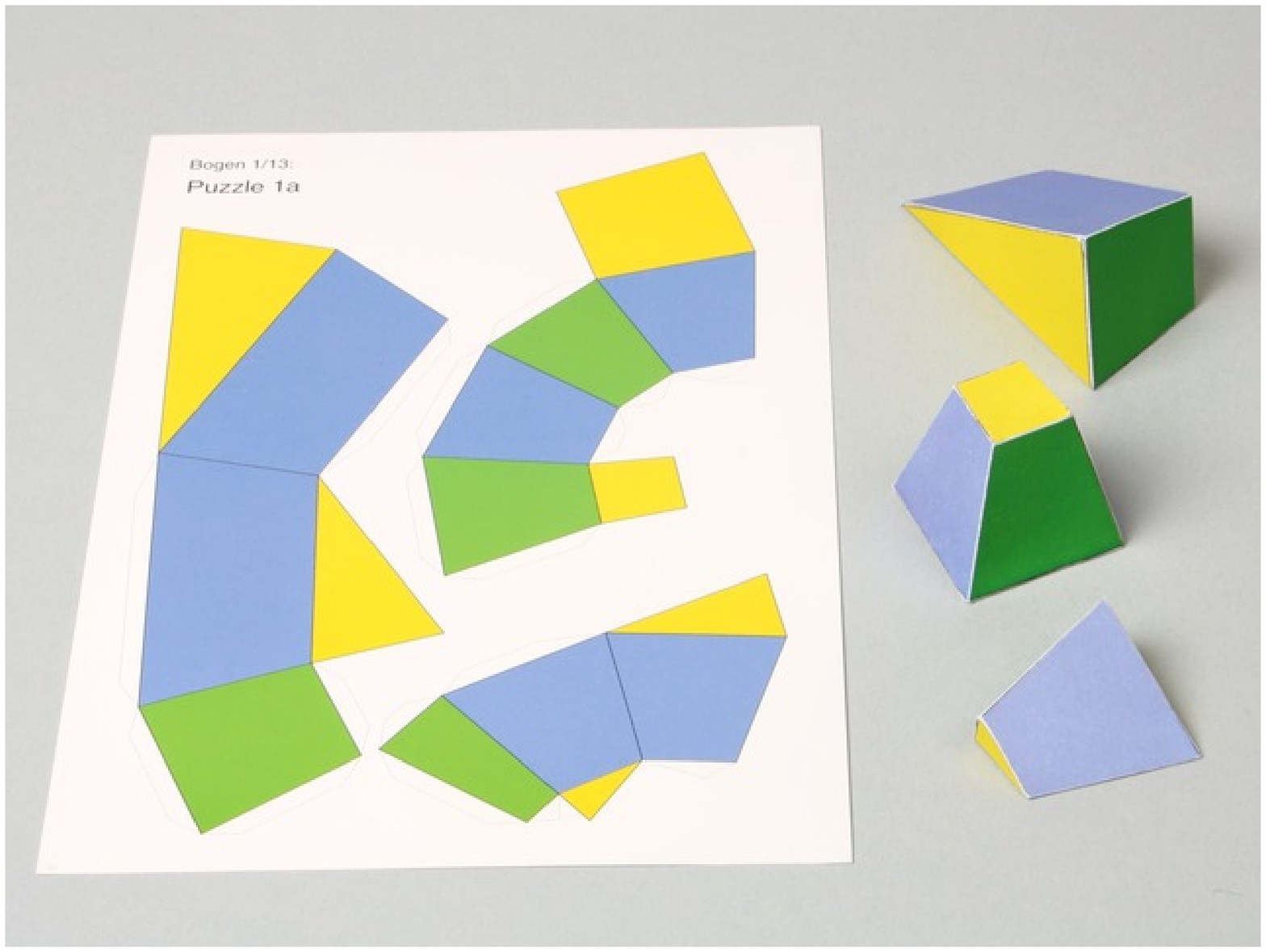}}\hfill
  \subfigure[]{\includegraphics[width=\hb]{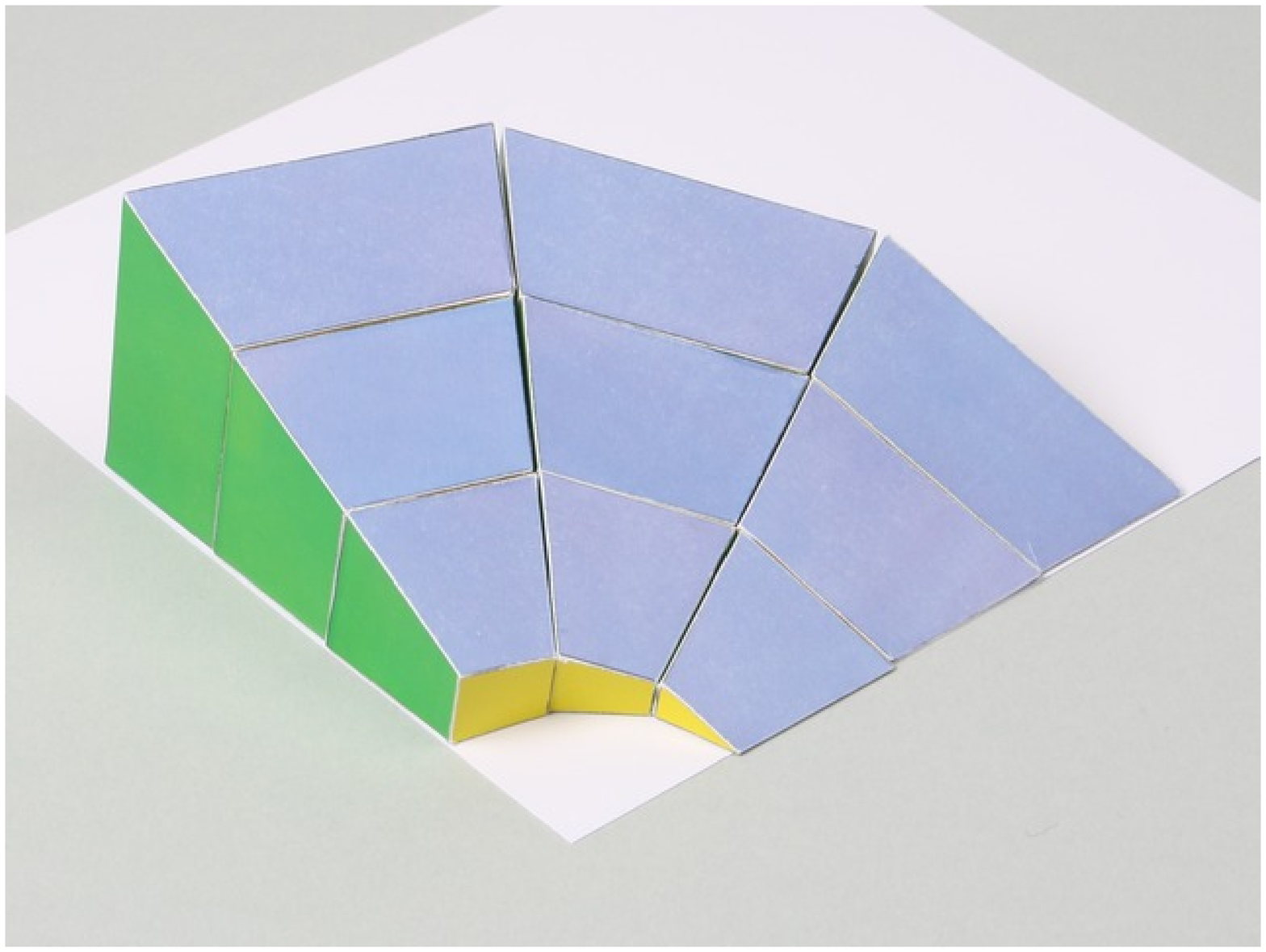}}\\
  \subfigure[]{\includegraphics[width=\hb]{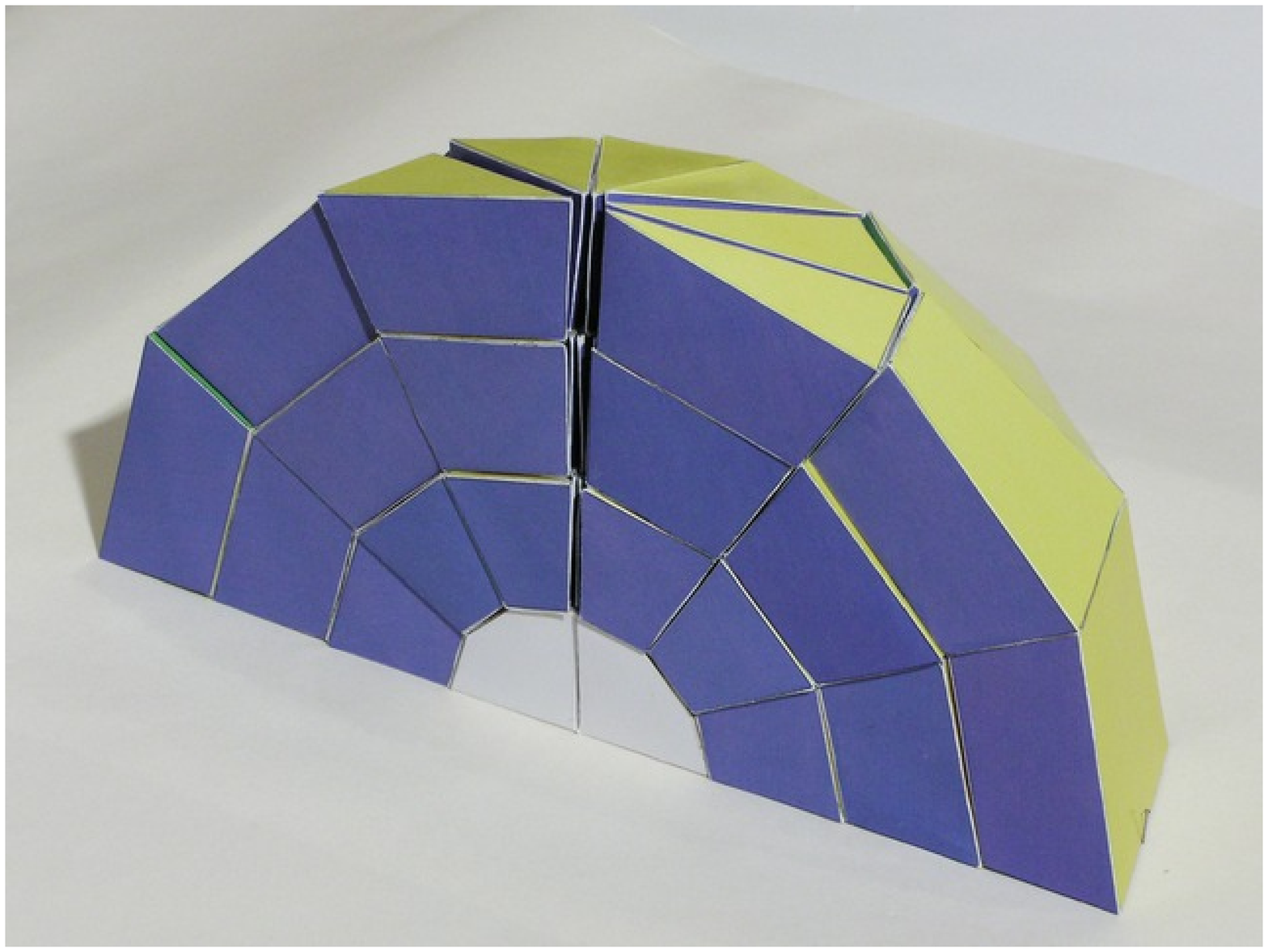}}\hfill
  \subfigure[]{\includegraphics[width=\hb]{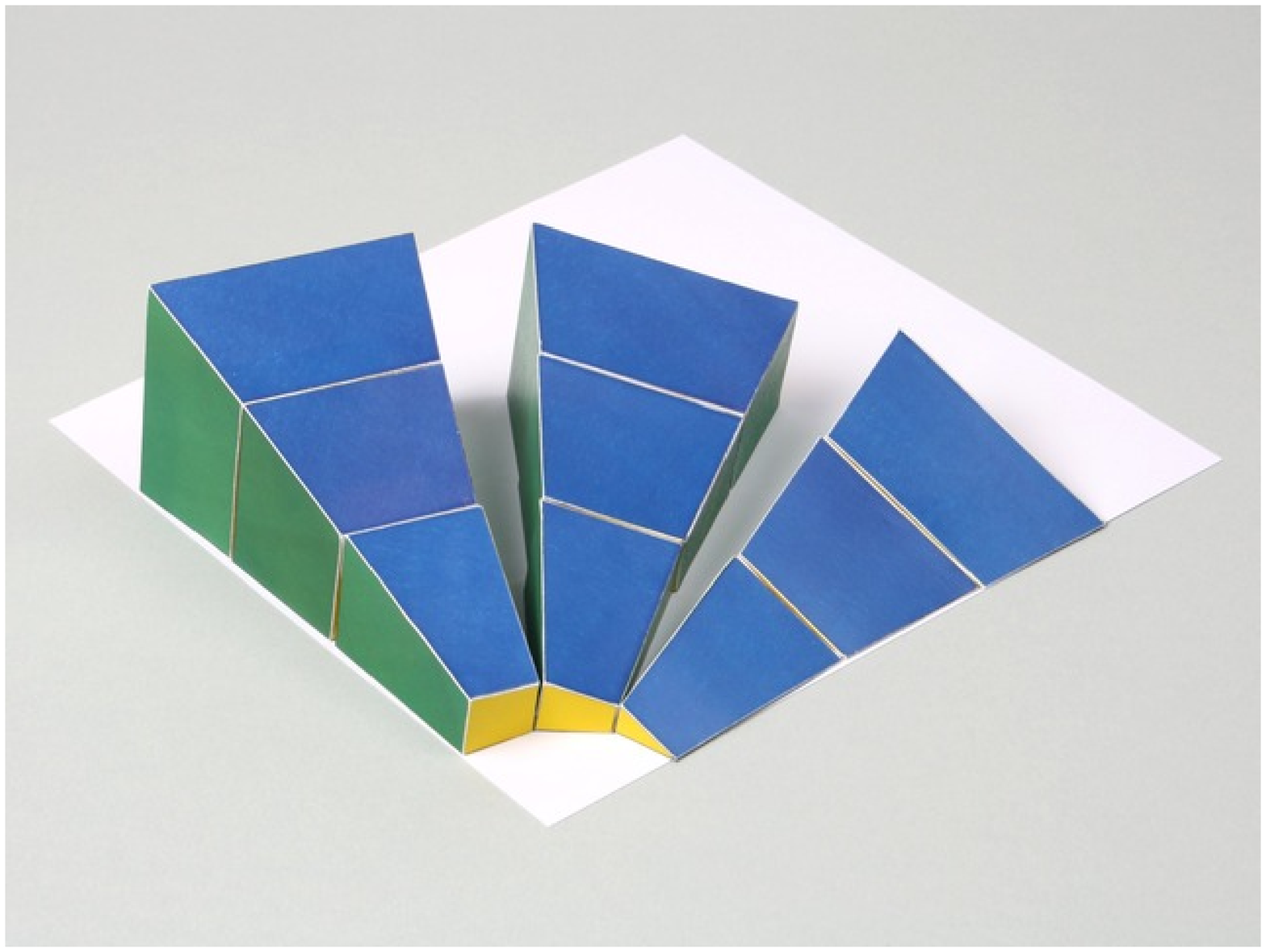}}
  \caption{\label{fig.schnitz}
  Sektormodelle: Bauteile (a), einzelner Schnitz (b)
  und Viertelkugel (c) des Modells \glqq ungekrümmter Raum\grqq,
  einzelner Schnitz des Modells \glqq Schwarzes Loch\grqq\ (d).
}
\end{figure}
Aus dem Bastelbogen
\glqq ungekrümmter Raum\grqq\ (Abb.~\ref{fig.bb-sl-3d} oben)
entstehen neun Klötzchen mit gelben, grünen und blauen Seitenflächen
(Abb.~\ref{fig.schnitz}(a)). Zusammengelegt (gelb auf gelb und grün auf
grün) haben sie die Form eines halbierten
Orangenschnitzes oder eines Tortenstücks
(Abb.~\ref{fig.schnitz}(b)). Nach Möglichkeit sollten mindestens drei
solche Schnitze gebaut werden. Stellt man diese zusammen, so
erkennt man, dass drei Schnitze eine Achtelkugel bilden
(Abb.~\ref{fig.schnitz}(c) zeigt die Erweiterung auf eine Viertelkugel
aus sechs Schnitzen), in deren Zentrum eine kleine Innenkugel
ausgespart ist. In Abb.~\ref{fig.schnitz}(c) befindet sich im
Inneren des Modells ein graues Klötzchen als Stütze, die das Aufstellen
ermöglicht (ein Bastelbogen für die Stütze steht online zur Verfügung,
siehe \citet{zah2014}).
Vierundzwanzig Schnitze ergeben das
vollständige Modell einer hohlen Kugel.
Die Computergrafik
in Abb.~\ref{fig.kloetze}(a) zeigt einen Blick auf das weitgehend
vollständige Sektormodell. Offensichtlich fügen sich alle Klötzchen
lückenlos aneinander, wie es bei einem ungekrümmten Raum zu
erwarten war.

Das zweite Sektormodell stellt ein Gebiet von gleicher Gestalt
dar: Wiederum eine innen hohle Kugel, die nach dem gleichen
Schema in 24~Schnitze zu je 9~Klötzchen eingeteilt ist. Hier aber
befindet sich im Zentrum der Kugel ein Schwarzes Loch, so dass der
Raum in dem betrachteten Gebiet stark gekrümmt ist. 
An dieser Stelle wird
der Grund für die innen ausgesparte Kugel deutlich: Sie
ist etwas größer als der Ereignishorizont des Schwarzen Lochs,
enthält also vollständig dessen Innenbereich.
Das Sektormodell kann nicht über den Ereignishorizont hinweg
fortgesetzt werden, weil der Innenbereich nicht durch ein statisches
Modell darstellbar ist.
Aus dem Bastelbogen
\glqq gekrümmter Raum\grqq\ (Abb.~\ref{fig.bb-sl-3d} unten)  
werden 9~Klötzchen 
gebaut und zu einem Schnitz zusammengelegt (Abb.~\ref{fig.schnitz}(d)).
Das weitgehend vollständige Modell ist in
der Computergrafik in Abb.~\ref{fig.kloetze}(b) zu sehen.
  Dass die Klötzchen des zweiten Modells sich
offensichtlich nicht lückenlos zu einer hohlen Kugel zusammenfügen
lassen, zeigt die innere Krümmung des Raums in diesem Gebiet an.

Lücken treten auf, wenn die Sektoren einer gekrümmten Fläche
in der Ebene ausgebreitet werden bzw. wenn die Klötzchen eines gekrümmten
Raums im un\-ge\-krümm\-ten Raum zusammengesetzt werden.
Befände sich dagegen ein Schwarzes Loch der passenden Masse im Zentrum
des Modells, würden die Klötzchen
{\em so wie sie sind} lückenlos zusammenpassen.
Ebenso wie die Flächenstücke lückenlos zusammenpassen,
wenn sie auf einer Fläche der passenden Krümmung ausgelegt werden.

\begin{figure}
  \subfigure[]{%
    \includegraphics[width=\hb]{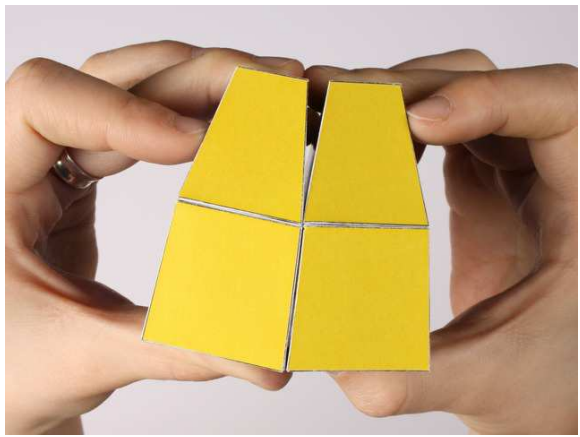}
    \label{fig.kaestchen-theta-phi}%
  }\hfill%
  \subfigure[]{%
    \includegraphics[width=\hb]{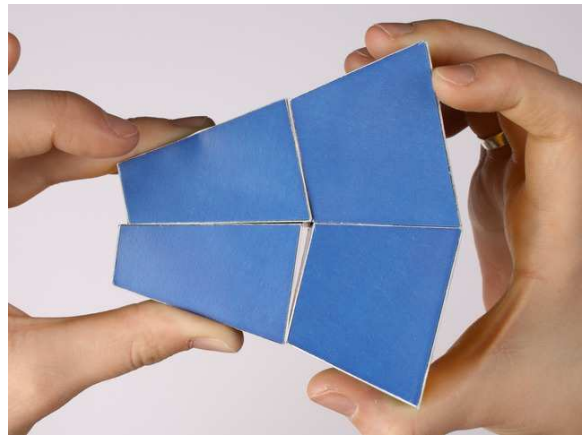}
    \label{fig.kaestchen-r-theta}%
  }\\%
  \subfigure[]{%
    \includegraphics[width=\hb]{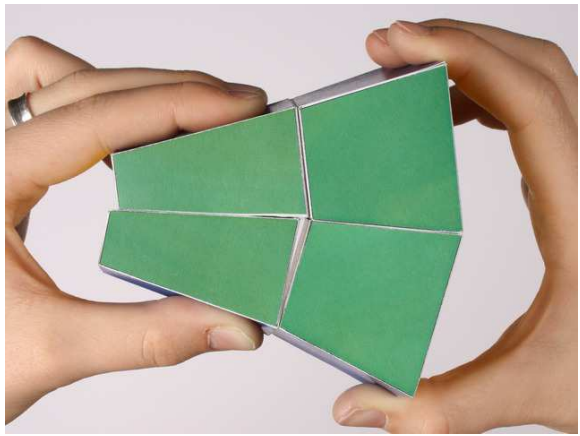}
    \label{fig.kaestchen-r-phi}%
  }\hfill%
  \subfigure[]{%
    \includegraphics[width=\hb]{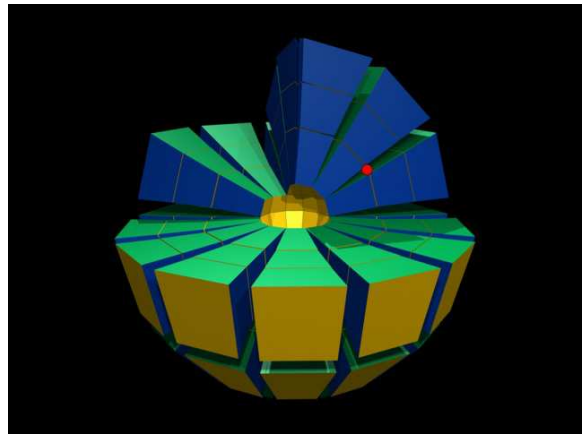}
    \label{fig.kaestchen-vertex}%
  }%
  \caption{\label{fig.kaestchen} 
    Krümmung im dreidimensionalen
    Raum. Der Vertex in der Bildmitte in (a), (b) und (c)
    ist in allen drei Teilbildern derselbe;
    in (d) ist er durch einen roten Punkt gekennzeichnet. 
    Um die radiale Kante ist die Krümmung positiv (a),
    um die tangentialen Kanten negativ (b, c).
}
\end{figure}

Anhand des Sektormodells wird nun das Vorzeichen der Krümmung untersucht.
Dies geschieht in Analogie zum zweidimensionalen Fall: In der
Darstellung von Abb.~\ref{fig.flaechepos-def}, \ref{fig.flaecheneg-def} werden jeweils vier Flächenstücke
um ihren gemeinsamen Eckpunkt angeordnet, 
um die Frage \glqq Aufreißen oder
Faltenwerfen\grqq\ zu beantworten.
Im
räumlichen Modell haben vier Klötzchen nicht einen Punkt, sondern
eine Kante gemeinsam. Ordnet man sie um diese Kante an, dann kann es
passieren, dass ein Schlitz offenbleibt (\glqq Aufreißen\grqq,
positive Krümmung) wie in Abb.~\ref{fig.kaestchen-theta-phi} 
oder aber dass das vierte Klötzchen in die
verbleibende offene Ecke nicht hineinpasst (\glqq Faltenwerfen\grqq,
negative Krümmung) wie in Abb.~\ref{fig.kaestchen-r-theta}.
Der springende Punkt ist nun, dass die drei Teilbilder (a), (b) und (c)
von Abb.~\ref{fig.kaestchen} zu drei 
verschiedenen Kanten {\em am selben Ort} gehören.
D.\,h. die Krümmung an ein und demselben Ort ist je nach
Richtung verschieden und kann auch ein verschiedenes Vorzeichen haben.
Die Krümmung ist in drei
Dimensionen also nicht eine Zahl, sondern eine Größe mit mehreren
Komponenten. 
Es ist eine der Stärken der Sektormodelle, dass sie
diese grundlegende Eigenschaft von gekrümmten 
Räumen mit mehr als
zwei Dimensionen verdeutlichen \citep{zah2008}.

Im Vergleich der beiden Modelle kann man 
ferner aufzeigen, dass die
Gesetze der euklidischen Geometrie in einem gekrümmten
Raum nicht gelten. Zum Beispiel der Zusammenhang zwischen Oberfläche
und Volumen eines Körpers: Die Kugelfläche, die die Modelle nach
außen begrenzt, hat in beiden Fällen denselben Flächeninhalt.
Man sieht das leicht ein, wenn man die Außenseiten einander
entsprechender Klötzchen aneinanderhält; sie sind gleich groß.
Dasselbe gilt für den Flächeninhalt der inneren Kugelfläche. Das
zwischen den beiden Flächen
eingeschlossene Volumen ist jedoch verschieden: Jedes Klötzchen im
Sektormodell \glqq Schwarzes Loch\grqq\
ist in radialer Richtung länger als
das entsprechende Klötzchen 
im Sektormodell \glqq ungekrümmter Raum\grqq,
d.\,h. die Oberfläche
mit demselben Flächeninhalt enthält ein größeres Volumen.
An dieser Stelle wird noch einmal auf 
die Flächenwesen verwiesen, die
in ihrer zweidimensionalen Welt
vergleichbare Feststellungen machen:
Wenn sie auf einem Hügel den Umfang und den Flächeninhalt eines Kreises
bestimmen, dann ist der Flächeninhalt größer als er es bei gleichem
Umfang in der Ebene wäre.

\subsection{Visualisierung der Sektormodelle}

\begin{figure}
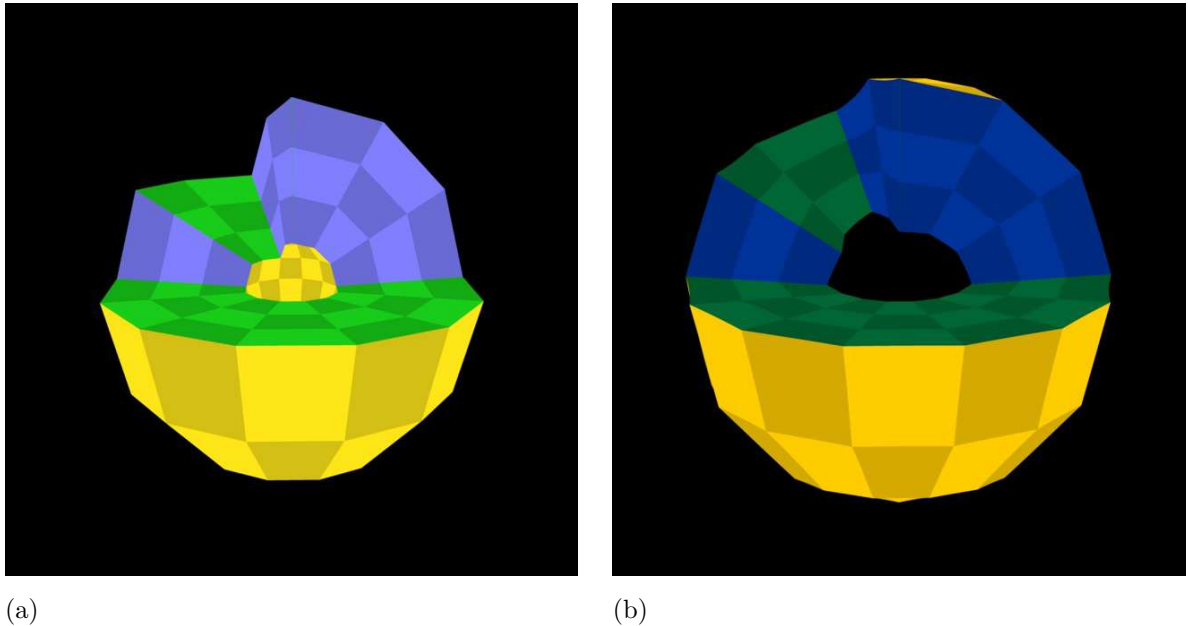

\subfigure[]{%
\includegraphics[width=\hb]{\bilder/figure7a.eps}
}%
\hfill
\subfigure[]{%
\includegraphics[width=\hb]{\bilder/figure7b.eps}
}
\caption{\label{fig.graum1-pdnw}
Anblick des Sektormodells
\glqq an seinem Ursprungsort\grqq:
(a) euklidischer Raum, (b) Schwarzschildraum.
}
\end{figure}

Nach Konstruktion füllt 
jedes Sektormodell
an seinem Ursprungsort 
den Raum lückenlos aus.
Dies gilt auch für das Sektormodell \glqq Schwarzes Loch\grqq,
dessen Sektoren im euklidischen Raum
nicht lückenlos zusammengesetzt werden können.
Man kann die Frage stellen, wie das Sektormodell
an seinem Ursprungsort um ein Schwarzes Loch denn aussehen würde,
wenn jedes einzelne Klötzchen die in
Abb.~\ref{fig.kloetze}(b) gezeigte Form hat
und sich diese Klötzchen lückenlos zu einer Kugel schließen.
Abb.~\ref{fig.graum1-pdnw}(b) zeigt diesen Anblick,
wobei die Anordnung der Klötzchen
dieselbe ist wie in Abb.~\ref{fig.kloetze}(b).
Die Bilder in Abb.~\ref{fig.graum1-pdnw}
wurden mit einem Raytracing-Verfahren berechnet, 
das von jedem Bildpunkt aus den eintreffenden Lichtstrahl
durch die gekrümmte Raumzeit des Schwarzen Lochs
zu seinem Ursprung zurückverfolgt \citep{zah1991}.
Wie erwartet sind zwischen den Klötzchen keine Lücken zu sehen%
\footnote{Um benachbarte Klötzchen im Bild unterscheiden zu können, 
wurden sie leicht unterschiedlich getönt.}.
Im Vergleich von Abb.~\ref{fig.graum1-pdnw}(b)
mit Abb.~\ref{fig.kloetze}(b) fällt neben
gewissen Verzerrungen insbesondere auf, dass
in der Visualisierung
die gelbe Innenfläche des Modells 
nicht sichtbar ist.
Diese Effekte sind 
eine Folge der gravitativen Lichtablenkung 
in der gekrümmten Raumzeit des Schwarzen Lochs.

\section{Die Sektormodelle}

\label{sec.modell}

\subsection{Sektormodelle gekrümmter Flächen}
\label{sec.2dmodell}

Die
in Abschnitt~\ref{sec.gekruemmte-flaechen} 
dargestellte
positiv gekrümmte Fläche  
ist eine Kugelkalotte zum Kugelradius~$R$.
Sie hat die Metrik
\begin{equation}
\label{eq.pos}
\ud s^2 = R^2 \ud \theta^2 + R^2 \cos^2\theta \, \ud\phi^2
\end{equation}
mit der konstanten Gaußschen Krümmung $K=1/R^2$.
Die negativ gekrümmte Fläche
ist eine Sattelfläche mit der Metrik
\begin{equation}
\label{eq.neg}
\ud s^2 = R^2 \ud \theta^2 + R^2 \cosh^2\theta \, \ud\phi^2
\end{equation}
und der konstanten Gaußschen Krümmung $K=-1/R^2$.

Zur Zerlegung in Sektoren 
(Abb.~\ref{fig.flaechepos-sk}, Abb.~\ref{fig.flaecheneg-sk})
werden Vertizes mit den Koordinaten
\begin{eqnarray*}
\theta_i &= -\frac{\pi}{6} + i\cdot \frac{\pi}{9} \qquad & i=0\dots 3,\\
\phi_j   &= j\cdot \frac{\pi}{9} \qquad & j=0\dots 3
\end{eqnarray*}
gewählt.
Die Längen der Sektorkanten
sind die Längen der raumartigen Geodäten zwischen benachbarten
Vertizes
und werden durch Integration des Linienelements längs der Geodäten
bestimmt.
Aus den Kantenlängen und der Bedingung,
dass die trapezförmigen Sektoren spiegelsymmetrisch sein sollen,
folgt eindeutig deren Form.
\begin{figure}
  \centering
  \includegraphics[width=0.3\textwidth]{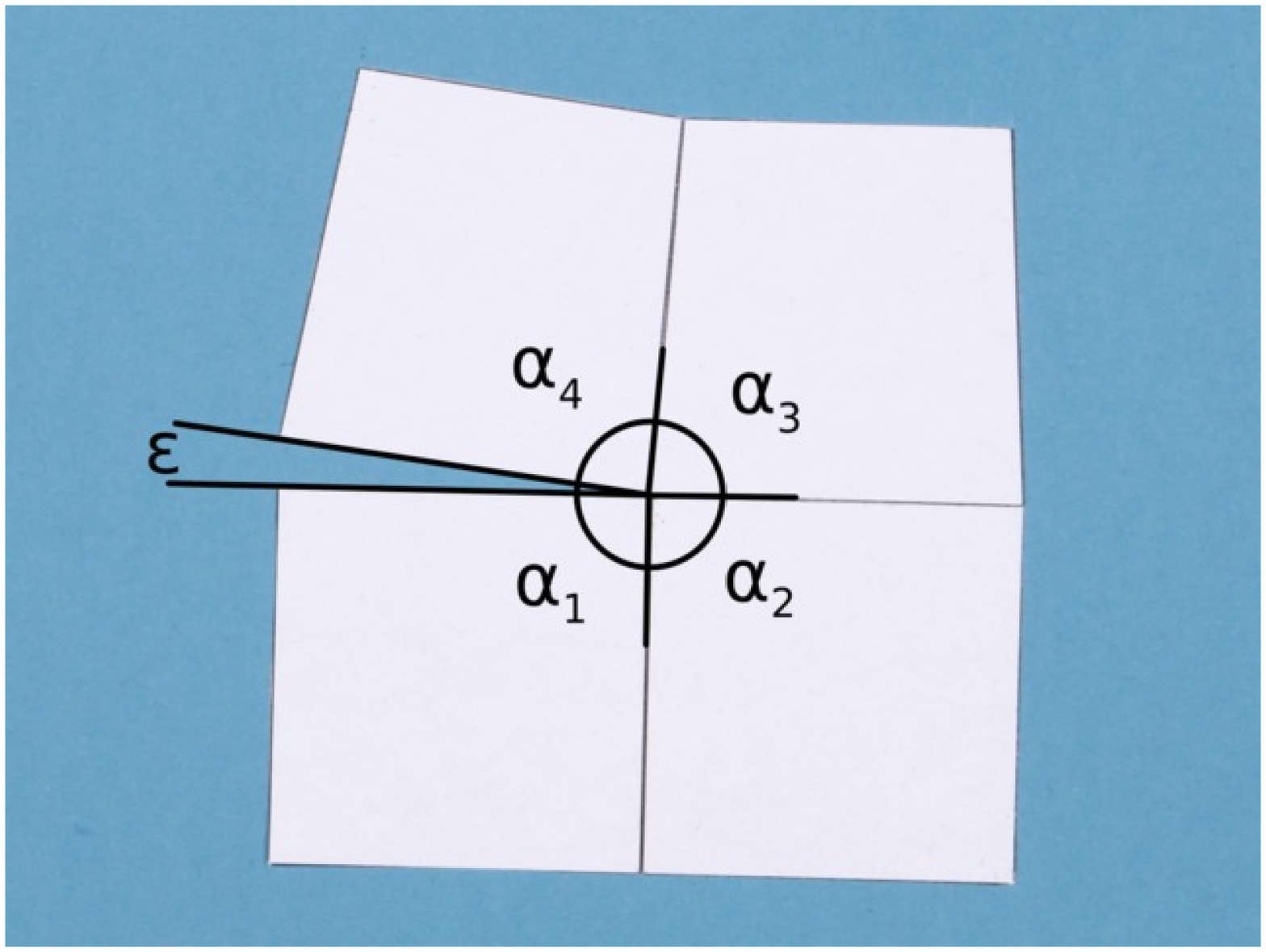}
  \caption{\label{fig.defizit}
Defizitwinkel $\epsilon$ an einem Vertex.
}
  \end{figure}

Der Defizitwinkel an einem Vertex 
(Abb.~\ref{fig.defizit})
ist definiert durch
\begin{equation}
\label{eq.def}
\epsilon = 2\pi - \sum_i \alpha_i,
\end{equation}
wobei die Winkel
$\alpha_i$ die Innenwinkel der an den Vertex angrenzenden Sektoren sind 
und sich die Summe über alle angrenzenden Sektoren erstreckt.
Bei zunehmender Feinheit der Zerlegung 
wird die lokale 
Gaußsche Krümmung der Fläche
durch
\begin{equation}
\label{eq.gauss}
K = \rho \epsilon
\end{equation}
approximiert,
wobei $\rho$ die lokale Dichte der Vertizes ist 
und $\epsilon$ der lokale Defizitwinkel
\citep{reg1961}.

\subsection{Räumliche Sektormodelle}
\label{sec.3dmodell}

In Abschnitt~\ref{sec.gekruemmte-raeume} 
werden 
Sektormodelle des euklidischen Raums
sowie der Umgebung eines Schwarzen Lochs vorgestellt.
Die Metriken sind
\begin{equation}
\label{eq.eukl}
\ud s^2 = \ud r^2 + r^2(\ud \theta^2 + \sin^2\theta \, \ud \phi^2)
\end{equation}
für den euklidischen Raum und
\begin{equation}
\label{eq.ssmr-krumm}
\ud s^2 = \left( 1 - \frac{\rs}{r} \right)^{-1} \ud r^2 + 
r^2(\ud \theta^2 + \sin^2\theta \, \ud \phi^2) 
\end{equation}
für die dreidimensionale raumartige Hyperfläche
der Schwarzschildraumzeit zu kon\-stan\-ter Schwarzschildzeit.
Hier ist $\rs = 2 G M / c^2$ der Schwarzschildradius zur Zen\-tral\-mas\-se $M$,
$G$ die Gravitationskonstante und $c$ die Vakuumlichtgeschwindigkeit.
Die als Online-Ressourcen erhältlichen Bastelbögen \citep{zah2014}
ergeben bei Ausdruck im Format DIN A4 das Sektormodell
für $\rs=2{,}68\, {\rm cm}$,
entsprechend einer Zentralmasse von drei Erdmassen.

Zur Zerlegung in Sektoren 
(Abb.~\ref{fig.schnitz})
werden Eckpunkte
mit den Koordinaten
\begin{eqnarray*}
r_i      &= i\cdot 1{,}25\, \rs \qquad & i=1\dots 4,\\
\theta_j &= j\cdot \frac{\pi}{6} \qquad & j=0\dots 6,\\
\phi_k   &= k\cdot \frac{\pi}{6} \qquad & k=0\dots 11
\end{eqnarray*}
gewählt.
Die Zerlegung
ist auf den Bereich außerhalb von $r=1{,}25\,\rs$
beschränkt. Innerhalb von $\rs$ ist keine statische raumartige
Hyperfläche definierbar.
Genau wie bei den Modellen gekrümmter Flächen
werden die Längen der Sektorkanten
durch Integration des Linienelements
längs raumartiger Geodäten zwischen benachbarten Gitterpunkten bestimmt.
Der Symmetrie der Aufteilung entsprechend
werden die Seitenflächen der Sektoren
durch gleichschenklige Trapeze dargestellt.
Die Form der Trapeze folgt 
aus den Kantenlängen und der Symmetriebedingung.
Ein vollständiges dreidimensionales Modell
besteht aus 216 Sek\-to\-ren.
Bei der gewählten symmetrischen Aufteilung
in 24 identische Schnitze zu je neun Sektoren
liefert ein Schnitz
bereits die vollständige 
Information über die Geometrie.
Die Bastelbögen enthalten
die neun Sektoren 
eines Schnitzes
entsprechend 
$i=1\ldots 4$, $j=0\ldots 3$ und $k=0\ldots 1$.
Die Sektormodelle des euklidischen Raums und des Schwarzschildraums
sind direkte Abbildungen der Metriken
und zeigen (im Rahmen der groben 
Diskretisierung
auch quantitativ
korrekt) deren geo\-me\-tri\-sche Eigenschaften.
Insbesondere ist
in radialer Richtung ein Klötzchen des Schwarzschildraums
entsprechend dem auf"|integrierten 
Metrikfaktor $\int \left( 1 - \rs/r \right)^{-1/2} \ud r$
länger als das entsprechende Klötzchen des euklidischen Raums.

Der Defizitwinkel um eine Kante ist durch
Gleichung~(\ref{eq.def}) definiert.
Beim eu\-kli\-di\-schen Raum ist er null,
beim Schwarzschildraum hängt er von der Orientierung der Kante ab
(Abb.~\ref{fig.kaestchen}).
In einem lokalen Orthonormalsystem
und mit einer Zerlegung längs der Koordinatenlinien
werden
bei zunehmender Feinheit der Zerlegung
die Komponenten $R^{\hat{j}}{}_{\hat{k}\hat{j}\hat{k}}$ 
des Krümmungstensors durch
\begin{equation}
  R^{\hat{j}}{}_{\hat{k}\hat{j}\hat{k}} = \rho \epsilon l
\end{equation}
approximiert,
wobei 
$\Ve_{\hat{i}}$, $\Ve_{\hat{j}}$, $\Ve_{\hat{k}}$
die Basisvektoren sind,
$\rho$ die Dichte der in $i$-Richtung verlaufenden Kanten,
$\epsilon$ ihr Defizitwinkel und $l$ ihre Kantenlänge
\citep{reg1961, mis}.

\section{Raumzeitliche Sektormodelle}

\label{sec.raumzeit}

\begin{figure}
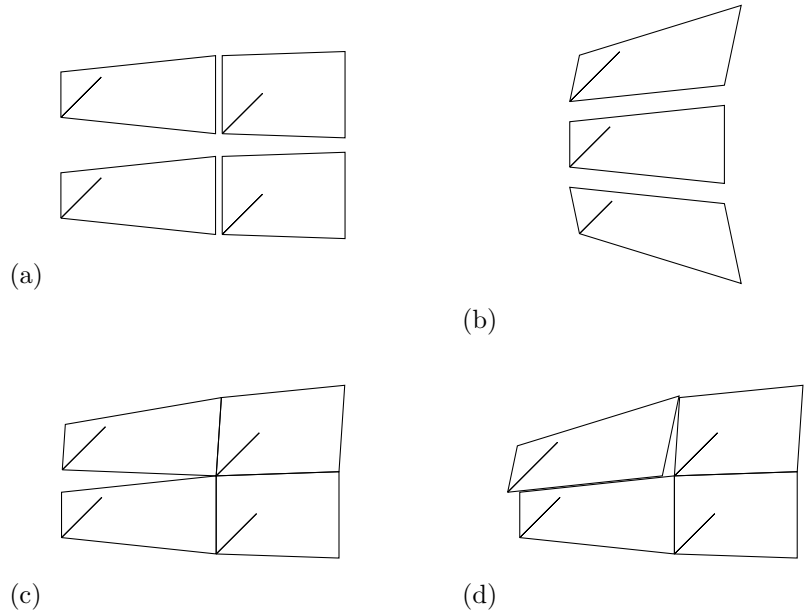


\centering

\subfigure[]{
\begin{minipage}{\db}
\centering
\includegraphics[scale=0.4]{\bilder/figure9a.eps}
\end{minipage}
\label{fig.rtsektoren-sekt-sym}
}%
\qquad
\subfigure[]{
\begin{minipage}{\db}
\centering
\includegraphics[scale=0.4]{\bilder/figure9b.eps}
\end{minipage}
\label{fig.rtsektoren-sektor-lt}
}

\subfigure[]{
\begin{minipage}{\db}
\centering
\includegraphics[scale=0.4]{\bilder/figure9c.eps}
\end{minipage}
\label{fig.rtsektoren-kruemm-3}
}%
\qquad
\subfigure[]{
\begin{minipage}{\db}
\centering
\includegraphics[scale=0.4]{\bilder/figure9d.eps}
\label{fig.rtsektoren-kruemm-2-l}
\end{minipage}
}

\caption{\label{fig.rtsektoren}
(a) 
Raumzeitliches Sektormodell für den $t$-$r$-Unterraum der
   Schwarz\-schild\-me\-trik.
   Die räumliche Koordinatenachse ver\-läuft ho\-ri\-zon\-tal,
    die zeitliche vertikal, die diagonalen Linien
   stellen die Lichtkegel dar.
(b) Verschiedene Darstellungen desselben raumzeitlichen Sektors.
   Sie gehen durch Lo\-rentz\-trans\-for\-ma\-tion ineinander über.
(c) Defizitwinkel in raumartiger Richtung: \glqq Aufreißen\grqq.
(d) Defizitwinkel in zeitartiger Richtung: \glqq Faltenwerfen\grqq. 
}
\end{figure}

Das Konzept der Annäherung einer gekrümmten Fläche 
oder eines gekrümmten Raums 
durch kleine ungekrümmte
Sektoren
lässt sich auf gekrümmte Raumzeiten über\-tra\-gen.
Die ungekrümmten Sektoren haben in diesem Fall Minkowskigeometrie.
Abb.~\ref{fig.rtsektoren} zeigt ein Beispiel für eine solche Zerlegung.
Die gekrümmte Raumzeit ist hier
der zweidimensionale $t$-$r$-Unterraum der Schwarzschildraumzeit
mit der Radialkoordinate $r$ als räumlicher und der Schwarzschildzeit $t$
als zeitlicher Koordinate. 
Seine Metrik ist
\begin{equation}                                                                 
\label{eq.ssm-rt}                                                            
\ud s^2 = - \left (1 - \frac{\rs}{r} \right) \ud t^2 +
            \left (1 - \frac{\rs}{r} \right)^{-1} \ud r^2
\end{equation}
mit dem Schwarzschildradius $\rs$. In diesem Abschnitt 
werden geometrische Einheiten verwendet, so dass die
Vakuumlichtgeschwindigkeit $c=1$ ist.

Zur Zerlegung in Sektoren werden Vertizes mit den Koordinaten
\begin{eqnarray*}
t_i  &= i\cdot 1{,}25\, \rs \qquad & i=0\dots 2,\\
r_j  &= j\cdot 1{,}25\, \rs \qquad & j=1\dots 3
\end{eqnarray*}
gewählt.
Die Geodäten zwischen den Vertizes sind teils raumartig, teils
zeitartig. In beiden Fällen werden
die Längen der Sektorkanten durch Integration des
Linienelements längs der Geodäten bestimmt.
Da die Metrik von der Koordinate $t$ unabhängig ist,
werden in $t$-Richtung identische Sektoren aneinandergefügt.
Jeder Sektor wird durch ein 
gleichschenkliges Trapez
dargestellt, dessen Grundseiten
die zeitartigen Kanten sind.
Mit dieser Symmetriebedingung wird die Form der Trapeze durch
die Kantenlängen festgelegt.
Abb.~\ref{fig.rtsektoren-sekt-sym} zeigt die
graphische Darstellung der Sektoren im Minkowskidiagramm.

Jeder Sektor hat Minkowskigeometrie und es
gelten die Aussagen der Speziellen Re\-la\-ti\-vi\-täts\-theo\-rie
über die Minkowskiraumzeit.
Dies beinhaltet insbesondere die Freiheit
zur Beschreibung von Ereignissen
ein beliebiges Inertialsystem zu wählen,
wobei der Wechsel zwischen Inertialsystemen
durch die Lorentztransformation be\-werk\-stel\-ligt wird.
In der graphischen Darstellung als Minkowskidiagramm
ist den beiden durch Lorentztransformation verbundenen
Bezugssystemen der Lichtkegel gemeinsam,
wäh\-rend
die Zeitachsen gegeneinander verdreht sind,
ebenso, in Gegenrichtung, die Raumachsen. 
Die in Abb.~\ref{fig.rtsektoren-sekt-sym} gezeigten Darstellungen
der Sektoren im Raum\-zeit\-dia\-gramm
entsprechen also jeweils der Wahl eines bestimmten Inertialsystems;
beim Wechsel in ein anderes Inertialsystem
ergibt sich durch Lorentztransformation
eine andere graphische Darstellung
im Minkowskidiagramm (Abb.~\ref{fig.rtsektoren-sektor-lt}).
Im raumzeitlichen Sinne bleiben Form und Symmetrieeigenschaften der
Sektoren dabei jedoch erhalten, da diese auf dem unter
Lorentztransformation invarianten Skalarprodukt basieren.
Um die Krümmung zu untersuchen, sollen nun
wiederum  Sektoren um einen gemeinsamen Vertex
zusammengelegt werden.
Im räum\-li\-chen Fall werden die Sektoren gedreht, 
um die Kanten benachbarter Sektoren zur Deckung zu bringen.
Im raumzeitlichen Fall tritt an die Stelle der Drehung
eine Lorentztransformation.
Man kann sich klarmachen, dass ein An\-ein\-an\-der\-fü\-gen durch
Drehung in der $t$-$r$-Ebene
nicht möglich ist, wenn man bedenkt, dass
die Lichtkegel der zusammengefügten Sektoren identisch
sein müssen.
Die oben geschilderte \"Anderung der Darstellung
beim Wechsel des Inertialsystems ermöglicht es,
die Kanten benachbarter Sektoren im Minkowskidiagramm
zur Deckung zu bringen
(Abb.~\ref{fig.rtsektoren-kruemm-3}).
Ist die zerlegte Raumzeit ein globaler Minkowskiraum,
dann lassen sich die Sektoren 
auf diese Weise
zu einem lückenlosen Ausschnitt der Raumzeit zusammenfügen.
Im obigen Beispiel verbleibt jedoch ein Defizitwinkel um den Vertex
(Abb.~\ref{fig.rtsektoren-kruemm-3}),
dies zeigt eine raumzeitliche Krümmung an.
Wenn, wie in diesem Beispiel,
zwischen zwei raumartigen Kanten
eine Lücke verbleibt,
ist die raumzeitliche Krümmung positiv,
im Falle einer Überlappung wäre sie negativ\footnote{%
Wird die Signatur  ($+$ $-$ $-$ $-$) gewählt,
dann hat die raumzeitliche Krümmung das
entgegengesetzte Vorzeichen:
Sie ist negativ (positiv), wenn zwischen zwei raumartigen
Kanten eine Lücke verbleibt (eine Überlappung auftritt).
}.
Setzt man die Sektoren so zusammen,
dass der Defizitwinkel zwischen zwei zeitartigen Kanten
liegt,
resultiert im obigen Beispiel eine Überlappung
(Abb.~\ref{fig.rtsektoren-kruemm-2-l}).
Die Lücke zwischen den raumartigen und die Überlappung der zeitartigen Kanten
entsprechen derselben Lorentztransformation,
die nötig wäre um die an die Lücke angrenzenden Kanten
zur Deckung zu bringen.

Ähnlich wie bei den gekrümmten Räumen
hängt auch hier der Defizitwinkel mit der ent\-spre\-chen\-den
Komponente des Krümmungstensors zusammen:
In einem lokalen Orthonormalsystem
und mit einer Zerlegung längs der Koordinatenlinien
wird
bei zu\-neh\-men\-der Feinheit der Zerlegung
die Komponente
$R^{\hat{t}}{}_{\hat{r}\hat{t}\hat{r}}$ 
durch
\begin{equation}
  R^{\hat{t}}{}_{\hat{r}\hat{t}\hat{r}} = \rho \alpha
\end{equation}
approximiert, wobei $\Ve_{\hat{t}}$ und $\Ve_{\hat{r}}$ 
die zeitartigen bzw. raumartigen Basisvektoren sind,
$\rho$ die lokale Flächendichte der Vertizes
und $\alpha$ 
mit $\beta=\tanh{\alpha}$ 
der lokale Ge\-schwin\-dig\-keits\-pa\-ra\-me\-ter der
dem Defizitwinkel entsprechenden Lorentztransformation
zur Geschwindigkeit~$\beta$.

\section{Fazit und Ausblick}

\label{sec.fazit}

Sektormodelle sind Anschauungsmodelle,
die zwei- und dreidimensionale gekrümmte Räume bzw. Raumzeiten
maßstabsgetreu darstellen.
Die Darstellung ist koordinatenfrei und daher der geometrischen
Anschauung unmittelbar zugänglich.
Wir haben gezeigt, wie mit Hilfe solcher Modelle
die Krümmung eines dreidimensionalen Raums sowie die
Krümmung einer Raumzeit auf anschauliche Weise dargestellt
werden kann. 
Dies ist unsere Antwort auf die erste der anfangs gestellten Fragen,
was man sich denn unter einem gekrümmten Raum vorzustellen habe.
Die Arbeit mit den Sektormodellen kann für sich allein stehen,
kann aber auch 
eine Ergänzung zur üblichen Einführung
des Riemannschen Krümmungstensors sein.

Der in Abschnitt~\ref{sec.unterricht} beschriebene Workshop 
zur Einführung
gekrümmter Räume anhand von Sektormodellen
wurde mit Studierenden,
Schüler/innen der Oberstufe und erwachsenen Interessenten
erprobt \citep{zah2004,kra2005,zah2010,zah2012}.
Der vorgestellte Ablauf 
sowie
die verwendeten Materialien sind
in einem mehrfach durchlaufenen Zyklus aus Erprobung und Überarbeitung
entstanden.
Seit einigen Jahren 
wird dieser Workshop
regelmäßig in der Lehramtsausbildung
an der Universität Hildesheim eingesetzt.

Im Vergleich mit dem in der Einleitung erwähnten Zugang zu gekrümmten
Räumen über die Eigenschaften von prototypischen gekrümmten Flächen,
bietet der
Einsatz von Sektormodellen eine erhebliche Erweiterung.
Gekrümmte zweidimensionale Flächen werden üblicherweise
eingebettet in einen dreidimensionalen Raum dargestellt. Eine
Darstellung gekrümmter dreidimensionaler Räume ist auf diese Weise
nicht möglich, da der dazu benötigte höherdimensionale Einbettungsraum
unserer Anschauung nicht zugänglich ist. Die Sektormodelle dagegen
benötigen
keine Zusatzdimension. Mit ihnen können wir als Raumwesen die
Geometrie des von uns bewohnten dreidimensionalen Raums untersuchen
und verstehen.
Darüber hinaus kann mit Sektormodellen der 
Tensorcharakter der Krümmung in Räumen von mehr als zwei Dimensionen
verdeutlicht werden. Diese Möglichkeit bieten weder prototypische
gekrümmte Flächen noch Ein\-bet\-tungs\-dia\-gram\-me.

Der Nutzen von Sektormodellen
geht 
über die Veranschaulichung der Krümmung weit hinaus,
wie in einem zweiten Teil
zu diesem Beitrag
gezeigt werden wird.
Sektormodelle erlauben es, 
weitere geometrische Konzepte
auf anschauliche Weise einzuführen 
(z.\,B.\ Paralleltransport, Geodäte)
und physikalische Phänomene mit zeich\-ne\-ri\-schen statt rechnerischen
Mitteln aus der Geometrie der Raumzeit zu folgern
(z.\,B.\ Teilchenbahnen, Rotverschiebung).
Andere Raumzeiten,
die z.\,B.\ Materie enthalten oder nicht statisch sind,
können auf die gleiche Weise beschrieben
werden wie hier die Schwarzschildraumzeit.
Anhand von Sektormodellen lassen sich also auch
die zweite und dritte der anfangs genannten drei Grundfragen,
nämlich die Bewegung der Teilchen sowie der
Zusammenhang zwischen Krümmung und Materiedichte
auf anschauliche Weise beantworten.